\documentclass[twocolumn,times,twocolappendix]{aastex63}
\usepackage{amsmath}
\usepackage{color}

\shorttitle{Systematic Study of AGN Clumpy Tori}
\shortauthors{Ogawa et al.}

\begin{document}

\title{Systematic Study of AGN Clumpy Tori with Broadband X-ray
Spectroscopy: Updated Unified Picture of AGN Structure}

\correspondingauthor{Shoji Ogawa}
\email{ogawa@kusastro.kyoto-u.ac.jp}

\author[0000-0002-5701-0811]{Shoji Ogawa}
\affil{Department of Astronomy, Kyoto University, Kitashirakawa-Oiwake-cho, Sakyo-ku, Kyoto 606-8502, Japan}

\author[0000-0001-7821-6715]{Yoshihiro Ueda}
\affiliation{Department of Astronomy, Kyoto University, Kitashirakawa-Oiwake-cho, Sakyo-ku, Kyoto 606-8502, Japan}

\author[0000-0002-0114-5581]{Atsushi Tanimoto}
\affiliation{Department of Physics, The University of Tokyo, Tokyo 113-0033, Japan}

\author[0000-0002-9754-3081]{Satoshi Yamada}
\affiliation{Department of Astronomy, Kyoto University, Kitashirakawa-Oiwake-cho, Sakyo-ku, Kyoto 606-8502, Japan}

\begin{abstract}

We present the results of a systematic, broadband X-ray spectral
analysis of nearby active galactic nuclei (AGNs) with the X-ray clumpy
torus model (XCLUMPY; \citealt{Tanimoto2019}). By adding 16 AGNs newly
analyzed in this paper, we study total 28 AGNs including unabsorbed and
absorbed AGNs taken from \citet{Ichikawa2015} and
\citet{Garcia2019}. This is the largest sample whose X-ray and infrared
spectra are analyzed by the clumpy torus models XCLUMPY and CLUMPY
\citep{Nenkova2008a,Nenkova2008b}, respectively. The relation between
the Eddington ratio and the torus covering factor determined from the
X-ray torus parameters of each object follows the trend found by
\citet{Ricci2017Nature} based on a statistical analysis. We confirm the
results by \citet{Tanimoto2020} that (1) the torus angular widths
determined by the infrared data are larger than those by the X-ray data
and that (2) the ratios of the hydrogen column density to V-band
extinction ($N_{\rm H}/A_{\rm V}$) along the line of sight in obscured
AGNs are similar to the Galactic value on average. 
Unobscured AGNs show apparently smaller line-of-sight $N_{\rm
H}/A_{\rm V}$ ratios than the Galactic one.
Our findings can be well explained by
an updated
unified picture of AGN structure including a dusty torus, dusty polar
outflows, and dust-free gas, where the inclination determines the X-ray
and optical classifications and observed torus properties in the X-ray
and infrared bands.

\end{abstract}


\section{Introduction}
\label{sec1}

To reveal the structure of an active galactic nucleus (AGN) ``torus'',
which surrounds the accreting supermassive black hole (SMBH), is
indispensable to understand AGN feedback and feeding processes and
thereby the origin of galaxy-SMBH co-evolution (see e.g.,
\citealt{RamosAlmeidaRicci2017} for a recent review). The torus consists of
dusty gas, which absorbs radiation from the nucleus and produces
reprocessed emission. Comparison of X-ray and infrared spectra of an AGN
gives us an important insight into the nature of the torus
\citep[e.g.,][]{Ricci2014A&A,Balokovic2018}. In X-rays, the torus
produces a reflection component accompanied by narrow fluorescence lines
such as Fe K$\alpha$ at 6.4~keV. X-rays trace all material including gas
and dust over a wide range of temperature, and hence are useful to probe
the overall matter distribution around the SMBH.
Whereas, the infrared data carry information on the properties of dust,
which mainly emits blackbody radiation with a temperature of $<$1500 K. 
Thus, by combining the X-ray and infrared spectra, we can investigate
both gas and dust properties. In such studies, it is desirable to apply
self-consistent models to both X-ray and infrared data in terms of the
torus geometry.

Many studies indicate that an AGN torus is not smooth but has a clumpy
structure (see \citealt{Tanimoto2019}). 
To analyse X-ray spectra of AGNs, \citet{Tanimoto2019} have developed
XCLUMPY, 
an X-ray spectral model
from a clumpy torus based on the Monte Carlo simulation for
Astrophysics and Cosmology (MONACO: \citealt{Odaka2016}) framework
(see also \citealt{Buchner2019} for a similar work , who assume the torus geometry like that of \citealt{Tanimoto2019}). 
The definition of clump distribution in XCLUMPY is the same as that in
the CLUMPY \citep{Nenkova2008a, Nenkova2008b} model, which has been
intensively used to analyse the infrared spectra of AGNs
\citep{Alonso-Herrero2011,Ichikawa2015,Garcia2019,Miyaji2019};
clumps are distributed according to power-law and
normal profiles in the radial and angular directions, respectively. 
This enables us to directly compare the results obtained from
X-ray data with XCLUMPY and those from infrared data with CLUMPY, and
hence constrain the geometry of gas and dust distributions.
In these models, 
the number density function $d(r,\theta,\phi)$ in units of pc$^{-3}$,
where $r$ is radius, $\theta$ is polar angle, and $\phi$ is azimuth, is
represented in the spherical coordinate system as:
\begin{eqnarray} 
d(r,\theta,\phi) = N \left(\frac{r}{r_{\mathrm{in}}}\right)^{-1/2} \exp{\left(-\frac{(\theta-\pi/2)^2}{\sigma^2}\right)}.
\end{eqnarray} 
where $N$ is the normalization, $r_{\mathrm{in}}$ is the inner radius of the torus, and $\sigma$ is the torus angular width. 
In XCLUMPY, the inner and outer radii of the torus, the radius of each
clump, the number of clumps along the equatorial plane, and the index of
radial density profile, are fixed at 0.05~pc, 1.00~pc, 0.002~pc, 10.0 and
0.50, respectively \citep{Tanimoto2019}. It has three variable torus
parameters: hydrogen column density along the equatorial plane
($N_{\mathrm{H}}^{\mathrm{Equ}}$), torus angular width ($\sigma_\mathrm{X}$), and
inclination angle ($i_\mathrm{X}$).

Utilizing XCLUMPY, several authors analyzed broadband X-ray spectra of
nearby AGNs: \citet{Ogawa2019} for IC~4329A and NGC~7469 (unobscured AGNs), 
\citet{Tanimoto2019} for Circinus~galaxy (obscured AGN),
\citet{Yamada2020} for UGC~2608 and NGC~5135 (obscured AGNs in luminous
infrared galaxies), and \citet{Tanimoto2020} for 10 obscured AGNs
(IC~5063, NGC~2110, NGC~3227, NGC~3281, NGC~5506, NGC~5643, NGC~5728,
NGC~7172, NGC~7582, NGC~7674).
All of these objects except UGC~2608 are selected from \citet{Ichikawa2015}, who studied their
infrared spectra with the CLUMPY code.
\citet{Ogawa2019} reported that the AGN tori in IC~4329A and NGC~7469 were
dust-rich compared with the Galactic intersteller medium (ISM), by
estimating the ratios of the hydrogen column density ($N_\mathrm{H}$) to V-band extinction ($A_\mathrm{V}$). 
Comparing the X-ray and infrared results of 12 absorbed AGNs
(including Circinus~galaxy and NGC~5135), \citet{Tanimoto2020} found
that the torus angular widths determined from the infrared spectra are
systematically larger than those from the X-ray data, and suggested that
this may be explained by contribution from dusty polar outflows to the
observed infrared flux. They also confirmed that a significant fraction
of AGNs has dust-rich circumnuclear environments. 
However, the samples
used in the previous studies are still limited in size, 
in order to construct a unified view of AGN tori 
by interpreting the results self-consistently.
In particular, including more unobscured (type-1) AGNs to the sample is
critical, because unobscured AGNs are generally viewed at lower inclination
angles than obscured (type-2) ones and hence could provide us with
independent information on the torus structure.

In this paper, we newly analyze the broadband X-ray spectra of 16 AGNs
with the XCLUMPY model in a systematic way. Combining them with the
sample studied by \citet{Tanimoto2020}, we make a sample of total 28 AGNs taken
from \citet{Ichikawa2015} and \citet{Garcia2019}, including 12 unobscured and 16 obscured AGNs. This is
the largest sample whose X-ray and infrared data are analyzed with
XCLUMPY and CLUMPY, respectively. Our aim is to establish a unified
picture of surrounding matter (gas and dust) around SMBHs in AGNs on the
basis of comparison between the X-ray and infrared results. This paper
is organized as follows: Section~\ref{sec2} describes the sample
selection. 
Sections~\ref{sec3} and \ref{sec4} describe the data reduction and X-ray
spectral analysis
for the 16 AGNs newly analyzed in this paper.
The torus properties of the combined sample (28 AGNs)
are discussed in section~\ref{sec5}.
In Section~\ref{sec6}, we interpret our results with an updated unified
picture of AGN structure. We adopt the cosmological
parameters of $(H_{0}, \Omega _{m}, \Omega_{\Lambda}) =
(70~\rm{km~s^{-1}~Mpc^{-1}}, 0.27, 0.73)$ and the solar abundances of
\citet{Anders&Grevesse1989} throughout the paper. Errors on spectral
parameters correspond to 90\% confidence limits for a single parameter of interest.

\section{Sample Selection}
\label{sec2}

Our sample finally consists of 28 AGNs
taken from a combined sample of
\citet{Ichikawa2015} and \citet{Garcia2019}\footnote{
Among the parent sample by \citet{Ichikawa2015} and \citet{Garcia2019}, we
exclude NGC~1365, NGC~1386, NGC~4151, and ESO~005--G004, which show strong
X-ray spectral variability among different observations 
(see \citealt{Ogawa2019} and \citealt{Tanimoto2020}). 
We also do not include NGC~1068 and NGC~4945 because of their too 
complex X-ray spectra, 
and NGC~4138 because it was not detected with the HXD.}, who utilized the CLUMPY code to fit
the nuclear infrared spectra and derived the torus parameters (e.g.,
V-band extinction along the equatorial plane, the torus angular width,
and the inclination).
Among the 28 objects, 2 unobscured AGNs and
12 obscured AGNs come from \citet{Ichikawa2015}, and 10 unobscured 
and 4 obscured ones from \citet{Garcia2019}\footnote{NGC~3081, MCG$-$5--23--16, and Centaurus~A
are included
in both papers; we refer to \citet{Garcia2019} for the infrared
results.}. 
In this paper, we newly analyze broadband X-ray spectra of 16 AGNs (12
unobsucred and 4 obscured AGNs) that were not analyzed in previos works
\citep{Tanimoto2019,Yamada2020,Tanimoto2020}.  Although the results of
IC~4329A and NGC~7469 were already reported in \citet{Ogawa2019}, we
reanalyze their X-ray spectra with a slightly different model to perform
a systematic analysis of the whole sample in a uniform way. The sample
is listed with their basic information in Table~\ref{tab-obj}.

\begin{deluxetable*}{lccccccc}
\tablewidth{\textwidth}
\tablecaption{Properties of the Targets \label{tab-obj}}
\tablehead{
Object & R.A. & Decl. & Redshift & Opt. Class & Group & X-ray Ref. & Infrared Ref.  \\
(1) & (2) & (3) & (4) & (5) & (6) & (7) & (8)
}
\startdata
NGC 2992 & 09h45m42.050s & $-$14d19m34.98s & 0.00771 & 1.9 & Unobscured & a & e \\
MCG$-$5--23--16 & 09h47m40.156s & $-$30d56m55.44s & 0.00849 & 2 & Unobscured & a & e \\
NGC 3783 & 11h39m01.762s & $-$37d44m19.21s & 0.00973 & 1.5 & Unobscured & a & e \\
UGC 6728 & 11h45m16.022s & $+$79d40m53.42s & 0.00652 & 1.2 & Unobscured & a & e \\
NGC 4051 & 12h03m09.614s & $+$44d31m52.80s & 0.00234 & 1 & Unobscured & a & e \\
NGC 4395 & 12h25m48.862s & $+$33d32m48.94s & 0.00106 & 1.8 & Unobscured & a & e \\
MCG$-$6--30--15 & 13h35m53.707s & $-$34d17m43.94s & 0.00775 & 1.2 & Unobscured & a & e \\
IC 4329A & 13h49m19.266s & $-$30d18m33.97s & 0.01605 & 1.2 & Unobscured & a & f \\
NGC 6814 & 19h42m40.644s & $-$10d19m24.57s & 0.00521 & 1.5 & Unobscured & a & e \\
NGC 7213 & 22h09m16.310s & $-$47d09m59.80s & 0.00584 & 1.5 & Unobscured & a & e \\
NGC 7314 & 22h35m46.191s & $-$26d03m01.68s & 0.00476 & 1.9 & Unobscured & a & e \\
NGC 7469 & 23h03m15.623s & $+$08d52m26.39s & 0.01632 & 1.2 & Unobscured & a & f \\ 
\hline
NGC 2110 & 05h52m11.381s & $-$07d27m22.36s & 0.00779 & 2 & Obscured & b & f \\
NGC 3081 & 09h59m29.539s & $-$22d49m34.60s & 0.00798 & 2 & Obscured & a & e \\
NGC 3227 & 10h23m30.5790s & $+$19d51m54.180s & 0.00386 & 1.5 & Obscured & b & f \\
NGC 3281 & 10h31m52.09s & $-$34d51m13.3s & 0.01067 & 2 & Obscured & b & f \\
NGC 4388 & 12h25m46.747s & $+$12d39m43.51s & 0.00842 & 2 & Obscured & a & e \\
Centaurus A & 13h25m27.6152s & $-$43d01m08.805s & 0.00183 & 2 & Obscured & a & e \\
NGC 5135 & 13h25m44.06s & $-$29d50m01.2s & 0.01369 & 2 & Obscured & c & f \\
Circinus Galaxy & 14h13m09.950s & $-$65d20m21.20s & 0.00145 & 2 & Obscured & d & f \\
NGC 5506 & 14h13m14.892s & $-$03d12m27.28s & 0.00618 & 1.9 & Obscured & b & f \\
NGC 5643 & 14h32m40.743s & $-$44d10m27.86s & 0.00400 & 2 & Obscured & b & f \\
NGC 5728 & 14h42m23.897s & $-$17d15m11.09s & 0.00935 & 2 & Obscured & b & f \\
NGC 6300 & 17h16m59.47s & $-$62d49m14.0s & 0.00370 & 2 & Obscured & a & e \\
IC 5063 & 20h52m02.34s & $-$57d04m07.6s & 0.01135 & 2 & Obscured & b & f \\
NGC 7172 & 22h02m01.891s & $-$31d52m10.80s & 0.00868 & 2 & Obscured & b & f \\
NGC 7582 & 23h18m23.500s & $-$42d22m14.00s & 0.00525 & 2 & Obscured & b & f \\
NGC 7674 & 23h27m56.724s & $+$08d46m44.53s & 0.02892 & 2 & Obscured & b & f \\
\enddata
\tablecomments{
The sample is divided into two subgroups with unobscured AGNs (top) and obscured ones (bottom), respectively. 
(1) Galaxy name. (2)--(3) Position (J2000) from the NASA/IPAC extragalactic database (NED). (4) Redshift from the NED. (5) Optical AGN classification from the NED. (6) Sub-group of AGNs based on X-ray obscuration. (7) Reference for the X-ray results delivered by the XCLUMPY model. (8) Reference for the infrared results delivered by the CLUMPY code.}
\tablerefs{(a) This work. (b) \cite{Tanimoto2020}. (c) \cite{Yamada2020}. (d) \cite{Tanimoto2019}. (e) \cite{Garcia2019}. (f) \cite{Ichikawa2015}.}
\end{deluxetable*}

\begin{deluxetable*}{llcccc}
\tablewidth{\textwidth}
\tablecaption{Summary of X-ray Observations \label{tab-obs}}
\tablehead{
Object         &
Satellite      &
ObsID          &
Start Date (UT) &
End Date (UT)   &
Exposure (ks)\tablenotemark{a}
}
\startdata
\multicolumn{6}{c}{Unobscured AGNs}\\
\hline
NGC 2992 & \textit{XMM-Newton}  & 0840920301 & 2019 May 09 21:17 & 2019 May 11 10:07 & 94 \\
 & \textit{NuSTAR} & 90501623002 & 2019 May 10 00:51 & 2019 May 11 10:21 & 57 \\
MCG$-$5--23--16 & \textit{Suzaku}  & 708021010 & 2013 Jun 01 22:06 & 2013 Jun 05 05:55 & 160 \\
 & \textit{NuSTAR} & 60001046002 & 2013 Jun 03 08:21 & 2013 Jun 07 20:21 & 157 \\
NGC 3783 & \textit{XMM-Newton} & 0780861001 & 2016 Dec 21 08:59 & 2016 Dec 22 00:24 & 39 \\
 & \textit{NuSTAR} & 80202006004 & 2016 Dec 21 10:41 & 2016 Dec 21 23:21 & 25 \\
UGC 6728 & \textit{Suzaku}  & 704029010 & 2009 Jun 06 13:33 & 2009 Jun 07 12:18 & 49 \\
NGC 4051 & \textit{XMM-Newton} & 0830430801 & 2018 Nov 09 09:48 & 2018 Nov 10 09:02 & 59 \\
 & \textit{NuSTAR} & 60401009002 & 2018 Nov 04 12:56 & 2018 Nov 11 14:46 & 303 \\
NGC 4395 & \textit{Suzaku}  & 702001010 & 2007 Jun 02 14:30 & 2007 Jun 05 07:09 & 102 \\
MCG$-$6--30--15 & \textit{XMM-Newton} & 069378120 & 2013 Jan 29 12:14 & 2013 Jan 31 01:24 & 95 \\
 & \textit{NuSTAR} & 60001047003 & 2013 Jan 30 00:11 & 2013 Feb 02 00:41 & 127 \\
IC 4329A & \textit{Suzaku}  & 707025010 & 2012 Aug 13 02:13 & 2012 Aug 14 10:53 & 118 \\ 
 & \textit{NuSTAR} & 60001045002 & 2012 Aug 12 16:06 & 2012 Aug 14 13:12 & 162 \\
NGC 6814 & \textit{Suzaku}  & 706032010 & 2011 Nov 02 16:46 & 2011 Nov 03 15:18 & 42 \\
NGC 7213 & \textit{Suzaku}  & 701029010 & 2006 Oct 22 05:34 & 2006 Oct 24 06:37 & 91 \\
NGC 7314 & \textit{XMM-Newton} & 0790650101 & 2016 May 14 13:06 & 2016 May 15 06:44 & 45 \\
 & \textit{NuSTAR} & 60201031002 & 2016 May 13 12:21 & 2016 May 15 18:51 & 100 \\
NGC 7469 & \textit{XMM-Newton} & 0760350201 & 2015 Jun 12 13:36 & 2015 Jun 13 14:50 & 91 \\
 & \textit{NuSTAR} & 60101001002 & 2015 Jun 12 18:41 & 2015 Jun 13 00:40 & 22 \\
\hline
\multicolumn{6}{c}{Obscured AGNs}\\
\hline
NGC 3081 & \textit{Suzaku}  & 703013010 & 2008 Jun 18 21:49 & 2008 Jun 19 19:33 & 44 \\
NGC 4388 & \textit{Suzaku}  & 800017010 & 2005 Dec 24 09:04 & 2005 Dec 27 06:00 & 122 \\
 & \textit{NuSTAR} & 60061228002 & 2013 Dec 27 06:46 & 2013 Dec 27 17:26 & 21 \\
Centaurus A & \textit{Suzaku}  & 708036010 & 2013 Aug 15 04:22 & 2013 Aug 15 10:14 & 11 \\
 & \textit{NuSTAR} & 60001081002 & 2013 Aug 06 13:01 & 2013 Aug 07 16:06 & 51 \\
NGC 6300 & \textit{Suzaku}  & 702049010 & 2007 Oct 17 12:20 & 2007 Oct 19 09:00 & 83 \\
 & \textit{NuSTAR} & 60261001004 & 2016 Aug 24 08:31 & 2016 Aug 24 20:51 & 22 \\
\enddata
\tablenotetext{a}{Based on the good time interval of XIS 0 for \textit{Suzaku}, EPIC-pn for \textit{XMM-Newton}, and FPMA for \textit{NuSTAR}.}
\end{deluxetable*}

\section{Observations and Data Reduction}
\label{sec3}

We analyze the broadband X-ray spectra of 16 AGNs 
that cover the energy band from $\sim$0.3~keV to several tens of ~keV 
observed with \textit{Suzaku} \citep{Mitsuda2007}, \textit{XMM-Newton} \citep{Jansen2001}, and/or \textit{NuSTAR}
\citep{Harrison2013}. The datasets we utilize are chosen in the following way. As for
unobscured AGNs with absorption column densities of $N_{\rm H} \lesssim 10^{22}$ cm$^{-2}$, 
we only use broadband data that were observed 
simultaneously (within a few days) in the energy bands above and below
10 keV, considering their fast spectral variability \citep[see e.g.,][]{Iso2016}.
If \textit{NuSTAR} data are available with simultaneous \textit{Suzaku} or
\textit{XMM-Newton} observations, we analyze \textit{NuSTAR}/Focal Plane Module (FPM) data
plus \textit{Suzaku}/X-ray imaging spectrometer (XIS) or \textit{XMM-Newton}/European
Photon Imaging Camera (EPIC) data; otherwise, we analyze \textit{Suzaku}/XIS and
\textit{Suzaku}/Hard X-ray Detector (HXD) data. As for obscured AGNs, we do not
require that the broadband data were simultaneously observed, and
analyze \textit{NuSTAR}/FPM and \textit{Suzaku}/XIS data. Since NGC~3081 has not been
observed with \textit{NuSTAR}, we utilize the \textit{Suzaku}/XIS and HXD data. The
observation log for the 16 AGNs analyzed in this paper is summarized in
Table~\ref{tab-obs}. Details of data reduction are described below.

\subsection{\textit{Suzaku} }

\textit{Suzaku} carried four X-ray CCD cameras called the XIS, which
covered an energy band below $\approx$10~keV. XIS 0, XIS 2, and XIS 3 are
frontside-illuminated CCDs (XIS-FI) and XIS 1 is a backside-illuminated
one (XIS-BI). \textit{Suzaku} also carried a collimated-type instrument
called the hard X-ray detector (HXD), which was sensitive to photons
above $\approx$10~keV. The HXD consisted of the PIN (10--70~keV) and GSO
(40--600~keV) detectors. We did not analyze any GSO data in this paper, 
because most of our targets were too faint to be detected with the GSO.

We reduced \textit{Suzaku} data in a standard way, utilizing heasoft version~6.26.1 and 
calibration database (CALDB) released on 2018
October 10 (XIS) and 2011 September 13 (HXD).
We reprocessed the unfiltered XIS event data with the \textsc{
aepipeline} script. The XIS source events were extracted from a circular
region with a radius of 3--4 arcmin (depending on the source flux) 
centered at the source position and
the background was taken from a source-free circular region with a radius of
2--3 arcmin. For NGC~4388, we utilized the same background spectrum as
used in \citet{Shirai2008}, which was produced from the data of Arp 220
observed on 2006 January 7--9, 
because there was a largely extended emission 
around the nucleus of NGC~4388. 
We generated the response matrix file (RMF) with \textsc{ xisrmfgen} and
ancillary response files (ARF) with \textsc{ xissimarfgen}
\citep{Ishisaki2007}. 
The spectra of XIS-FIs were co-added, in order to improve the statistics.
We utilized only the data of XIS-FIs, whose effective
area in the iron-K band was larger than that of XIS-BI.

The unfiltered HXD-PIN data were also reprocessed by using \textsc{
aepipeline}.
We made the spectrum of the non-X-ray background (NXB)
using the ``tuned'' background event files
\citep{Fukazawa2009}\footnote{Since the background event file for the
NGC~6814 observation seems to contain an unignorable systematic error,
we utilized the night-earth occultation data as the NXB for this
target.}. 
We added a simulated spectrum of the cosmic X-ray background to the spectrum of NXB.
In the spectral
analysis, only the 16--40~keV range was utilized, where the source flux
is ensured to be brighter than 3\% of the NXB level (the maximum
systematic error in the 15--70~keV range; \citealt{Fukazawa2009}).

\subsection{\textit{XMM-Newton}}

\textit{XMM-Newton} carries three X-ray CCD cameras, one EPIC/pn 
and two EPIC/MOS cameras.
We did not use the
data of MOS cameras, which have much smaller effective area than the pn camera.
We analyzed the pn data
using the Science Analysis Software (SAS) version~17.0.0 and 
calibration file (CCF) released on 2018 June 22. 
We reprocessed the 
data with the \textsc{ epproc} script. The source spectra were extracted
from a circular region with a radius of 40 arcsec centered at the source
peak, and the background from a source-free circular region with a
50 arcsec radius in the same CCD chip. 
The RMF and ARF were generated with
\textsc{rmfgen} and \textsc{arfgen}, respectively.

\subsection{\textit{NuSTAR}}

\textit{NuSTAR} carries two FPMs (FPMA and FPMB), which cover an energy
range of 3--79~keV.  
The FPMs data were analyzed with HEAsoft v6.26.1 and
CALDB released on 2019 October 12. 
The spectra were extracted from a
circular region with a 50--75 arcsec radius (depending on the source
flux) centered at the source peak, and the background was taken from a
nearby source-free circular region with a 60--75 arcsec radius. 
The source spectra, background spectra, RMF, and ARF of the two FPMs
are combined with the \textsc{addascaspec}.

\section{X-ray Spectral Analysis}
\label{sec4}

We perform a simultaneous spectral fit to the \textit{Suzaku}/XIS
(0.5--10~keV), \textit{Suzaku}/HXD-PIN (16--40~keV),
\textit{XMM-Newton}/EPIC-pn (0.3--10~keV), and/or \textit{NuSTAR}/FPMs
(3--70~keV) spectra. The combination of the spectra we adopt 
depends on each target, as explained in Section~\ref{sec3}.  
The spectra folded with the energy responses are plotted in 
the left sides of
Figure~\ref{fig1} 
and~\ref{fig2} for unobscured and obscured AGNs, respectively.
Different spectral models are adopted for unobscured and obscured
AGNs. In both cases, we utilize the XCLUMPY model to represent the
reflection component from the torus. The details of the models are
described below.

\subsection{Model 1: Unobscured AGNs}
\label{sec4.1}

A typical broadband X-ray spectrum of an unobscured AGN consists of (1)
a direct component from the nucleus, which is well represented by a
power-law with an exponential high energy cutoff, (2) reflection
components from the accretion disk (if any) and torus accompanied by
fluorescence emission lines, and (3) a soft excess component. The
spectrum is often subject to absorption by ionized (warm) and/or cold
gas in the line of sight. To explain a hump structure peaked around 30~keV 
and an apparently broad iron-K emission line feature, two major
distinct models have been proposed: one invoking a strong relativistic
reflection component from the inner accretion disk \citep[e.g.,][]{Tanaka1995} and the
other invoking variable partial absorbers \citep[e.g.,][]{Miyakawa2012}. The discussion
which model is physically correct has long been controversial. This
paper does not aim to answer this long-standing question, because our
main goal is to constrain the torus structure. In fact, \citet{Ogawa2019}
obtained similar results on the torus parameters by adopting either of
the two models.

For simplicity, in this paper, we adopt the latter model (partial
covering model). The whole model for unbscured AGNs is expressed in the
XSPEC \citep{Arnaud1996} terminology as follows (Model~1):

\begin{eqnarray}
    \mathrm{Model\ 1} & = & \mathsf{const * phabs} \nonumber\\
    & * &\mathsf{(zphabs * cabs * WA * WA * WA} \nonumber\\
    & * &\mathsf{(zcutoffpl + compTT) } \nonumber\\
    & + &\mathsf{ atable\{xclumpy\_R.fits\} + atable\{xclumpy\_L.fits\}} \nonumber\\
    & + & \mathsf{zgauss)} \nonumber\\
\end{eqnarray}

\begin{enumerate}
\renewcommand{\labelenumi}{(\arabic{enumi})}

\item The \textsf{const} and  \textsf{phabs} terms represent the
cross-calibration constant ($C$)
and the Galactic absorption, respectively.
When we analyze \textit{NuSTAR} data, we set $C=1$
for the XIS-FI or EPIC-pn data as the reference, and make
it free for the \textit{NuSTAR}/FPMs. That for the \textit{Suzaku}/HXD-PIN is fixed 
at 1.16 or 1.18 depending on the target position in the detector
coordinates, based on the calibration using the Crab Nebula. We fix
the hydrogen column density of the Galactic absorption 
($N^{\mathrm{Gal}}_{\mathrm{H}}$) to the total Galactic $HI$ and $H_{2}$ value given by
\citet{Willingale2013}.

\item The  \textsf{zcutoffpl} term represents the direct component (cutoff
power-law), the  \textsf{compTT} term the soft excess (thermal
comptonization model by \citealt{Titarchuk1994} is adopted).
The cutoff energy is fixed at 300~keV, a canonical value for nearby
AGNs \citep{Dadina2008}. 
The photoelectric absorption and Compton scattering are taken into account 
with the  \textsf{zphabs} and  \textsf{cabs} models, respectively.
Its line-of-sight
column density ($N_{\mathrm{H}}^{\mathrm{LOS}}$) is self-consistently determined by the torus parameters 
with the equation:
\begin{equation}
\label{nhlos}
N_{\mathrm{H}}^{\mathrm{LOS}} = N_{\mathrm{H}}^{\mathrm{Equ}} \exp{\left(-\frac{(i_\mathrm{X}-90^{\circ})^2}{\sigma_\mathrm{X} \,^2}\right)}.
\end{equation}
Note that this condition was not considered in the analysis of \citet{Ogawa2019}.
Adopting the model by \citet{Iso2016}, we consider 
three layers of absorption by ionized matter (\textsf{WA}), 
one is a full absorber and two are partial absorbers. 
To model the ionized absorbers, we generate a table
model by running XSTAR version 2.54a \citep{Kallman2001,Bautista2001}
for different values of the 
ionization parameter ($\xi$) and hydrogen column density ($N_{\mathrm{H}}$). 
We adopt the same grid as that used by \citet{Miyakawa2012}, where
$\xi$ and $N_{\mathrm{H}}$ range over $0.1 < \log \xi /\mathrm{erg\,cm\,s}^{-1}< 5$  
and $20 < \log N_{\mathrm{H}} /\mathrm{cm}^{-2} < 25$ 
with 20 logarithmic intervals, respectively.
We assume that the ionized gas has the solar abundances, a temperature of
10$^5$~K, a density of 10$^{12}$~cm$^{-3}$, and a turbulent
velocity of 200~km~s$^{-1}$, and that the incident spectrum 
is a power-law with a photon index of 2.0.
	 
\item The table models (\textsf{atable\{xclumpy\_R.fits\}} and \textsf{
    atable\{xclumpy\_L.fits\}}) correspond to the reflection continuum
    and fluoresence emission lines from the torus, respectively, based
    on the XCLUMPY model. The parameters are the power-law normalization
      at 1 keV, photon index, cutoff
    energy, equatorial hydrogen column density $(N^{\rm Equ}_{\rm H})$,
    torus angular width $( \sigma_\mathrm{X})$, and inclination angle $(i_\mathrm{X})$. 
    We link the 
    photon index, normalization, and cutoff energy to those
    of the direct component. 
    In unobscured AGNs, the observed
    equvalent width of the narrow iron-K$\alpha$ line mainly constrains
    the torus
    parameters \citep{Ogawa2019}. To avoid degeneracy among them, we fix $i_\mathrm{X}$ at $45^{\circ}$ for all
    objects as a typical value of unobscured AGNs (see Appendix~\ref{A2}).
    
\item The \textsf{zgauss} term represents an additional emission line 
      feature bellow 1~keV.

\end{enumerate}

\subsection{Model 2: Obscured AGNs}

We adopt the same spectral model as in \citet{Tanimoto2020} for obscured AGNs.
Compared with Model 1, we ignore ionized absorbers because absorption by
the torus is dominant. Instead, we consider (1) an unabsorbed scattered
component by a surrounding gas of the power-law continuum and (2)
optically-thin thermal emission from the host galaxy, both of which
become unignorable at low energy bands. This model (Model~2) is expressed
in the XSPEC terminology as follows:

\begin{eqnarray}
    \mathrm{Model\ 2} & = & \mathsf{const1 * phabs} \nonumber\\
    & * &\mathsf{(const2 * zphabs * cabs * zcutoffpl } \nonumber\\
    & + &\mathsf{const3 * zcutoffpl + atable\{xclumpy\_R.fits\} } \nonumber\\
    & + &\mathsf{ const4 * atable\{xclumpy\_L.fits\} + apec)} \nonumber\\
\end{eqnarray}

\begin{enumerate}
\renewcommand{\labelenumi}{(\arabic{enumi})}

\item The \textsf{const1} and \textsf{phabs} terms represent the
cross-calibration constant and the Galactic absorption, respectively
(the same as in Model~1; see Section~\ref{sec4.1}). 

\item The \textsf{zcutoffpl} term represents the direct component. 
The \textsf{const2} factor accounts for its possible time
variability between two different observation epochs, which is fixed to
unity for \textit{Suzaku}/XIS. The
\textsf{zphabs * cabs} represents the photoelectric absorption and
Compton scattering by the torus, whose line-of-sight column density is
determined by equation~\ref{nhlos}.

\item The \textsf{const3} factor gives the scattering fraction ($f_\mathrm{scat}$).
The parameters of \textsf{zcutoffpl} are linked to those of the direct
component.

\item The table models of XCLUMPY 
(\textsf{atable\{xclumpy\_R\\.fits\}} and
\textsf{const4 * atable\{xclumpy\_L.fits\}}) correspond to the
reflection continuum and emission line components from the torus,
respectively.
To account for possible systematic uncertainties (due to the 
simplified assumption of the geometry and metal abundances), 
\textsf{const4} is multiplied to the latter table. The photon index,
normalization, and cutoff energy are linked to those of the direct 
component. The values of $N^\mathrm{Equ}_\mathrm{H}$, $\sigma_\mathrm{X}$
,and $i_\mathrm{X}$ are left free.
    
\item The \textsf{apec} term represents optically thin thermal emission
from the host galaxy.  In NGC~3081 and NGC~4388, two different
      temperature components are required.

\end{enumerate}

\begin{figure*}
\plottwo{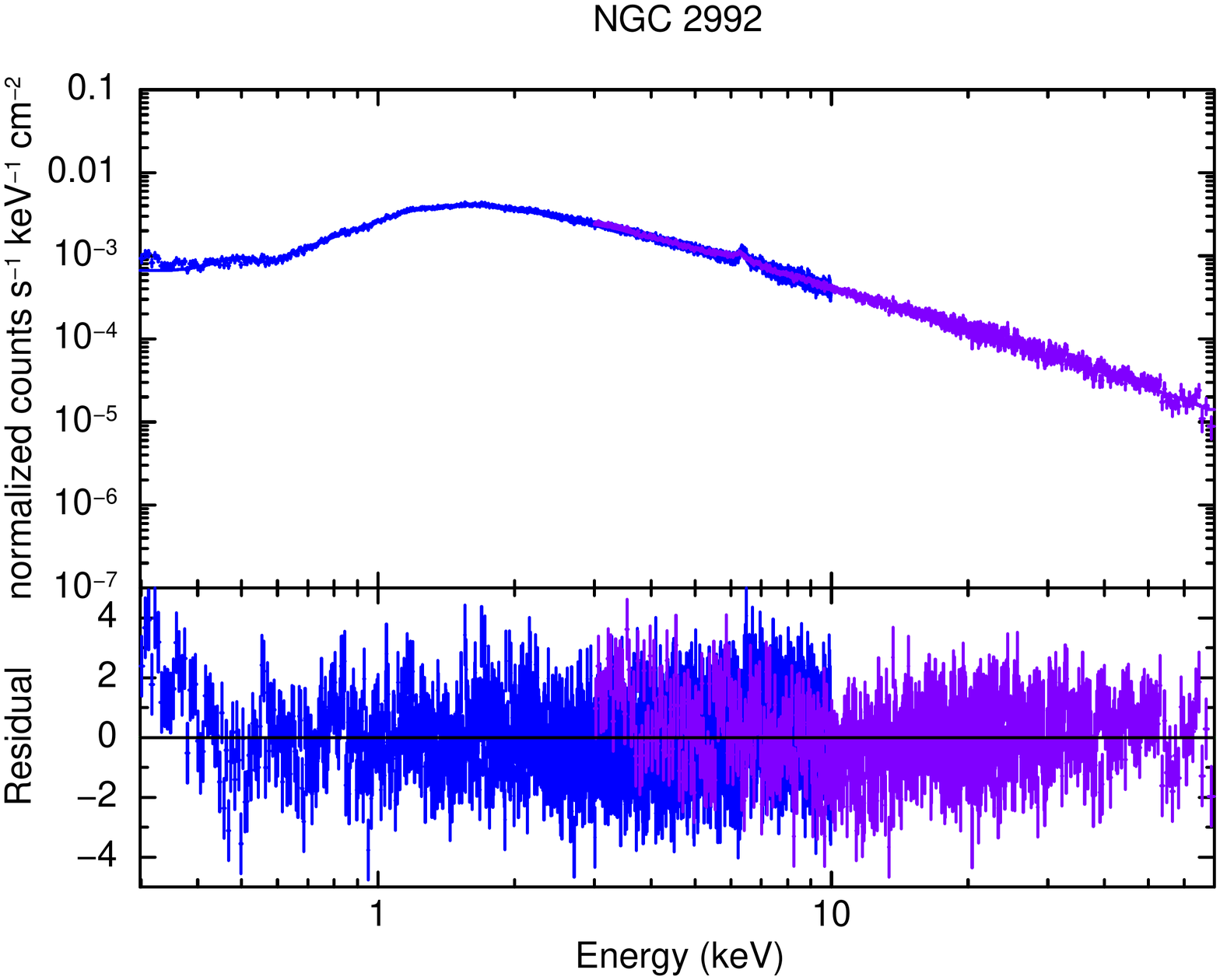}{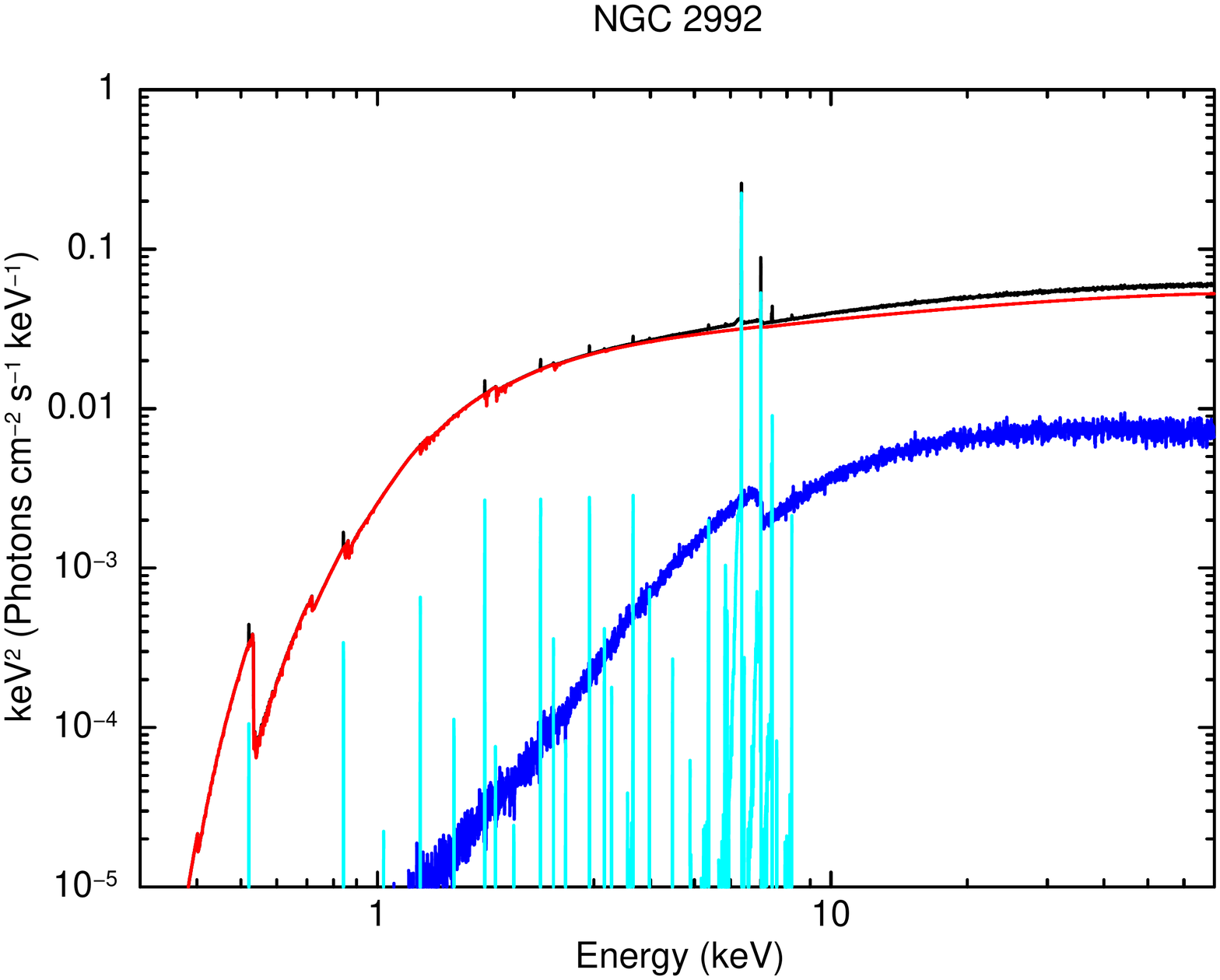}
\plottwo{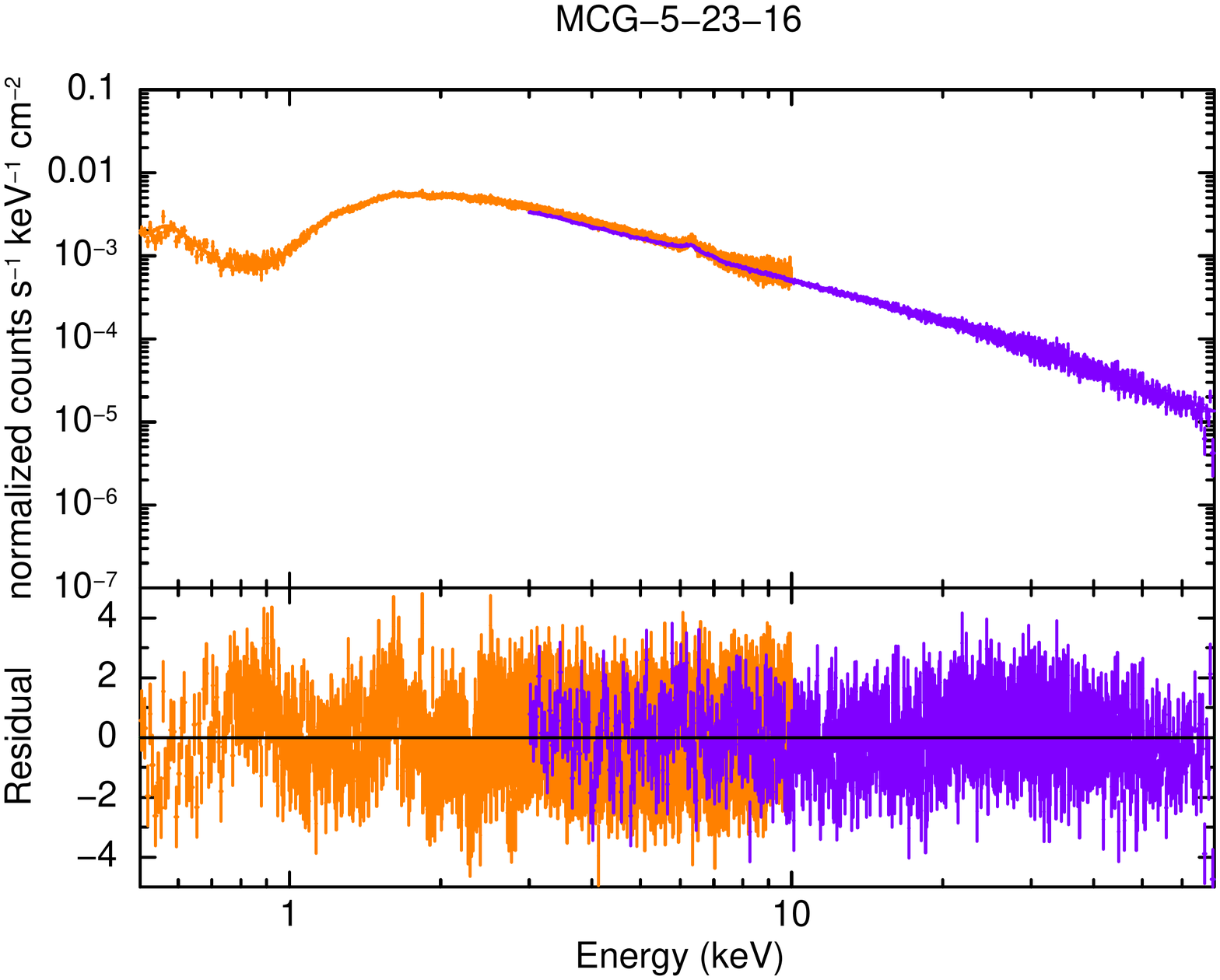}{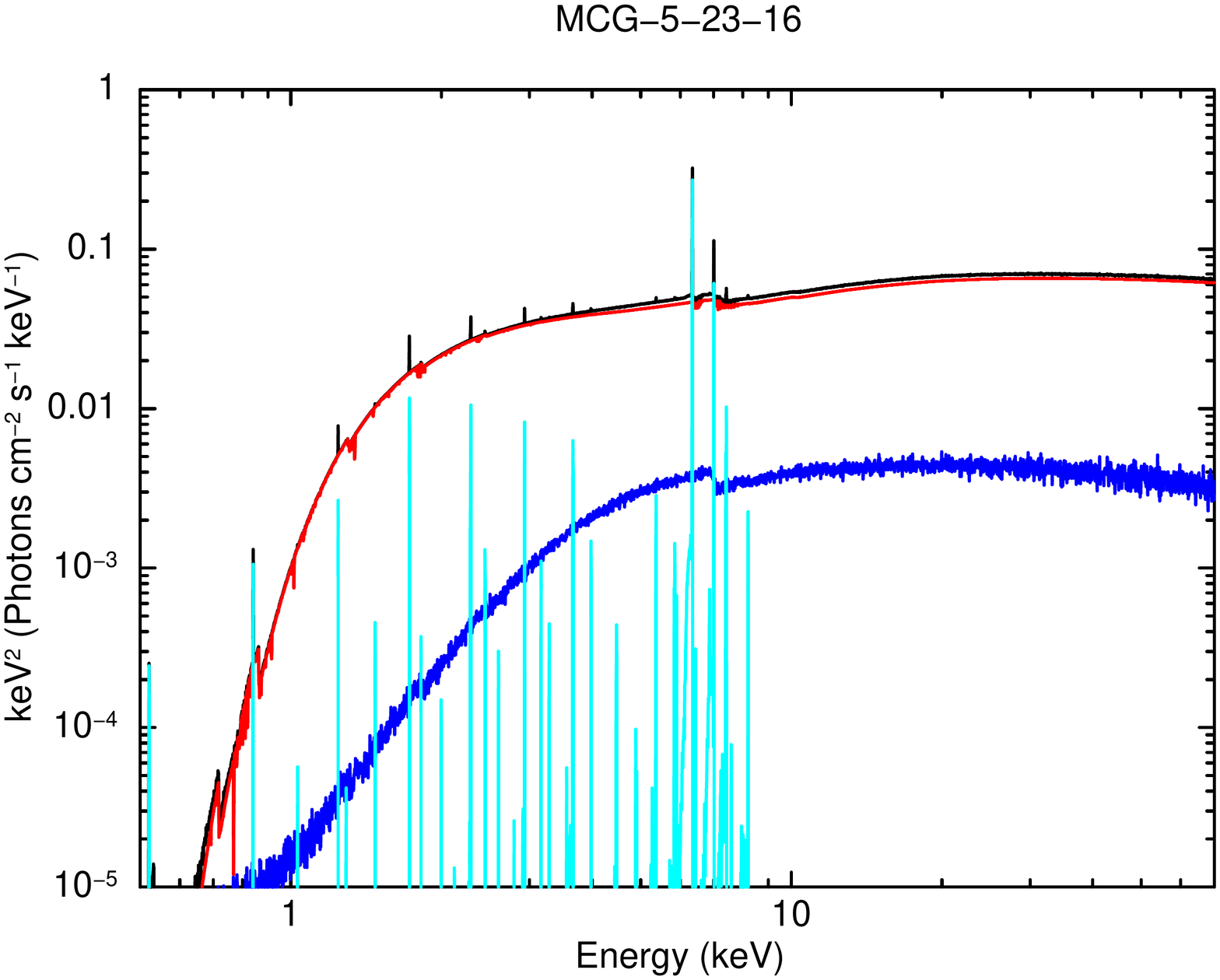}
\plottwo{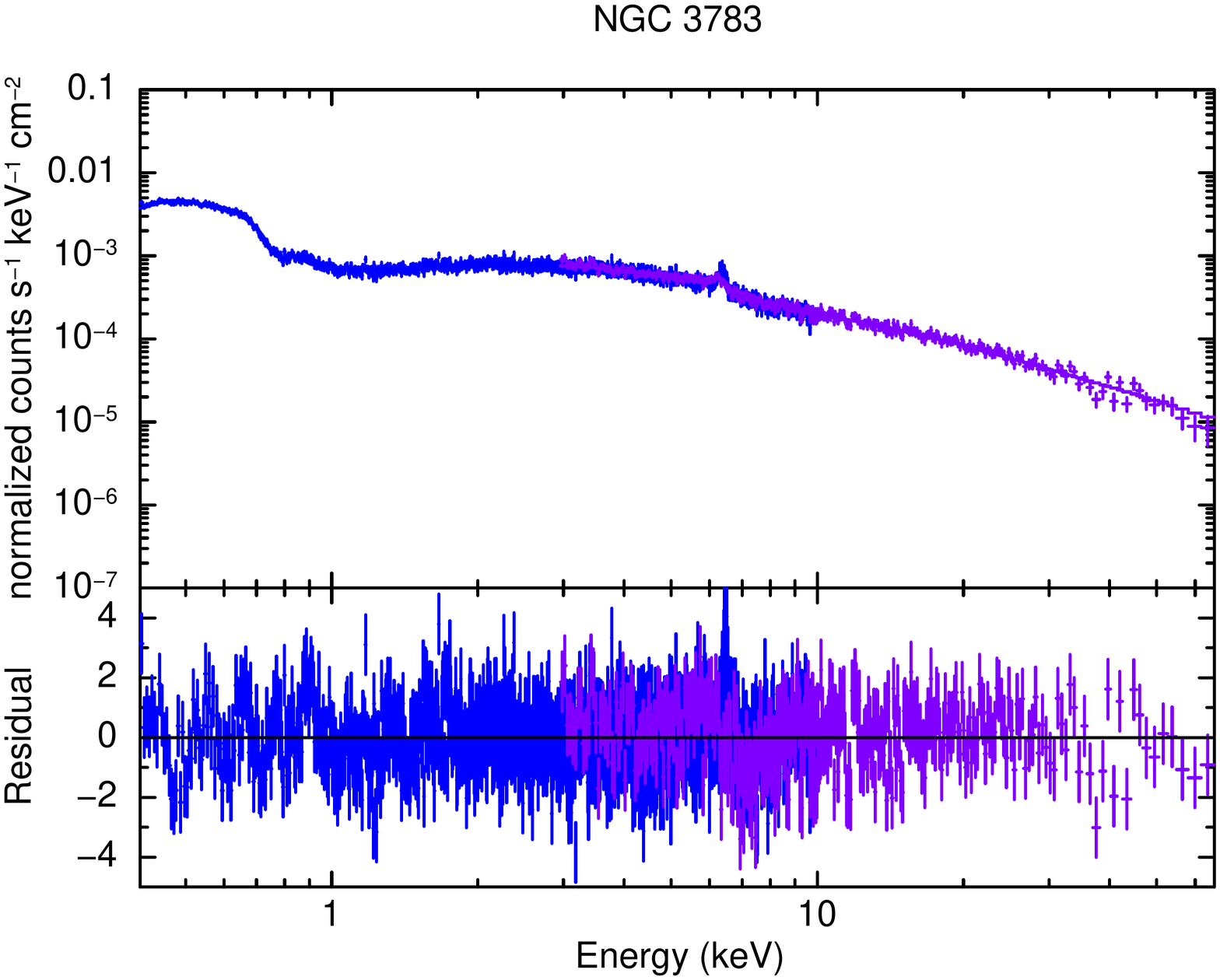}{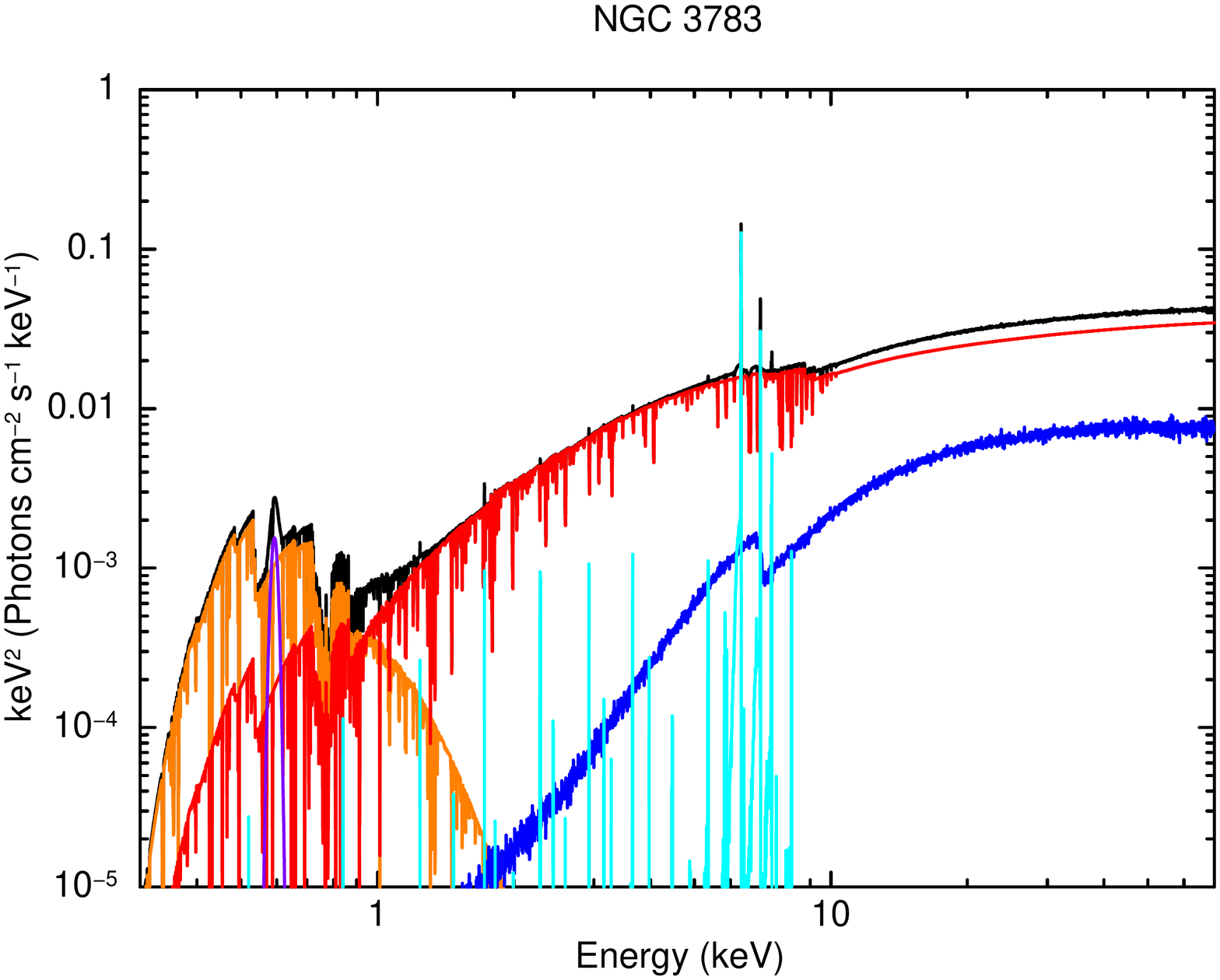}
\caption{
Left: Observed broadband spectra of the unobscured AGNs folded with the energy
responses. The best-fit models are overplotted. In the upper panels, 
the spectra of \textit{Suzaku}/XIS (orange crosses),
\textit{Suzaku}/HXD-PIN (red crosses), \textit{XMM-Newton}/EPIC-pn (blue crosses), and \textit{NuSTAR}/FPMs (purple
crosses) are plotted. Solid curves
represent the best-fit models. In the lower panels, the fitting 
residuals in units of 1$\sigma$ error are shown.
Right: The best-fit models in units of $E I_E$ (where $I_E$ is the energy flux
at the energy $E$). 
The solid lines show the total (black), direct component (red), reflection continuum from the torus (blue), emission lines from the torus (light blue), soft excess (orange), and additional emission line (purple).\\
}
\label{fig1}
\end{figure*}

\begin{figure*}
\plottwo{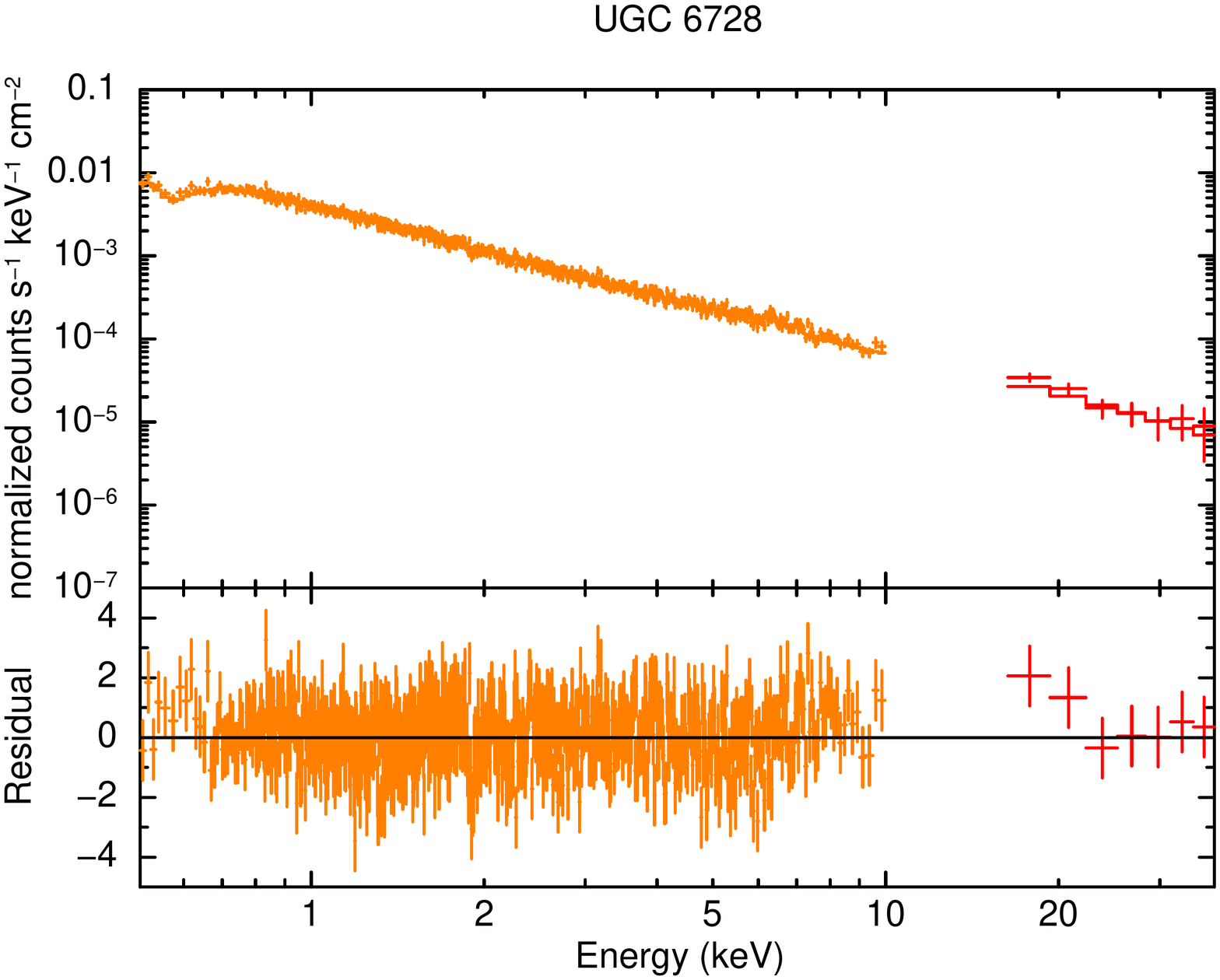}{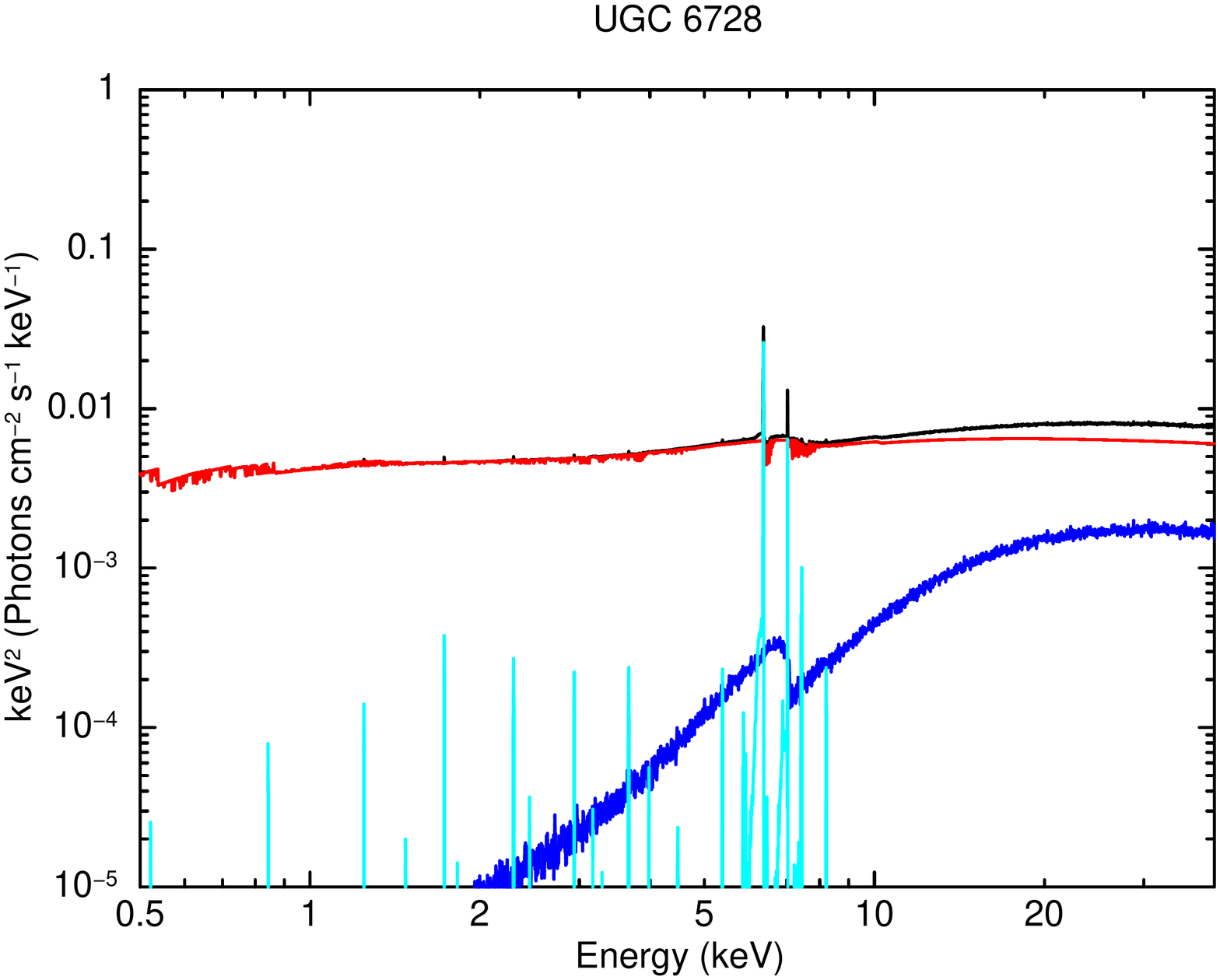}
\plottwo{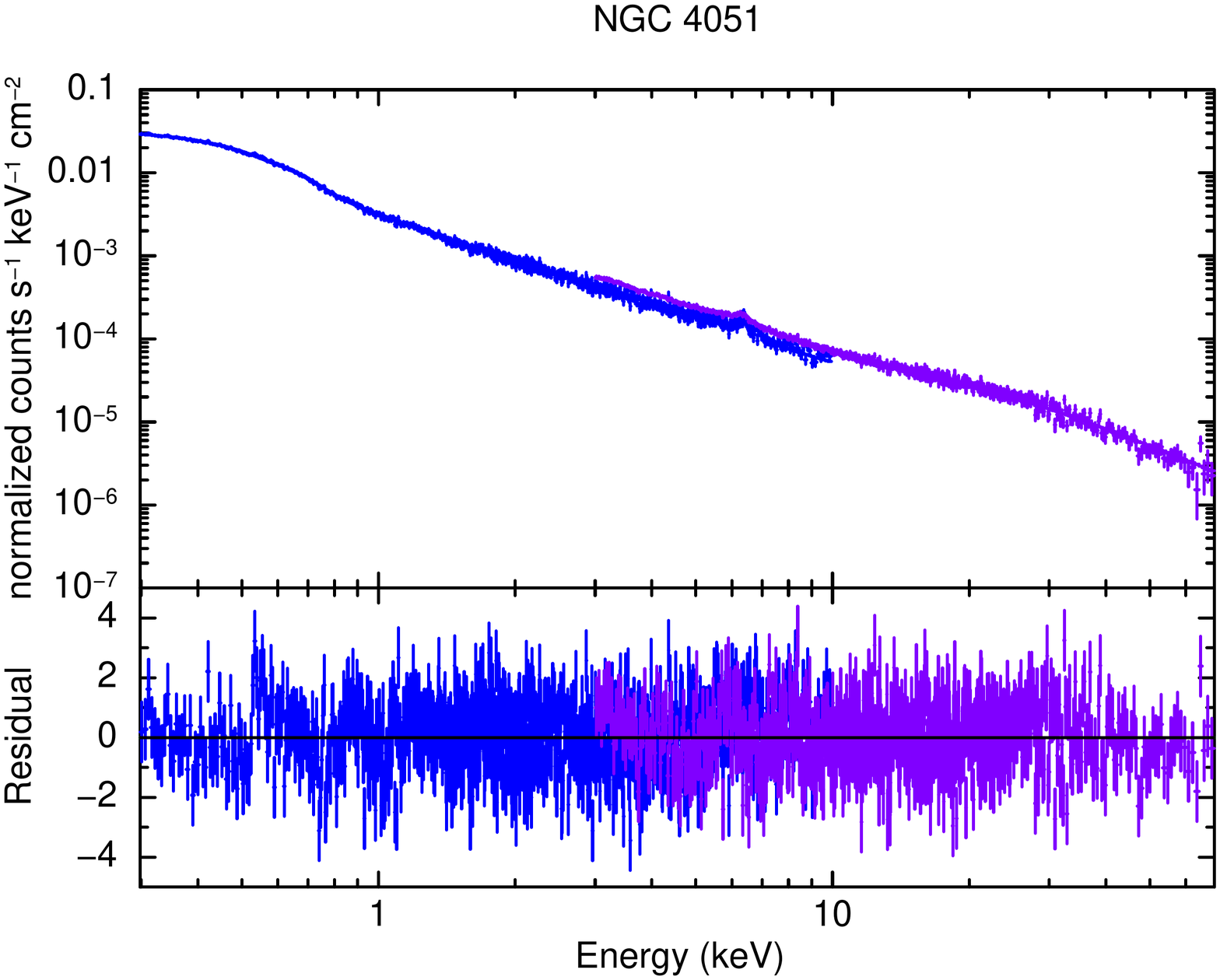}{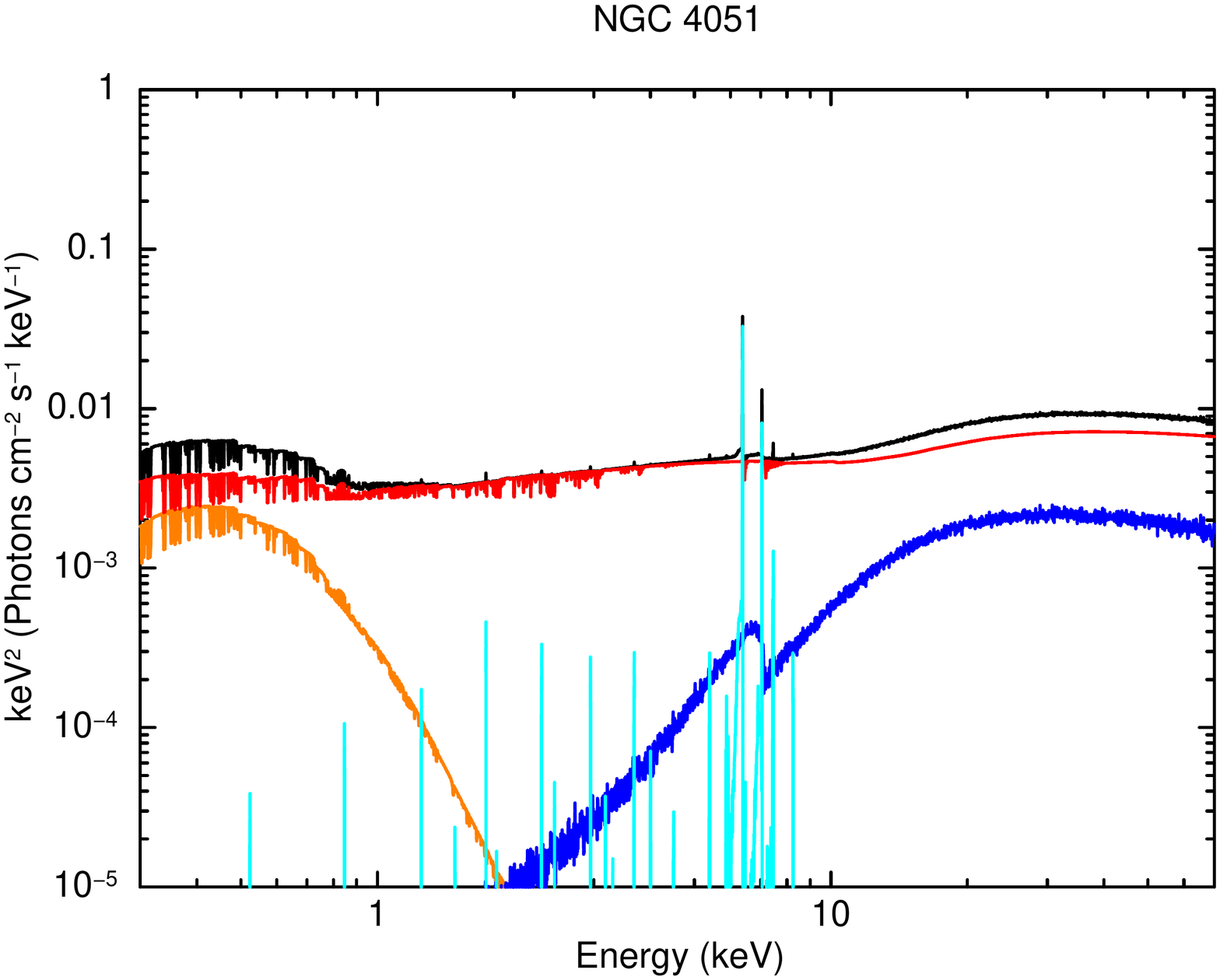}
\plottwo{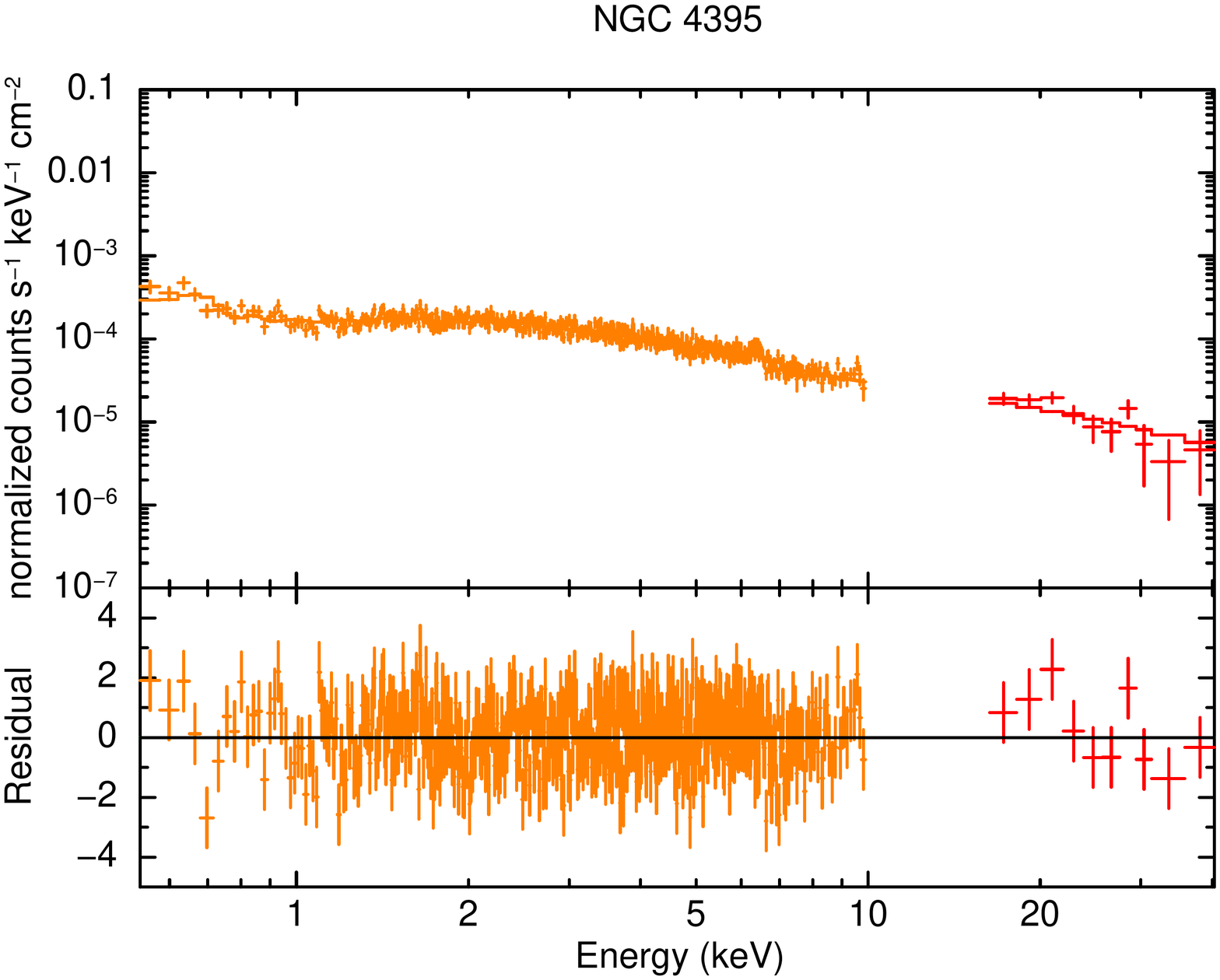}{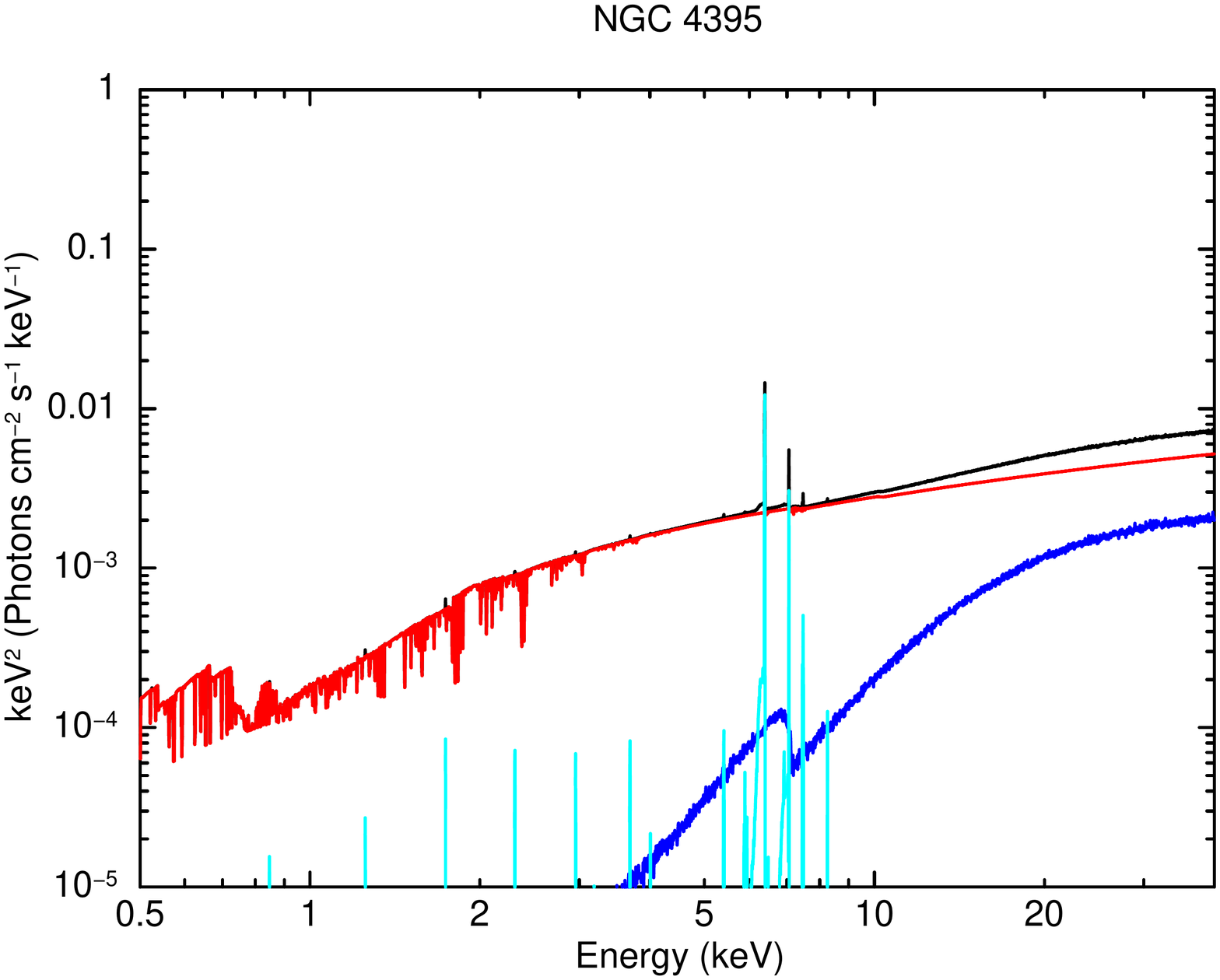}
\setcounter{figure}{0}
\caption{
Continued.
}
\end{figure*}

\begin{figure*}
\plottwo{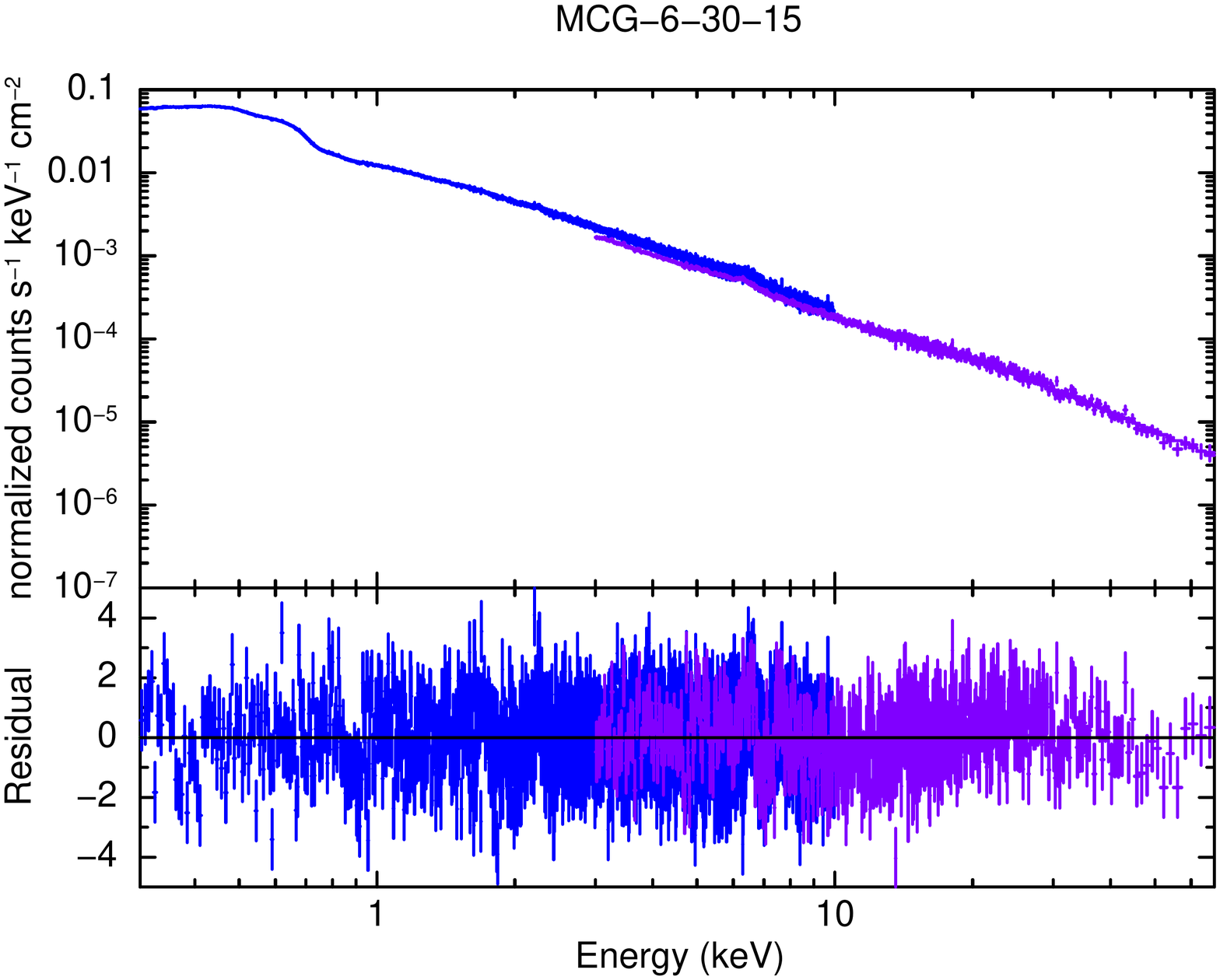}{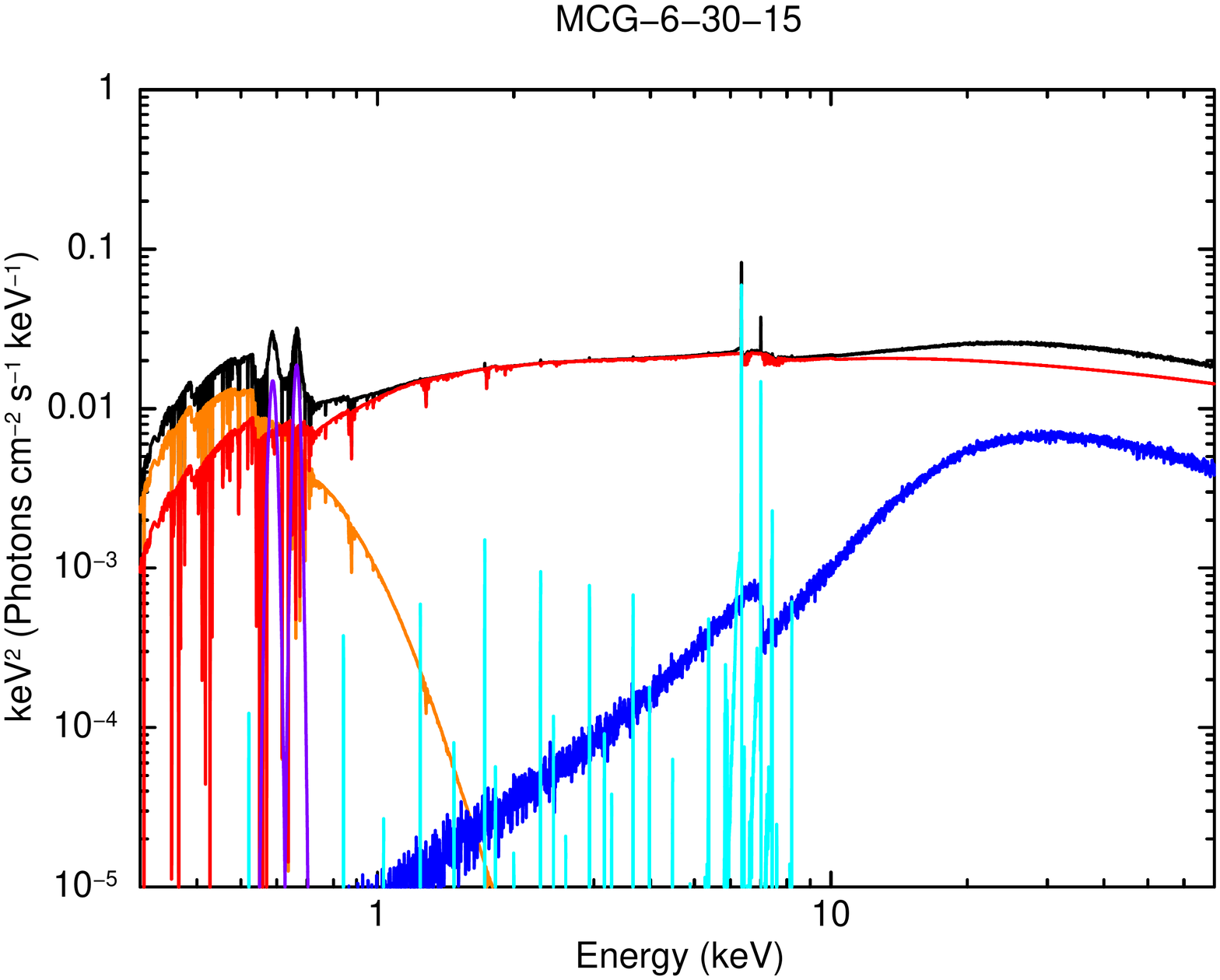}
\plottwo{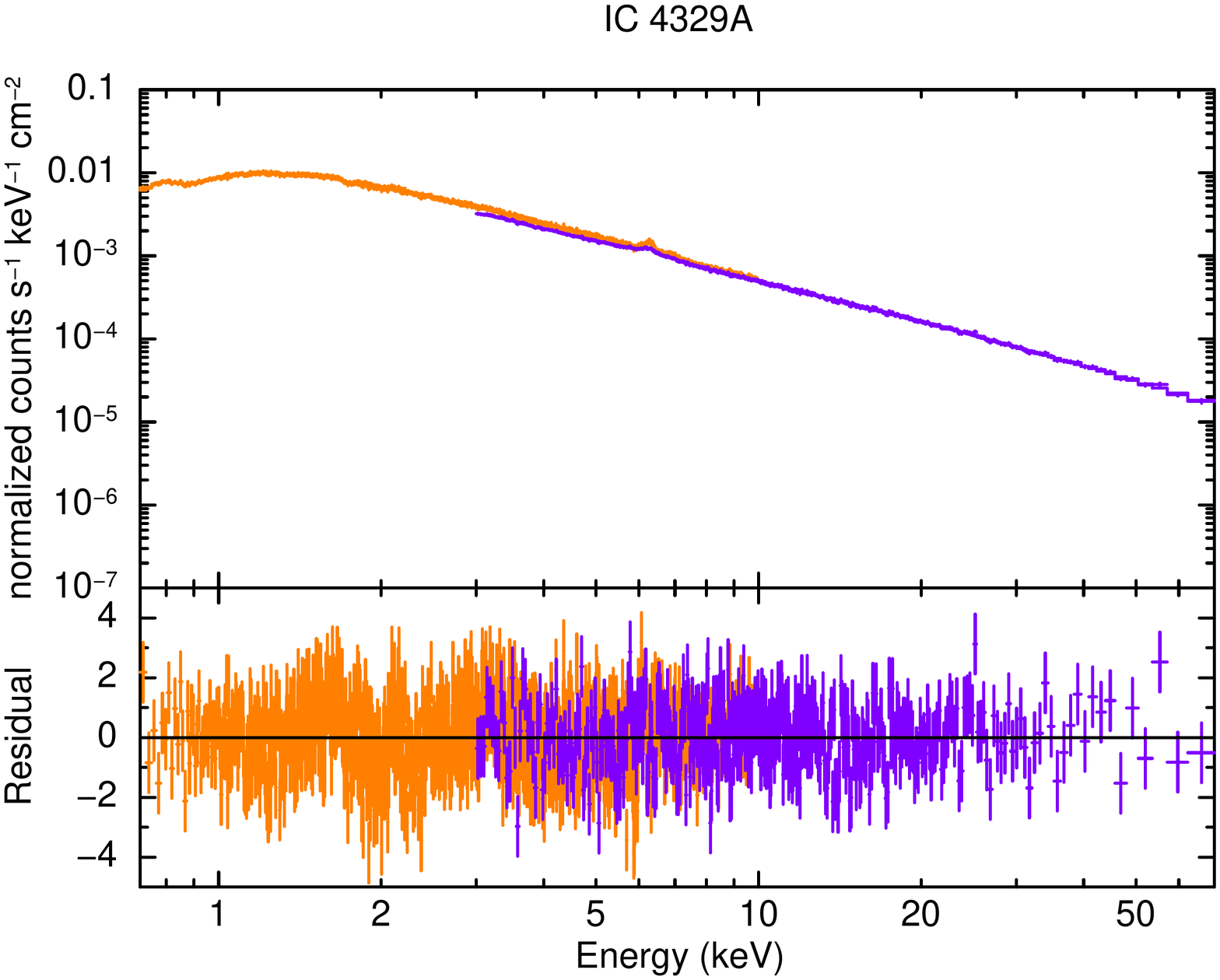}{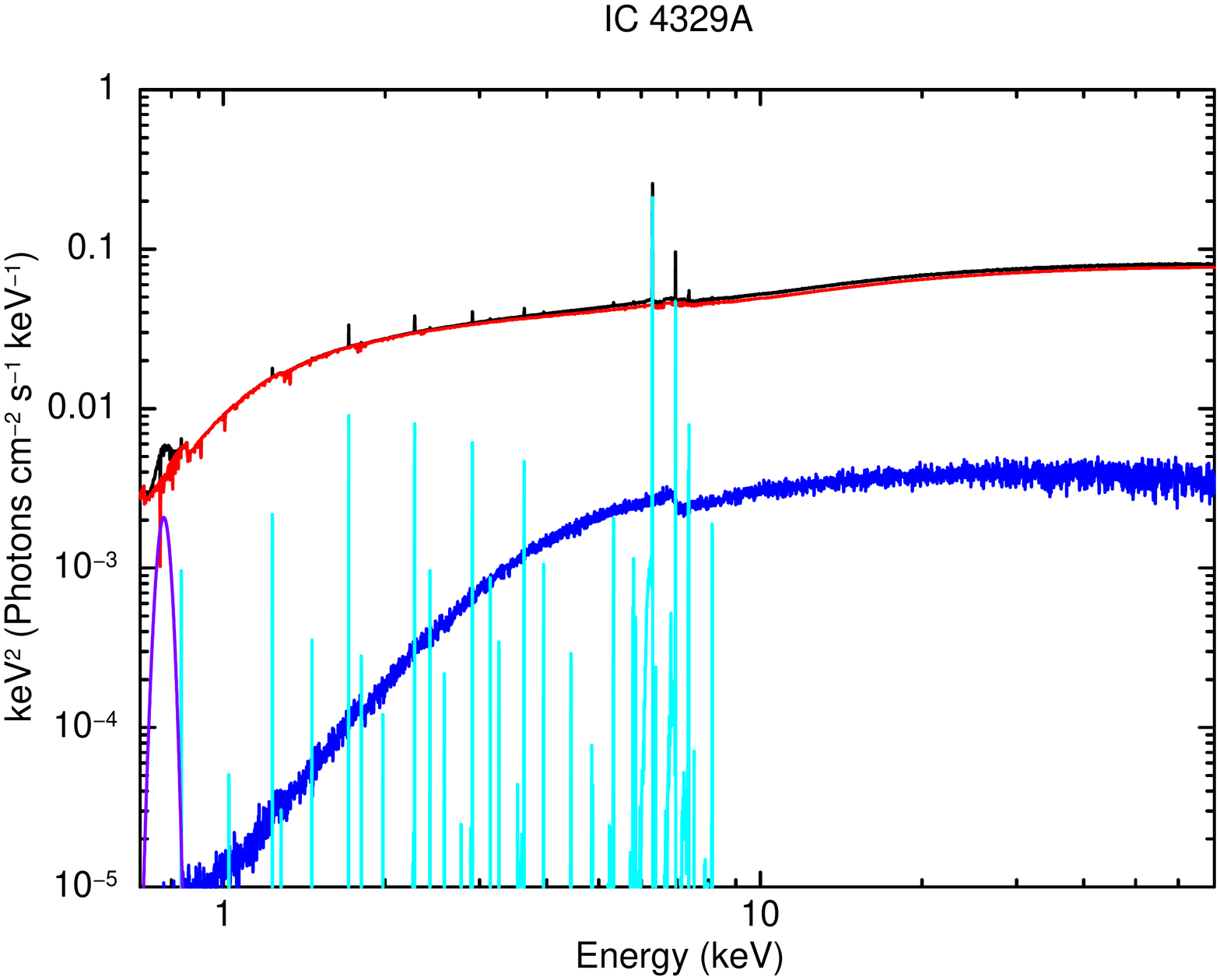}
\plottwo{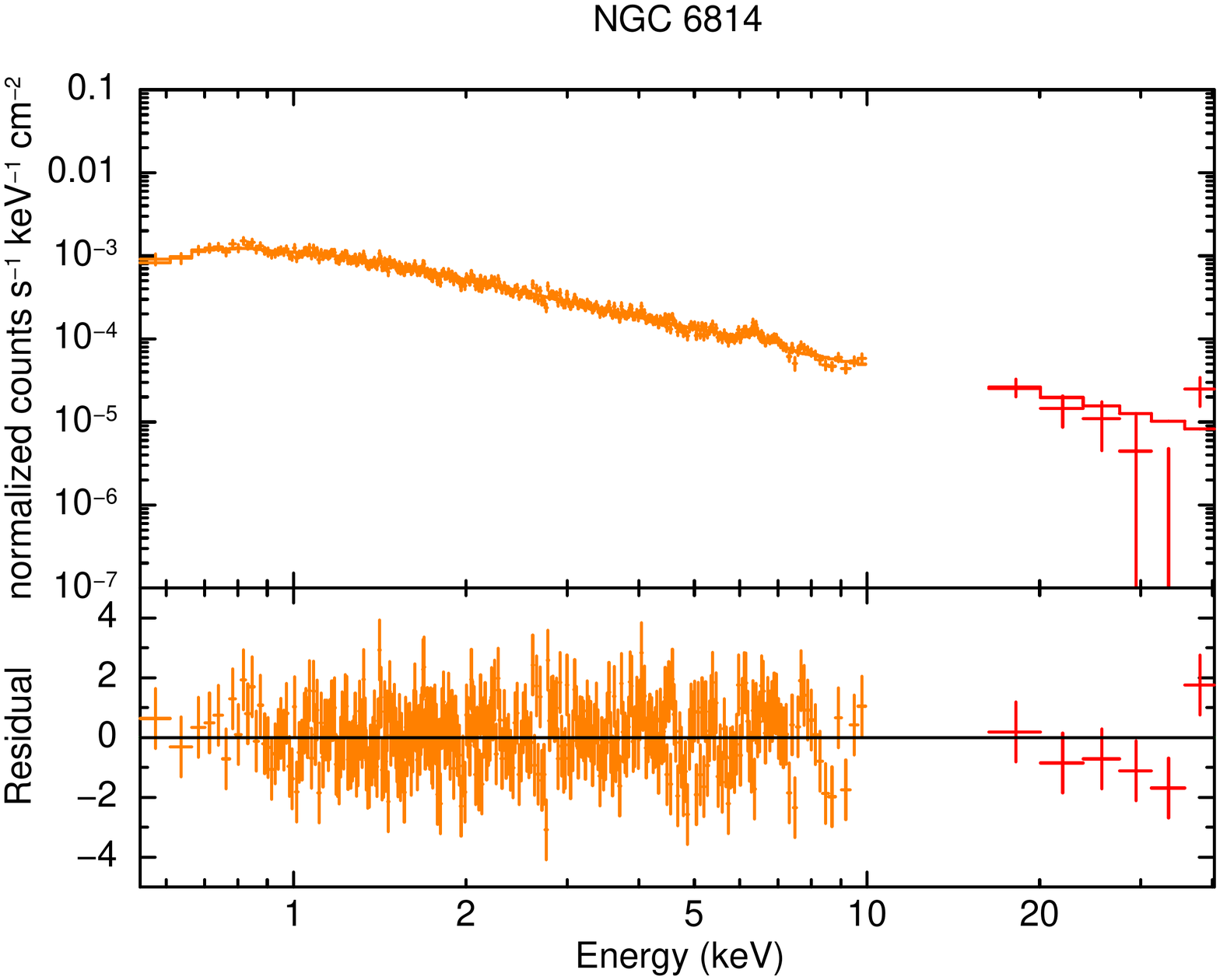}{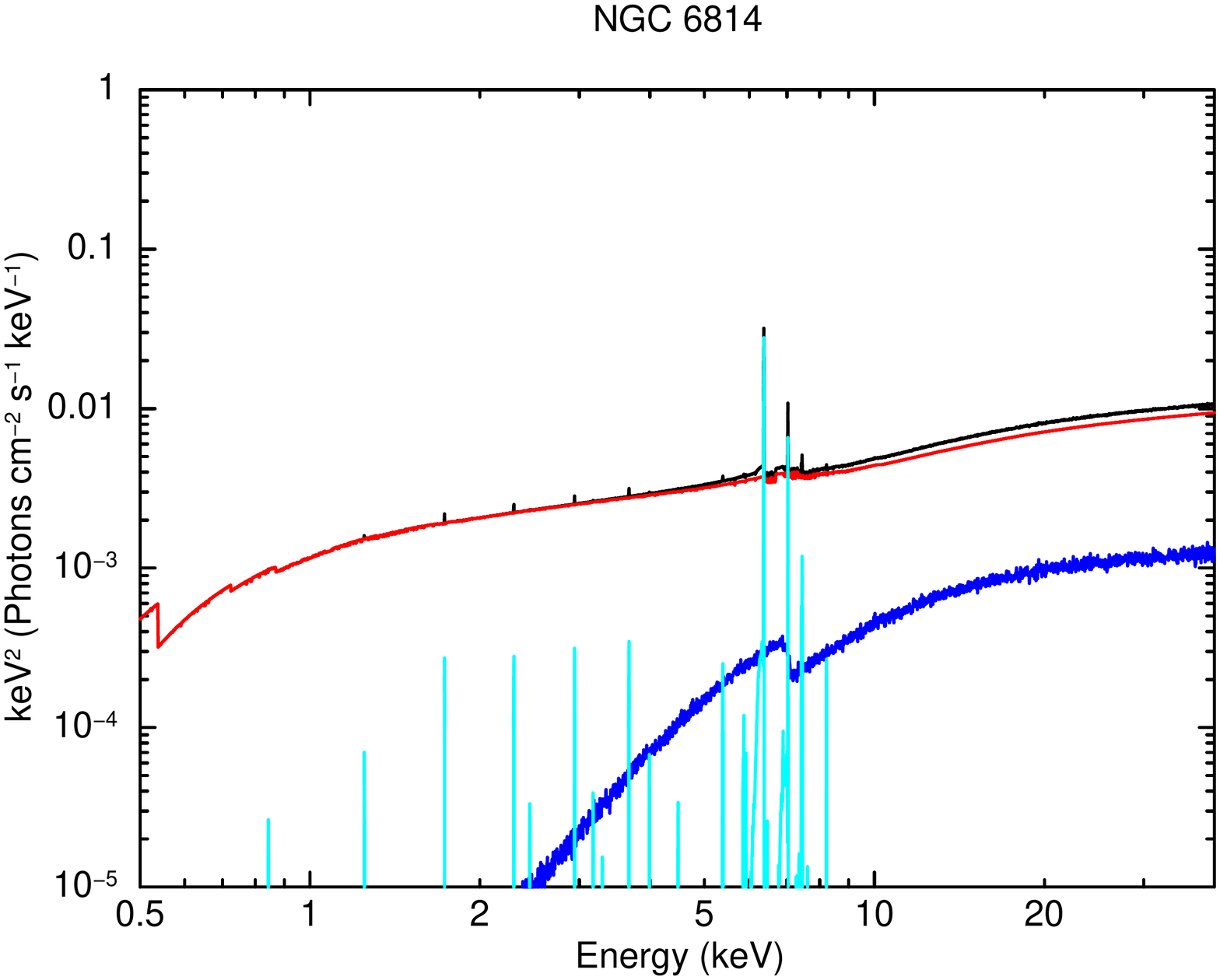}
\setcounter{figure}{0}
\caption{
Continued.
}
\end{figure*}

\begin{figure*}
\plottwo{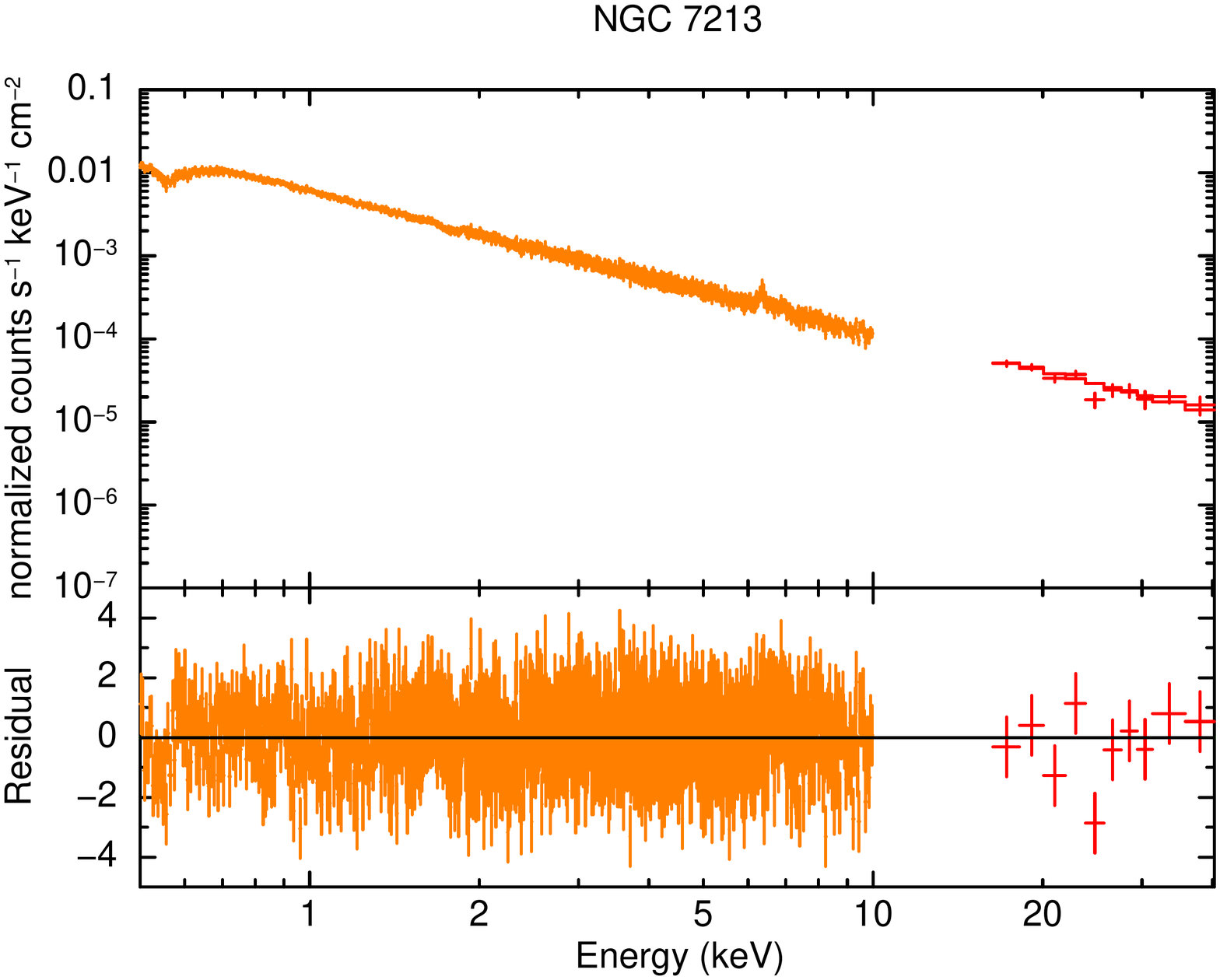}{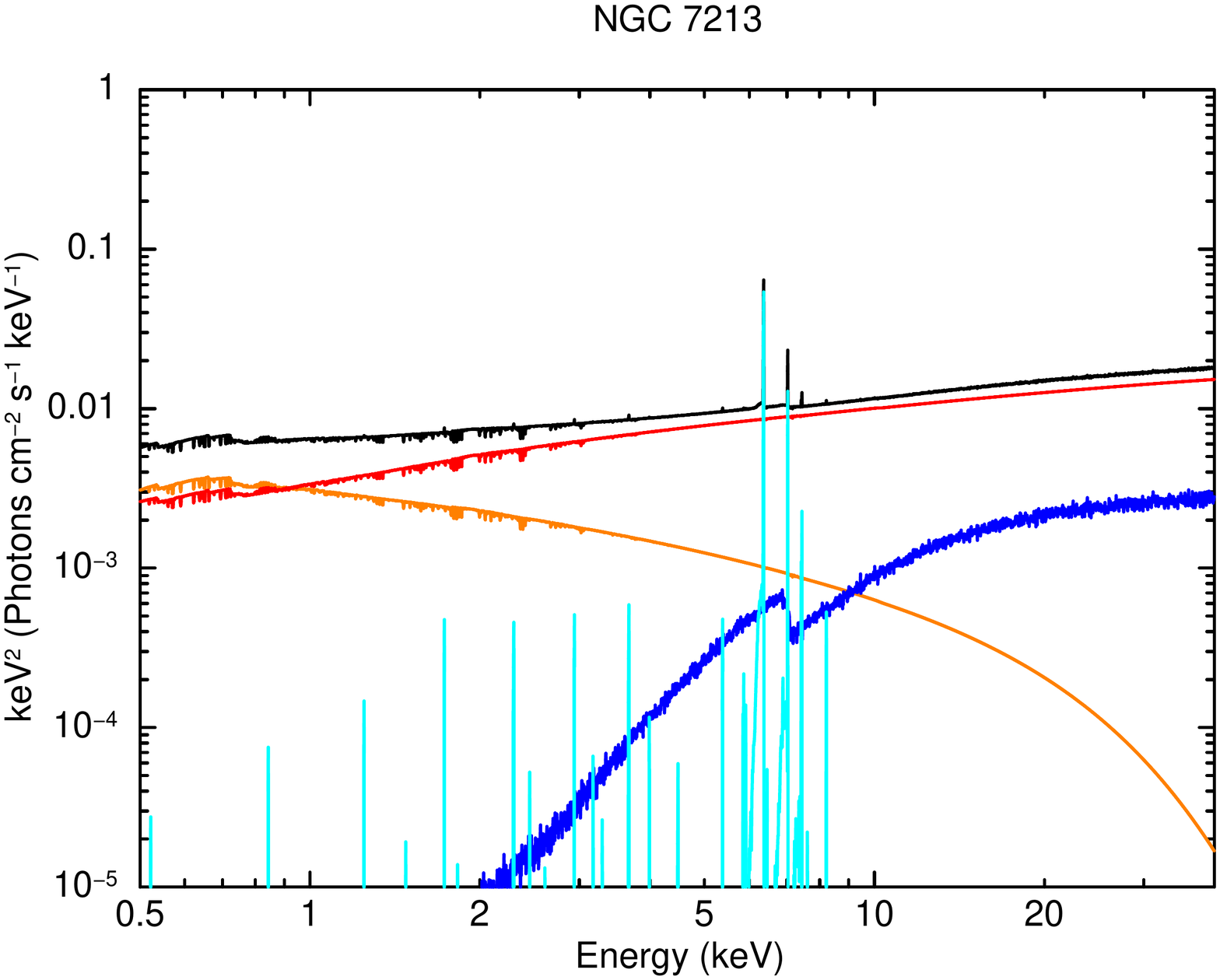}
\plottwo{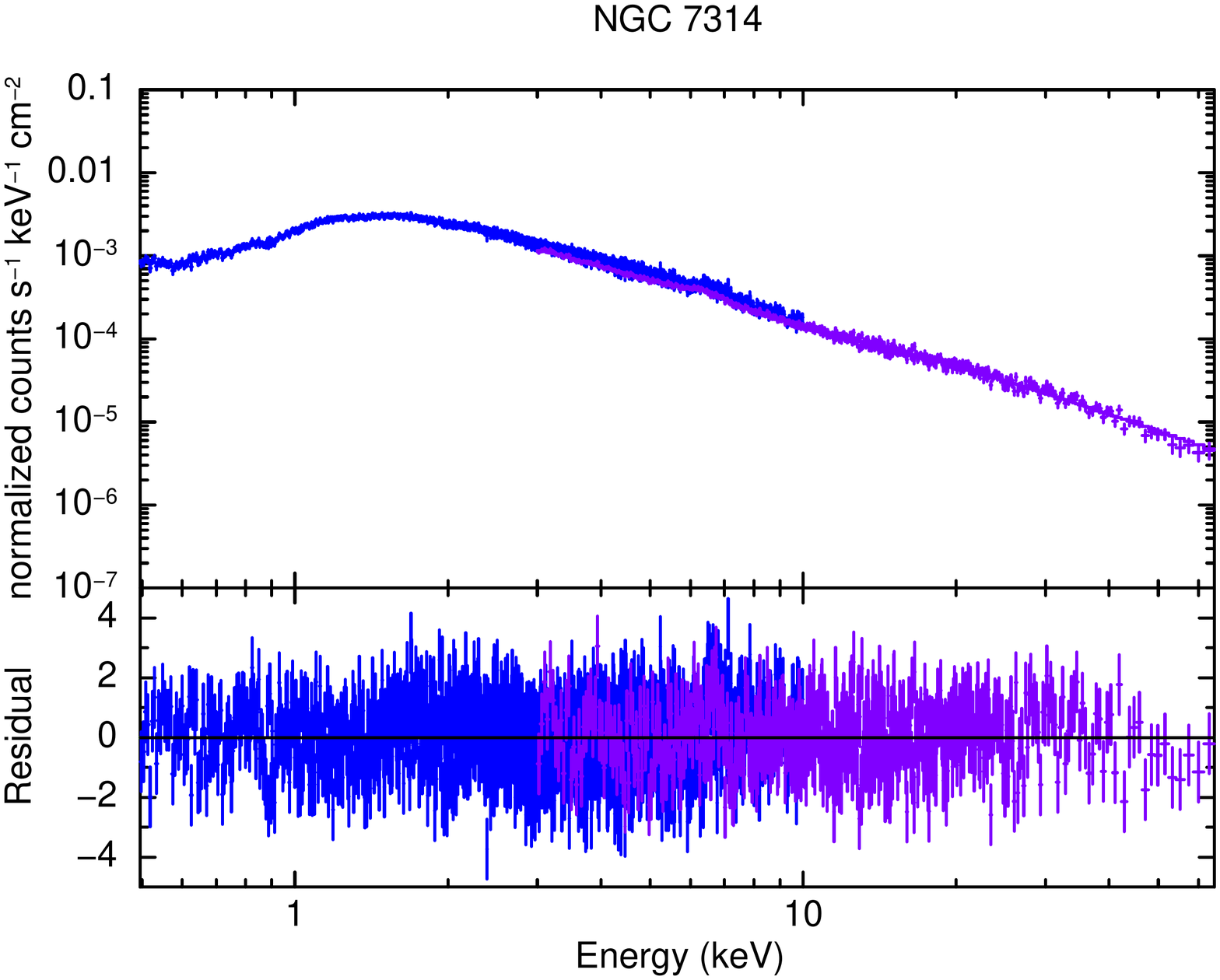}{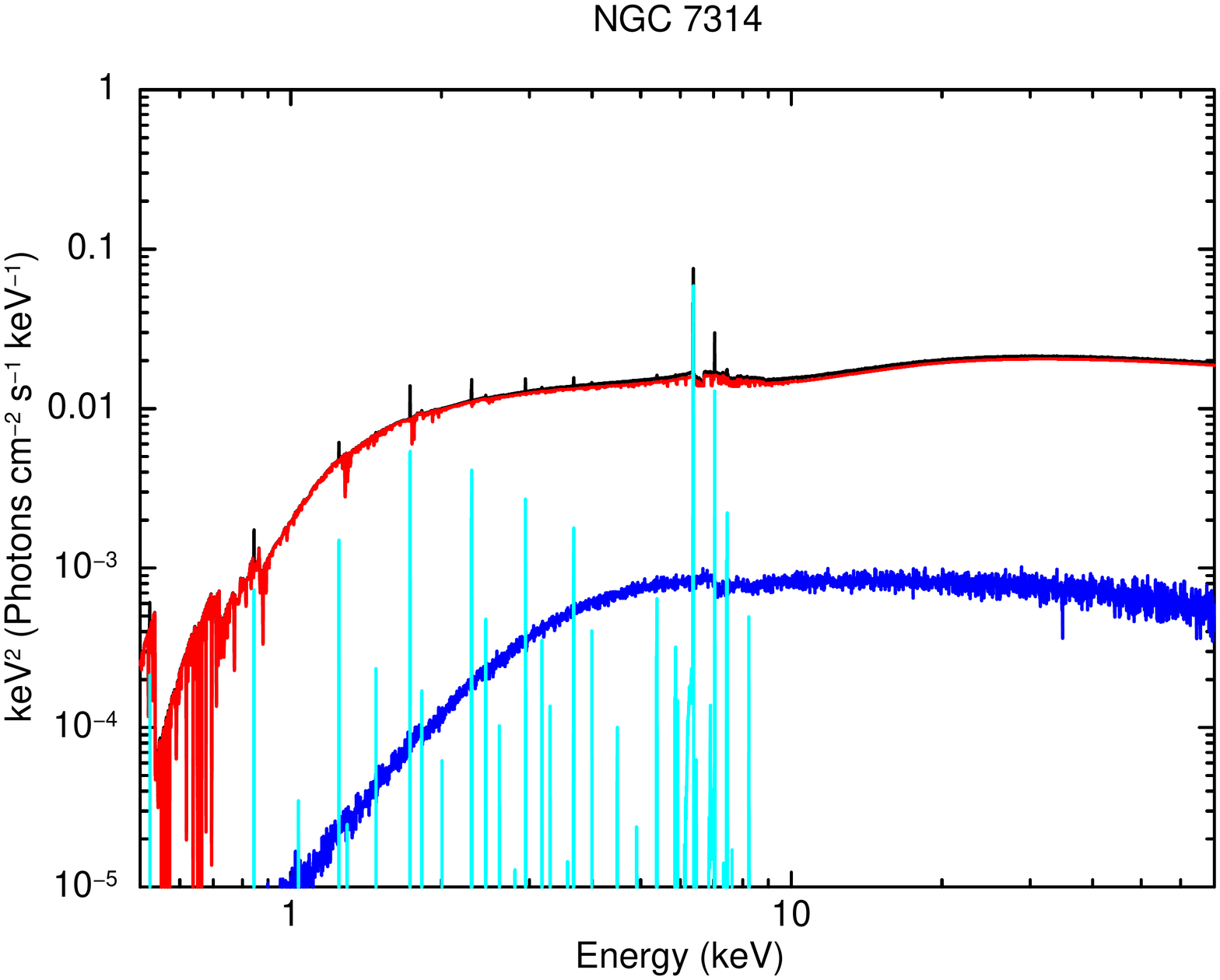}
\plottwo{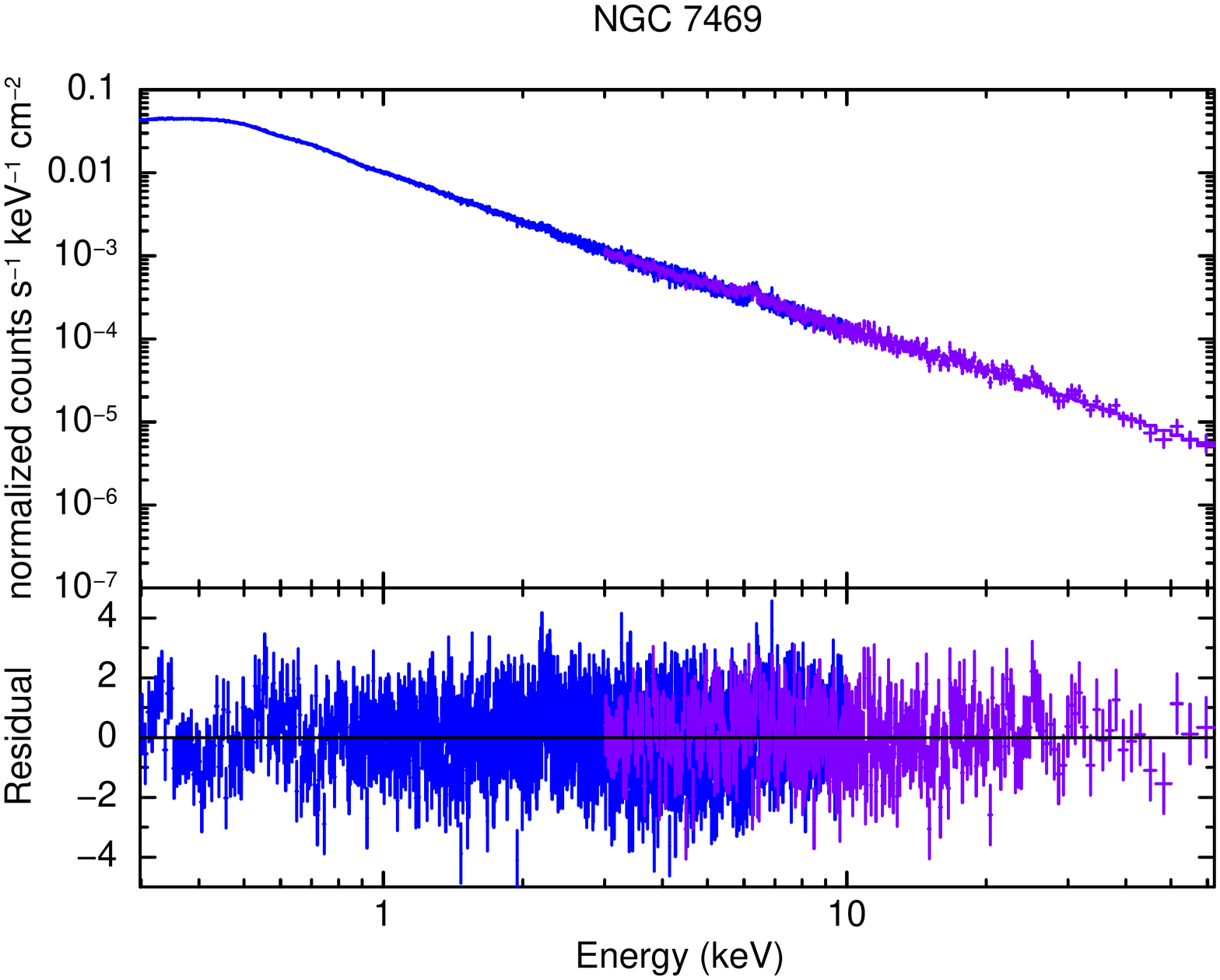}{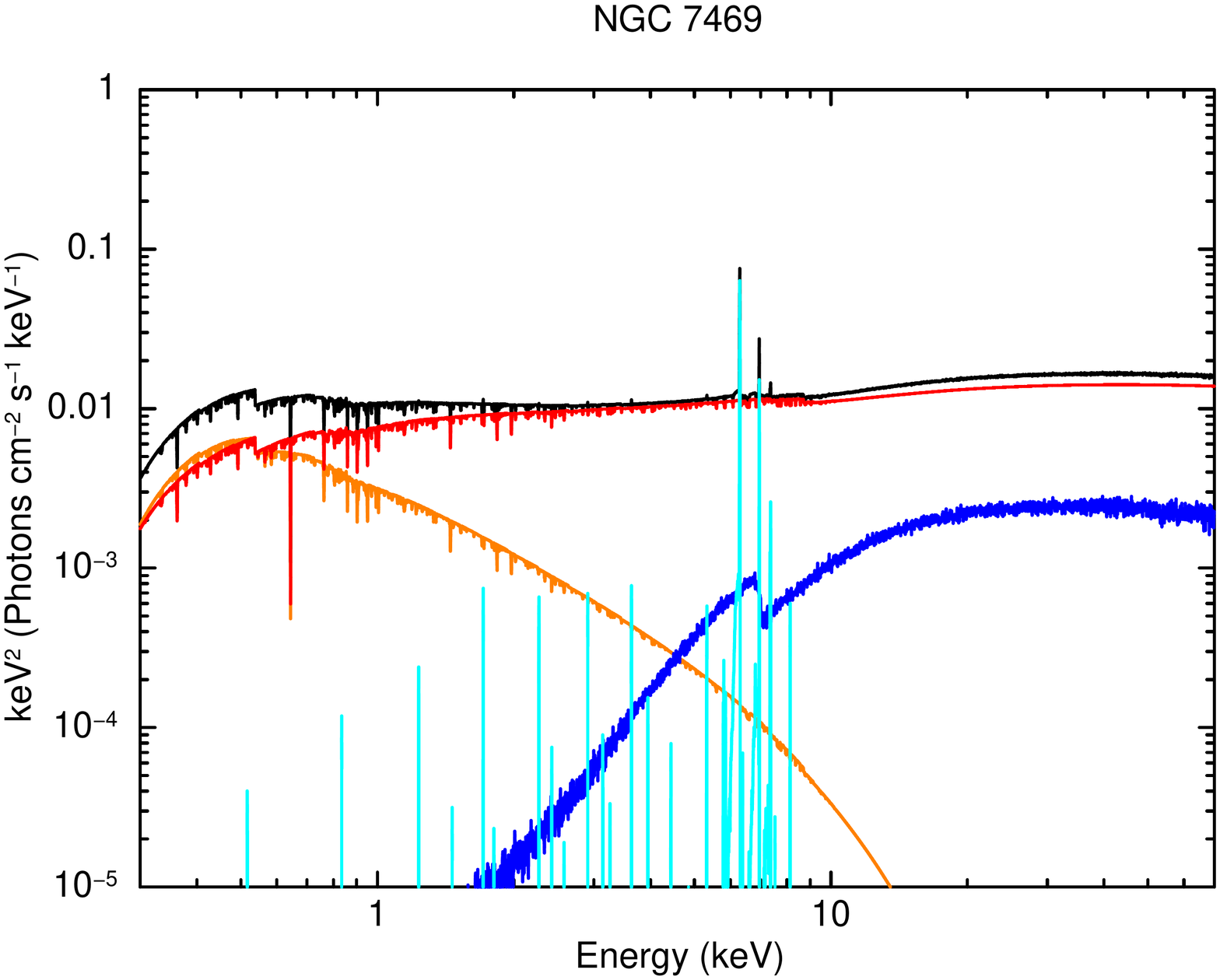}
\setcounter{figure}{0}
\caption{
Continued.
}
\end{figure*}

\begin{figure*}
\plottwo{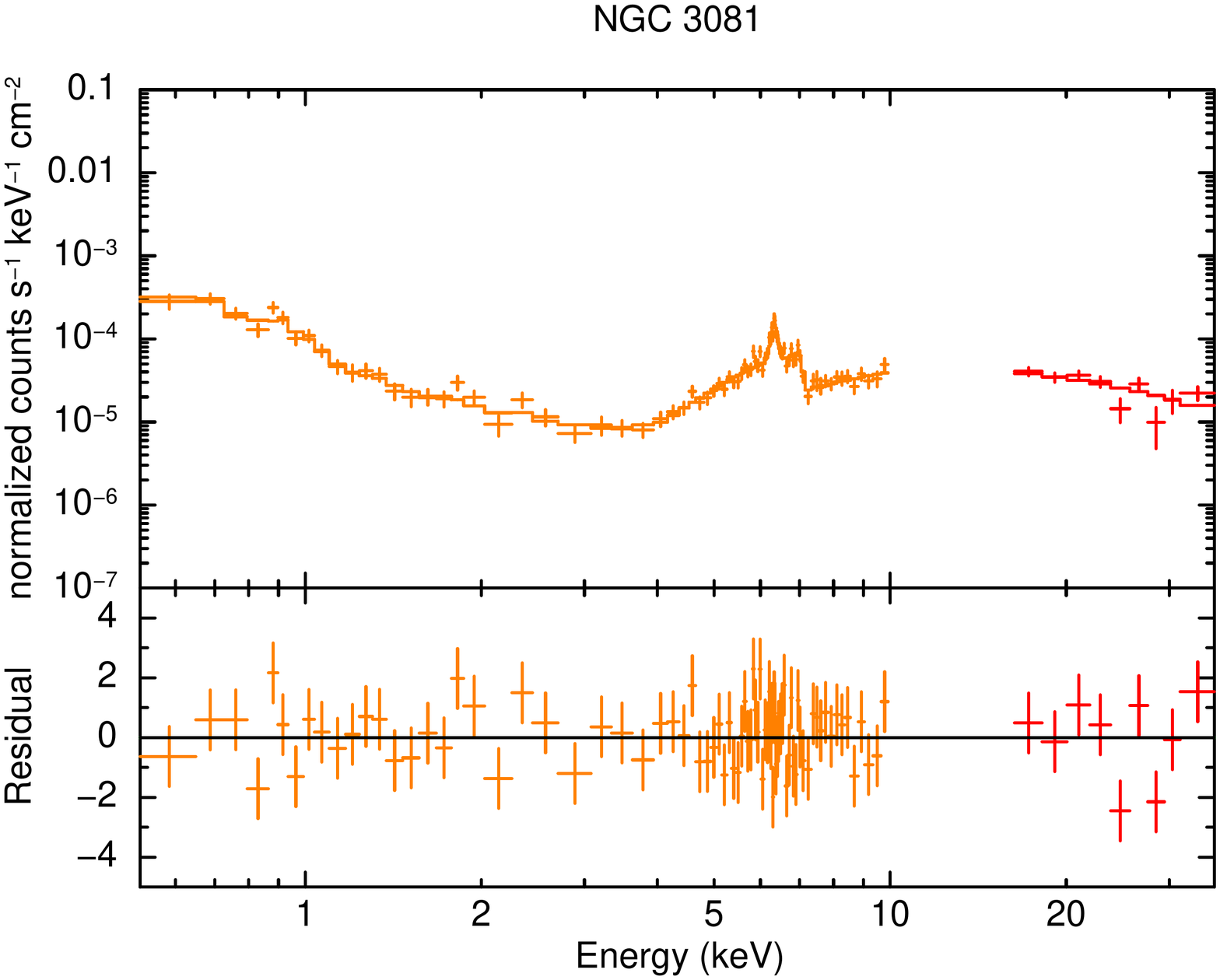}{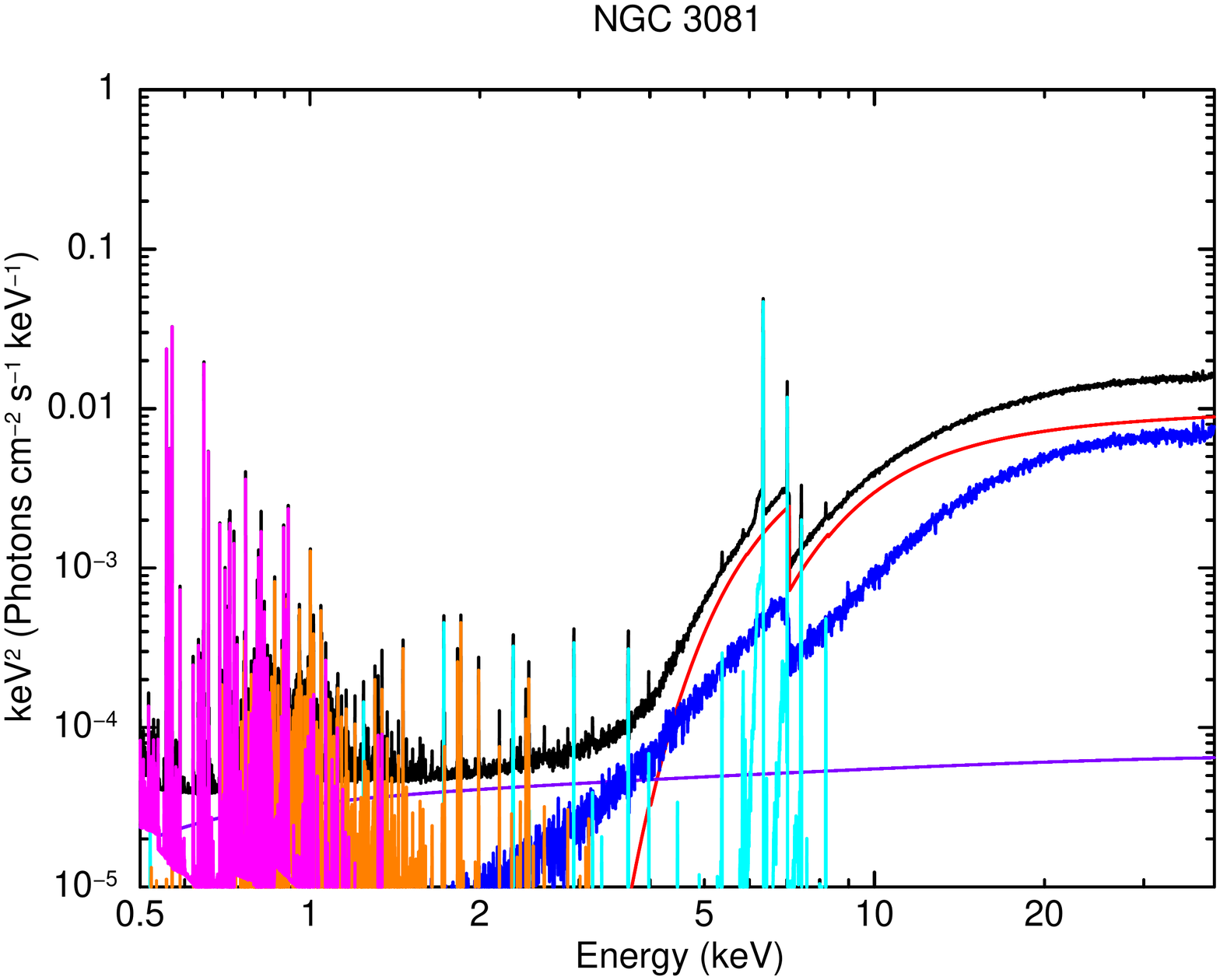}
\plottwo{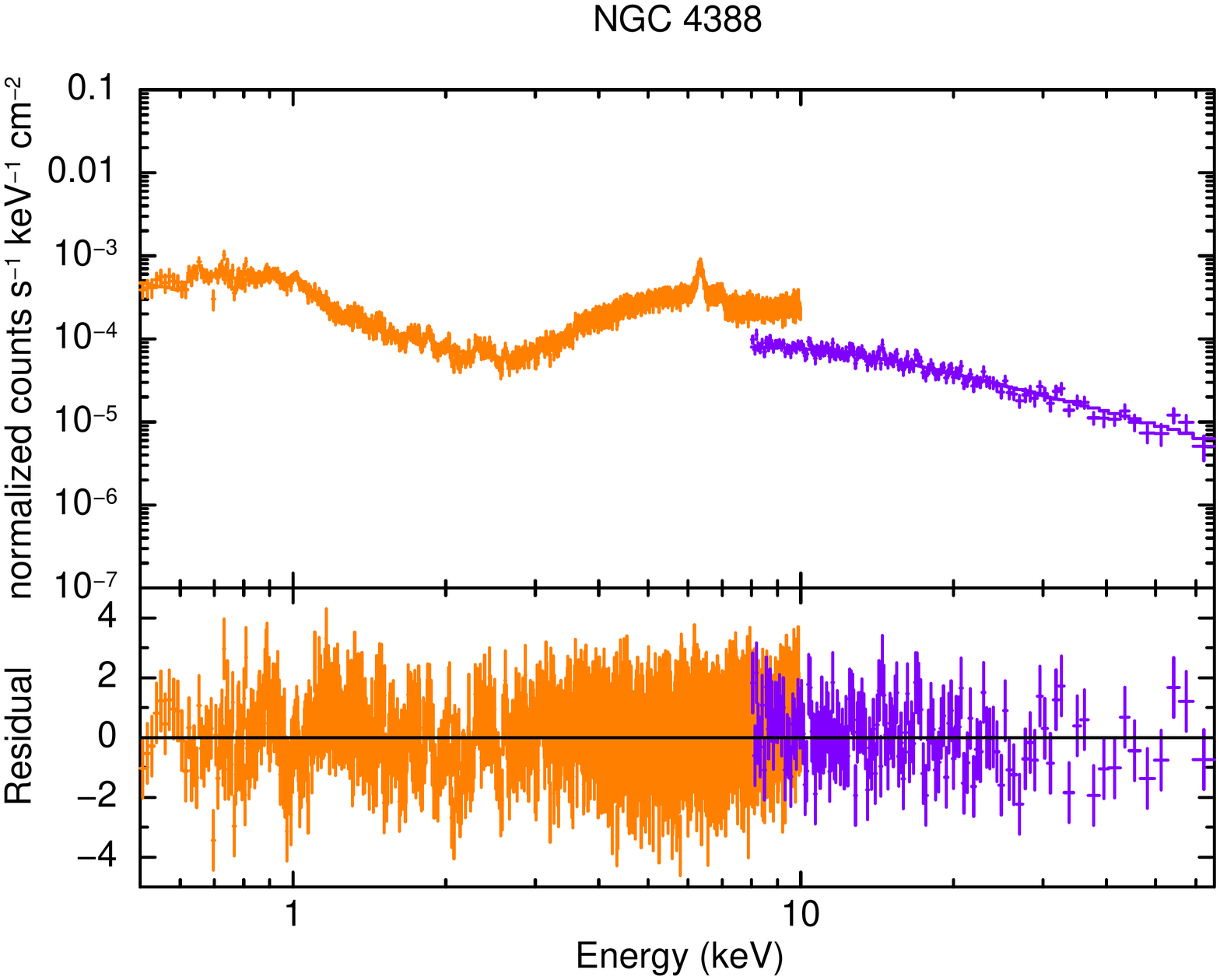}{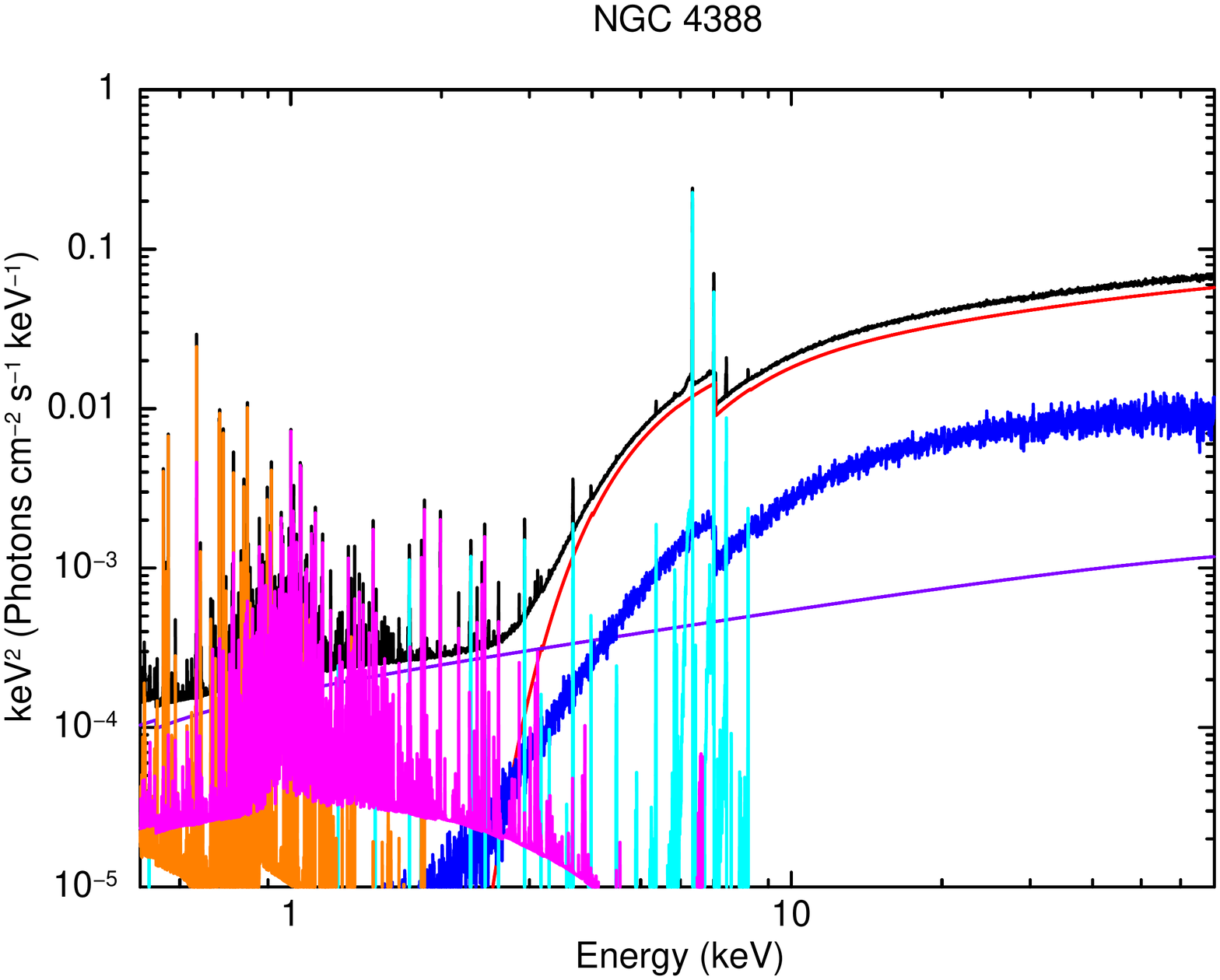}
\plottwo{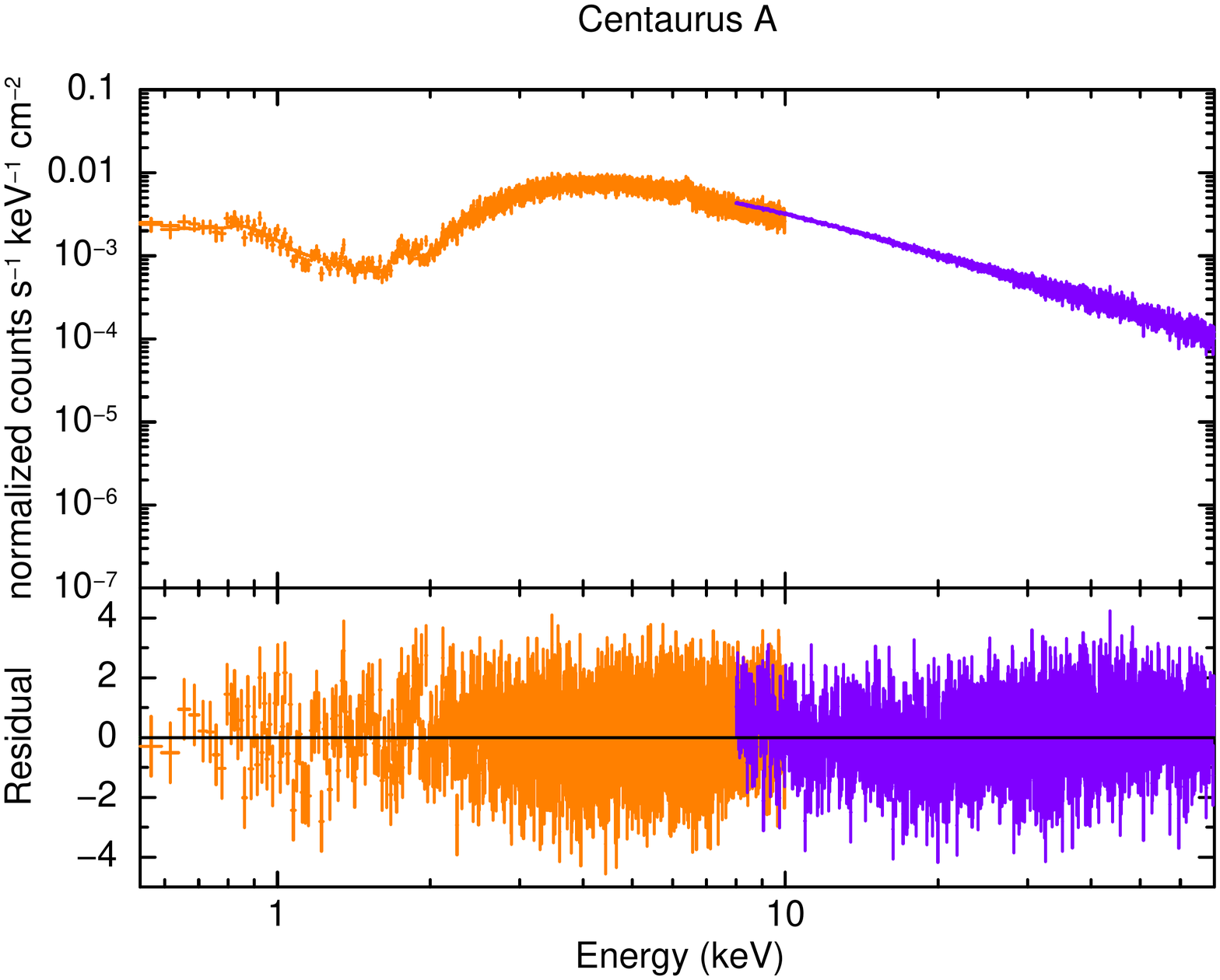}{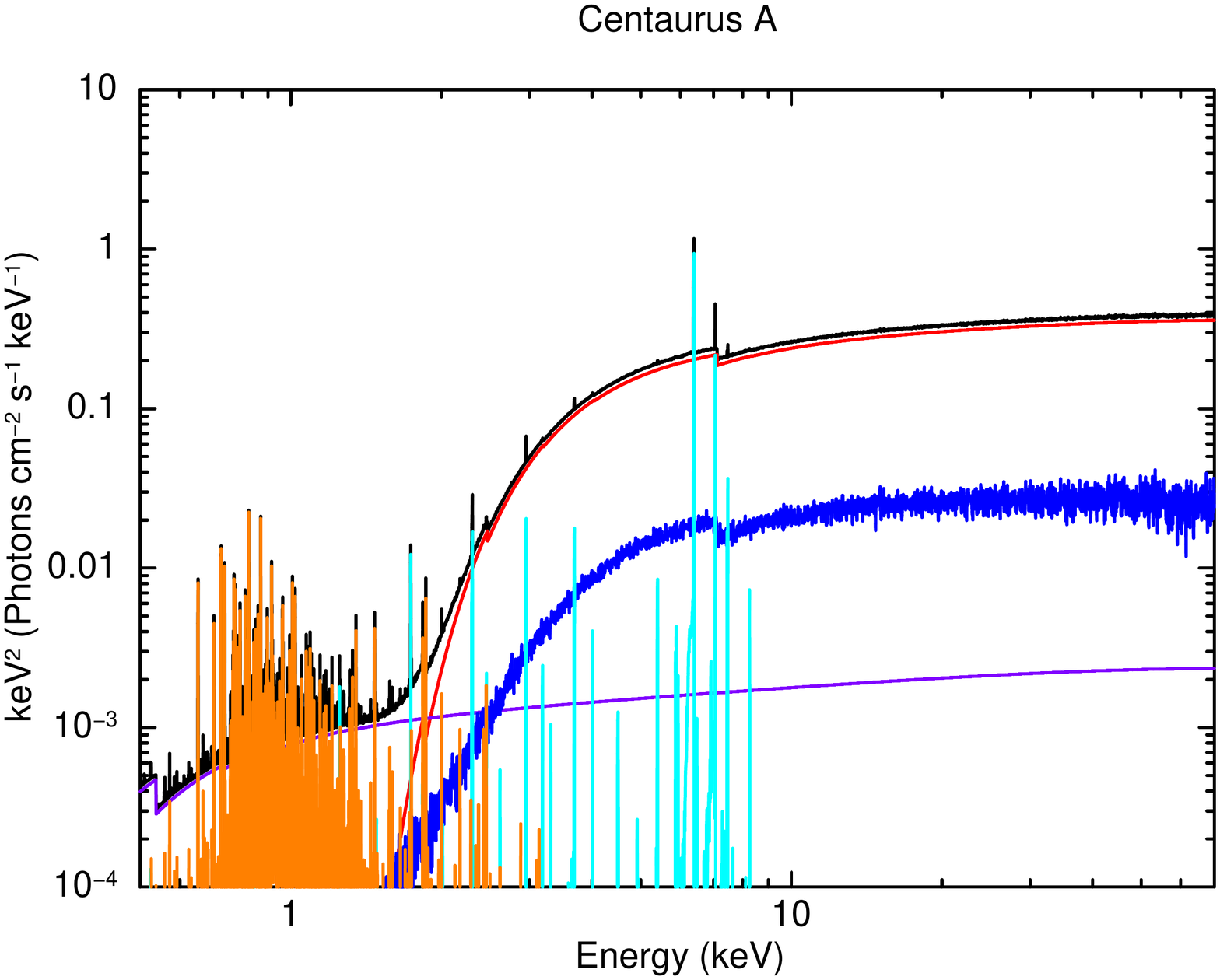}
\plottwo{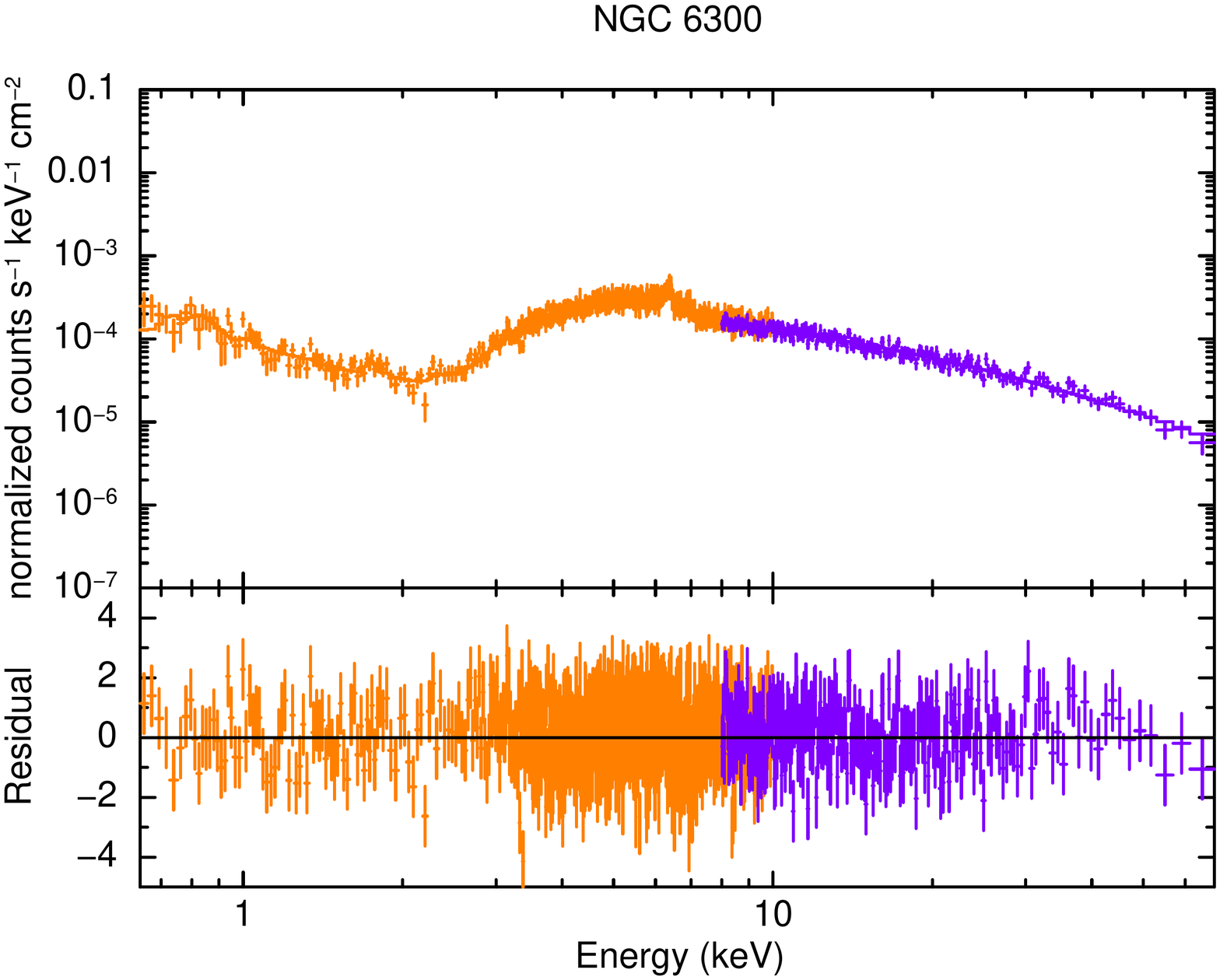}{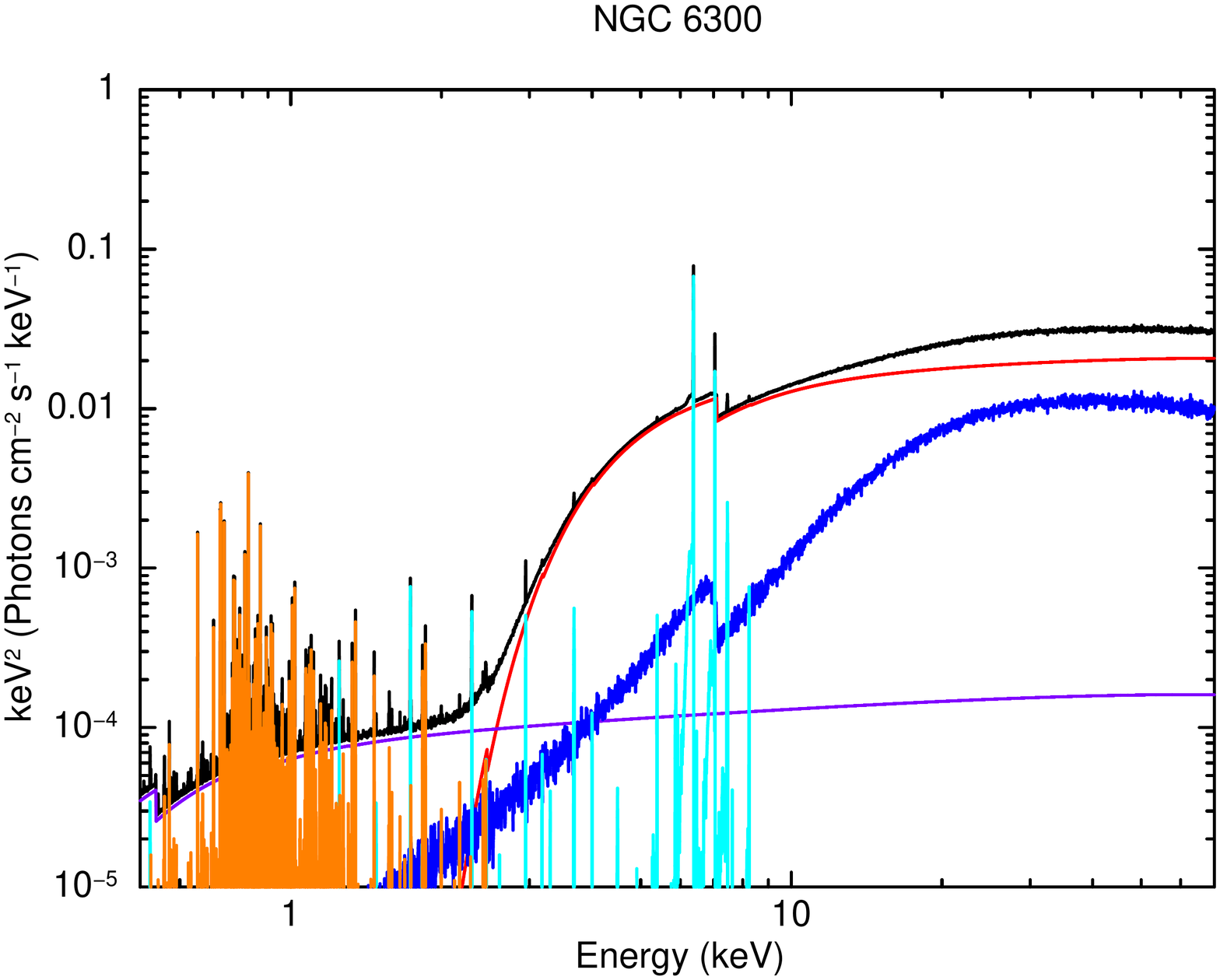}
\caption{
Left: Observed broadband spectra of obscured AGNs folded with the energy
responses. The best-fit models are overplotted. In the upper panels, 
the spectra of \textit{Suzaku}/XIS (orange crosses),
\textit{Suzaku}/HXD-PIN (red crosses), and \textit{NuSTAR}/FPMs (purple
crosses) are plotted. Solid curves
represent the best-fit models. In the lower panels, the fitting 
residuals in units of 1$\sigma$ error are shown.
Right: The best-fit models in units of $E I_E$ (where $I_E$ is the energy flux
at the energy $E$). 
The solid lines show the total (black), direct component (red), reflection continuum from the torus (blue), emission lines from the torus (light blue), optically thin thermal plasma (orange and magenta), and scattered component (purple).
}
\label{fig2}
\end{figure*}

\begin{longrotatetable}
\begin{deluxetable*}{lllllllllll}
\tabletypesize{\scriptsize}
\tablewidth{\textwidth}
\tablecaption{Best-fit Parameters of unobscured AGNs\label{Best_fit_UAGN}}
\tablehead{
(1)           &(2)                 &(3)                &(4)                &(5)                &(6)                &(7)                				&(8)                				&(9)                         &(10)	&\\
Object        &\textsf{phabs}  &\textsf{zcutoffpl}  &\textsf{compTT}    &\textsf{WA1}\tablenotemark{a}   &\textsf{WA2}\tablenotemark{a}   &\textsf{WA3}\tablenotemark{a}   &\textsf{XCLUMPY}   				&\textsf{zgauss}   				&\textsf{const}                &\\
&$N^{\mathrm{Gal}}_{\mathrm{H}}$&$\Gamma$            &$kT_{\rm bb}$      &$N_{\rm H} $       &$N_{\rm H} $       &$N_{\rm H} $       &$N_{\mathrm{H}}^{\mathrm{Equ}}$	&$E_{\mathrm{line}}$, $\sigma_{\mathrm{line}}$, $K_{\mathrm{line}}$&$C_{\mathrm{FPMs/XIS}}$   &$\chi^2/$dof  \\
&                    &$K_{\rm P} $        &$kT_{\rm P}$       &$\log \xi $        &$\log \xi $        &$\log \xi $        &$\sigma $          				&                   &$C_{\mathrm{FPMs/pn}}$   &$\chi^2_{\mathrm{red}}$\\
&                    &                    &$ \tau $           &$C_{\rm frac}$     &$C_{\rm frac}$     &$C_{\rm frac}$     &$ i $              				&                   &$C_{\mathrm{HXD/XIS}}$		 &\\
&                    &                    &$K_{\rm S} $       &                   &                   &                   &$N_{\mathrm{H}}^{\mathrm{LOS}}$	&                   &                            & 
}
\startdata
NGC 2992&$5.89$\tablenotemark{c}&$1.71^{+0.01}_{-0.01}$&\nodata&\nodata&$0.61^{+0.14}_{-0.18}$&$1.58^{+0.16}_{-0.07}$&$0.97^{+0.06}_{-0.10}$&\nodata&\nodata&$3140.6/2478$\\
&&$19.6^{+0.2}_{-0.3}$&\nodata&\nodata&$0.10^{+0.09}_{-0.00}$\tablenotemark{d}&$0.15^{+0.23}_{-0.05}$\tablenotemark{d}&$19.0^{+0.2}_{-0.1}$&&$1.12^{+0.01}_{-0.01}$&$1.27$\\
&&&\nodata&\nodata&$0.53^{+0.05}_{-0.04}$&$0.53$\tablenotemark{b}&$45$\tablenotemark{c}&&\nodata&\\
&&&\nodata&&&&$36.4^{+1.4}_{-1.4}$&&&\\
\hline
MCG$-$5-23-16&$12.0$\tablenotemark{c}&$1.95^{+0.01}_{-0.01}$&\nodata&$0.11^{+0.05}_{-0.05}$&$221^{+10}_{-8}$&\nodata&$0.25^{+0.02}_{-0.02}$&\nodata&$1.00^{+0.01}_{-0.01}$&$4096.8/3436$\\
&&$63.3^{+2.0}_{-2.1}$&\nodata&$1.66^{+0.22}_{-0.11}$&$2.65^{+0.03}_{-0.04}$&\nodata&$26.3^{+0.3}_{-0.3}$&&\nodata&$1.19$\\
&&&\nodata&$1$\tablenotemark{c}&$0.34^{+0.01}_{-0.01}$&\nodata&$45$\tablenotemark{c}&&\nodata&\\
&&&\nodata&&&&$137^{+1}_{-1}$&&&\\
\hline
NGC 3783&$14.1$\tablenotemark{c}&$1.64^{+0.03}_{-0.02}$&$93.8^{+4.2}_{-5.2}$&$1.72^{+0.13}_{-0.16}$&$111^{+48}_{-17}$&$9.27^{+0.98}_{-0.56}$&$2.08^{+1.03}_{-0.35}$&$0.60^{+0.01}_{-0.01}$, $10$\tablenotemark{c}, $3.23^{+0.41}_{-0.41}$&\nodata&$1714.0/1391$\\
&&$9.68^{+0.66}_{-0.44}$&$2.02^{+7.61}_{-0.02}$\tablenotemark{d}&$1.53^{+0.06}_{-0.05}$&$3.43^{+0.15}_{-0.05}$&$1.13^{+0.15}_{-0.06}$&$16.7^{+0.5}_{-0.4}$&&$1.12^{+0.01}_{-0.01}$&$1.23$\\
&&&$0.010^{+0.427}_{-0.000}$\tablenotemark{d}&$1$\tablenotemark{c}&$0.68^{+0.02}_{-0.02}$&$0.68$\tablenotemark{b}&$45$\tablenotemark{c}&&\nodata&\\
&&&$16.6^{+11.1}_{-6.6}$&&&&$14.0^{+5.2}_{-1.5}$&&&\\
\hline
UGC 6728&$5.55$\tablenotemark{c}&$2.03^{+0.01}_{-0.02}$&\nodata&\nodata&$60.8^{+10.8}_{-10.5}$&\nodata&$6.74^{+6.14}_{-3.45}$&\nodata&\nodata&$668.0/652$\\
&&$7.97^{+0.67}_{-0.51}$&\nodata&\nodata&$2.54^{+0.10}_{-0.05}$&\nodata&$11.6^{+1.3}_{-1.6}$\tablenotemark{d}&&\nodata&$1.02$\\
&&&\nodata&\nodata&$0.39^{+0.04}_{-0.04}$&\nodata&$45$\tablenotemark{c}&&$1.16$\tablenotemark{c}&\\
&&&\nodata&&&&$<$0.44&&&\\
\hline
NGC 4051&$1.20$\tablenotemark{c}&$2.01^{+0.02}_{-0.02}$&$96.7^{+5.8}_{-4.6}$&\nodata&$500^{+0}_{-63}$\tablenotemark{d}&$10.5^{+0.8}_{-0.6}$&$6.01^{+1.04}_{-0.90}$&\nodata&\nodata&$1522.7/1313$\\
&&$8.92^{+0.43}_{-0.56}$&$2.08^{+12.94}_{-0.08}$\tablenotemark{d}&\nodata&$1.62^{+1.24}_{-0.85}$&$2.16^{+0.02}_{-0.02}$&$12.0^{+0.7}_{-2.0}$\tablenotemark{d}&&$1.44^{+0.01}_{-0.01}$&$1.16$\\
&&&$0.85^{+0.61}_{-0.84}$\tablenotemark{d}&\nodata&$0.43^{+0.04}_{-0.02}$&$0.43$\tablenotemark{b}&$45$\tablenotemark{c}&&\nodata&\\
&&&$1.20^{+0.08}_{-0.67}$&&&&$<$0.22&&&\\
\hline
NGC 4395&$1.93$\tablenotemark{c}&$1.50^{+0.03}_{-0.00}$\tablenotemark{d}&\nodata&\nodata&$5.05^{+0.56}_{-0.39}$&\nodata&$7.44^{+7.20}_{-3.53}$&\nodata&\nodata&$508.5/473$\\
&&$0.94^{+0.06}_{-0.02}$&\nodata&\nodata&$1.74^{+0.10}_{-0.09}$&\nodata&$14.5^{+0.6}_{-0.7}$&&\nodata&$1.08$\\
&&&\nodata&\nodata&$0.84^{+0.03}_{-0.03}$&\nodata&$45$\tablenotemark{c}&&$1.18$\tablenotemark{c}&\\
&&&\nodata&&&&$5.15^{+3.53}_{-3.21}$&&&\\
\hline
MCG$-$6-30-15&$4.74$\tablenotemark{c}&$2.16^{+0.02}_{-0.01}$&$77.8^{+1.9}_{-1.8}$&$0.24^{+0.03}_{-0.02}$&$115^{+8}_{-8}$&$2.54^{+0.18}_{-0.16}$&$18.8^{+7.3}_{-1.5}$&$0.59^{+0.01}_{-0.01}$, $10$\tablenotemark{c}, $15.7^{+0.7}_{-0.8}$&\nodata&$2621.4/2083$\\
&&$36.2^{+2.0}_{-1.7}$&$2.00^{+2.28}_{-0.00}$\tablenotemark{d}&$0.36^{+0.03}_{-0.02}$&$2.71^{+0.02}_{-0.02}$&$0.10^{+0.23}_{-0.00}$\tablenotemark{d}&$13.8^{+0.1}_{-0.2}$&$0.67^{+0.01}_{-0.01}$, $10$\tablenotemark{c}, $13.9^{+0.6}_{-0.6}$&$0.91^{+0.01}_{-0.01}$&$1.26$\\
&&&$0.010^{+0.455}_{-0.000}$\tablenotemark{d}&$1$\tablenotemark{c}&$0.26^{+0.02}_{-0.02}$&$0.26$\tablenotemark{b}&$45$\tablenotemark{c}&&\nodata&\\
&&&$29.4^{+2.3}_{-14.6}$&&&&$4.57^{+0.72}_{-0.74}$&&&\\
\hline
IC 4329A&$5.60$\tablenotemark{c}&$1.77^{+0.01}_{-0.01}$&\nodata&$0.073^{+0.042}_{-0.036}$&$319^{+31}_{-23}$&\nodata&$0.26^{+0.02}_{-0.02}$&$0.79^{+0.01}_{-0.01}$, $20$\tablenotemark{c}, $2.21^{+0.28}_{-0.28}$&$0.98^{+0.01}_{-0.01}$&$1523.2/1248$\\
&&$37.6^{+0.9}_{-0.8}$&\nodata&$1.56^{+0.19}_{-0.23}$&$2.96^{+0.01}_{-0.01}$&\nodata&$22.1^{+0.2}_{-0.2}$&&\nodata&$1.22$\\
&&&\nodata&$1$\tablenotemark{c}&$0.21^{+0.01}_{-0.01}$&\nodata&$45$\tablenotemark{c}&&\nodata&\\
&&&\nodata&&&&$40.7^{+1.0}_{-1.3}$&&&\\
\hline
NGC 6814&$14.8$\tablenotemark{c}&$1.60^{+0.05}_{-0.03}$&\nodata&\nodata&$298^{+202}_{-170}$\tablenotemark{d}&\nodata&$1.44^{+4.28}_{-0.92}$&\nodata&\nodata&$368.0/332$\\
&&$2.54^{+1.34}_{-0.79}$&\nodata&\nodata&$2.96^{+0.29}_{-0.35}$&\nodata&$12.1^{+3.9}_{-2.1}$\tablenotemark{d}&&\nodata&$1.11$\\
&&&\nodata&\nodata&$0.34^{+0.22}_{-0.09}$&\nodata&$45$\tablenotemark{c}&&$1.16$\tablenotemark{c}&\\
&&&\nodata&&&&$<$2.2&&&\\
\hline
NGC 7213&$1.11$\tablenotemark{c}&$1.63^{+0.10}_{-0.05}$&$123^{+8.1}_{-10.0}$&\nodata&$4.27^{+0.46}_{-0.61}$&\nodata&$1.82^{+1.45}_{-0.36}$&\nodata&\nodata&$1874.9/1613$\\
&&$4.53^{+1.33}_{-0.88}$&$4.75^{+1.10}_{-2.36}$&\nodata&$1.76^{+0.08}_{-0.05}$&\nodata&$13.8^{+0.9}_{-1.4}$&&\nodata&$1.16$\\
&&&$2.69^{+7.87}_{-0.94}$&\nodata&$0.25^{+0.05}_{-0.04}$&\nodata&$45$\tablenotemark{c}&&$1.16$\tablenotemark{c}&\\
&&&$0.33^{+0.12}_{-0.17}$&&&&$0.36^{+1.09}_{-0.33}$&&&\\
\hline
NGC 7314&$1.59$\tablenotemark{c}&$1.99^{+0.02}_{-0.01}$&\nodata&$0.33^{+0.04}_{-0.04}$&$1.10^{+0.22}_{-0.18}$&$500^{+0}_{-8}$\tablenotemark{d}&$0.16^{+0.06}_{-0.02}$&\nodata&\nodata&$1945.0/1698$\\
&&$23.0^{+1.4}_{-0.8}$&\nodata&$0.64^{+0.20}_{-0.10}$&$0.10^{+0.43}_{-0.00}$\tablenotemark{d}&$3.17^{+0.03}_{-0.04}$&$24.2^{+0.6}_{-1.1}$&&$0.96^{+0.01}_{-0.01}$&$1.15$\\
&&&\nodata&$1$\tablenotemark{c}&$0.38^{+0.02}_{-0.02}$&$0.38$\tablenotemark{b}&$45$\tablenotemark{c}&&\nodata&\\
&&&\nodata&&&&$49.5^{+2.0}_{-2.4}$&&&\\
\hline
NGC 7469&$5.39$\tablenotemark{c}&$1.86^{+0.02}_{-0.02}$&$80.6^{+2.7}_{-2.0}$&$0.16^{+0.05}_{-0.04}$&$500^{+0}_{-13}$\tablenotemark{d}&\nodata&$2.06^{+1.22}_{-0.47}$&\nodata&\nodata&$1696.2/1561$\\
&&$10.1^{+0.6}_{-0.5}$&$2.00^{+1.97}_{-0.00}$\tablenotemark{d}&$2.39^{+0.07}_{-0.06}$&$3.31^{+0.06}_{-0.05}$&\nodata&$12.4^{+0.8}_{-1.3}$&&$1.13^{+0.01}_{-0.01}$&$1.09$\\
&&&$3.41^{+6.86}_{-0.46}$&$1$\tablenotemark{c}&$0.18^{+0.03}_{-0.03}$&\nodata&$45$\tablenotemark{c}&&\nodata&\\
&&&$2.63^{+0.13}_{-2.21}$&&&&$<$0.15&&&\\
\enddata
\tablecomments{
(1) Galaxy name. 
(2) Hydrogen column density of the Galactic absorption in [$10^{20}$~cm$^{-2}$].
(3) Photon index of the direct component, and the normalization at 1~keV in [$10^{-3}$~photons~keV$^{-1}$~cm$^{-2}$~s$^{-1}$]. 
(4) Input soft photon temperature in [eV], plasma temperature in [keV], plasma optical depth, and normalization of soft excess component at 1~keV in [$10^{-2}$~photons~keV$^{-1}$~cm$^{-2}$~s$^{-1}$].
(5)--(7) Hydrogen column density of a ionized absorber in [$10^{22}$~cm$^{-2}$], its logarithmic ionization parameter ($\xi$ in [$\rm erg~cm~s^{-1}$]), and its covering fraction.
(8) Hydrogen column density along the equatorial plane in [$10^{24}$~cm$^{-2}$], torus angular width in [degree], inclination angle of the observer in [degree], and hydrogen column density along the line of sight in [$10^{20}$~cm$^{-2}$].
(9) Energy of the emission line in [keV], its line width in [eV], and its normalization at 1~keV in [$10^{-4}$~photons~keV$^{-1}$~cm$^{-2}$ s$^{-1}$]. 
(10) Relative normalization of \textit{NuSTAR}/FPMs to \textit{Suzaku}/XIS, relative normalization of \textit{NuSTAR} to \textit{XMM-Newton}/EPIC-pn, and relative normalization of \textit{Suzaku}/HXD to \textit{Suzaku}/XIS.
}
\tablenotetext{a}{The \textsf{WA1} model is a full absorber, and the \textsf{WA2} and \textsf{WA3} models represent partial absorbers.}
\tablenotetext{b}{The value is linked to that of \textsf{WA2}.}
\tablenotetext{c}{Fixed.}
\tablenotetext{d}{The parameter reaches a limit of its allowed range.}
\end{deluxetable*}
\end{longrotatetable}

\begin{longrotatetable}
\begin{deluxetable*}{lllllllll}
\tablecaption{Best-fit Parameters of obscured AGNs\label{Best_fit_OAGN}}
\tablehead{
(1)&(2)&(3)&(4)&(5)&(6)&(7)&(8)&\\
Object&\textsf{phabs}&\textsf{zcutoffpl}&\textsf{apec}&\textsf{XCLUMPY}&\textsf{const3}&\textsf{const4}&\textsf{const(1,2)}&\\
&$N^{\mathrm{Gal}}_{\mathrm{H}}$&$\Gamma$&$kT_1 $&$N_{\mathrm{H}}^{\mathrm{Equ}}$&$f_{\rm scat} $&$N_{\rm Line} $&$C_{\mathrm{TIME}}$&$\chi^2/$dof \\
&&$K_{\rm P} $&$K_{\mathrm{A1}} $&$\sigma $&&&$C_{\mathrm{FPMs/XIS}}$&$\chi^2_{\mathrm{red}}$\\
&&&$kT_2 $&$ i $&&&$C_{\mathrm{HXD/XIS}}$&         \\ 
&&&$K_{\mathrm{A2}} $&$N_{\mathrm{H}}^{\mathrm{LOS}}$&&&&         
}
\startdata
NGC 3081&$4.63$\tablenotemark{a}&$1.81^{+0.69}_{-0.25}$\tablenotemark{b}&$0.99^{+0.27}_{-0.20}$&$5.65^{+10.20}_{-4.69}$&$0.42^{+0.40}_{-0.26}$&$1$\tablenotemark{a}&$1$\tablenotemark{a}&$96.9/82$\\
&&$0.90^{+0.84}_{-0.45}$&$2.15^{+0.78}_{-0.41}$&$23.1^{+12.5}_{-6.3}$&&&\nodata&$1.18$\\
&&&$0.19^{+0.04}_{-0.03}$&$57.5^{+24.6}_{-10.0}$&&&$1.18$\tablenotemark{c}\tablenotemark{a}&\\
&&&$19.9^{+14.6}_{-8.5}$&$7.77^{+1.22}_{-0.76}$&&&&\\
\hline
NGC 4388&$2.87$\tablenotemark{a}&$1.50^{+0.01}_{-0.00}$\tablenotemark{b}&$0.30^{+0.02}_{-0.02}$&$0.90^{+0.14}_{-0.19}$&$1.69^{+0.05}_{-0.06}$&$1.68^{+0.34}_{-0.30}$&$0.26^{+0.02}_{-0.02}$&$2071.9/1974$\\
&&$1.07^{+0.03}_{-0.03}$&$9.50^{+1.15}_{-1.16}$&$19.7^{+4.0}_{-3.8}$&&&$1$\tablenotemark{a}&$1.05$\\
&&&$1.21^{+0.03}_{-0.03}$&$69.5^{+2.7}_{-2.4}$&&&\nodata&\\
&&&$13.8^{+1.0}_{-1.0}$&$3.02^{+0.05}_{-0.05}$&&&&\\
\hline
Centaurus A&$11.7$\tablenotemark{a}&$1.76^{+0.01}_{-0.01}$&$0.78^{+0.07}_{-0.07}$&$0.18^{+0.08}_{-0.03}$&$0.62^{+0.04}_{-0.05}$&$0.70^{+0.15}_{-0.08}$&$1.30^{+0.01}_{-0.01}$&$3095.0/2919$\\
&&$17.1^{+0.4}_{-0.2}$&$31.1^{+5.1}_{-5.2}$&$49.7^{+4.3}_{-8.5}$&&&$1$\tablenotemark{a}&$1.06$\\
&&&\nodata&$51.6^{+6.8}_{-3.1}$&&&\nodata&\\
&&&\nodata&$0.99^{+0.01}_{-0.01}$&&&&\\
\hline
NGC 6300&$10.6$\tablenotemark{a}&$1.79^{+0.05}_{-0.06}$&$0.59^{+0.16}_{-0.33}$&$9.78^{+6.21}_{-2.88}$&$0.68^{+0.10}_{-0.08}$&$1$\tablenotemark{a}&$0.95^{+0.03}_{-0.03}$&$1128.5/1059$\\
&&$1.23^{+0.11}_{-0.14}$&$2.55^{+0.69}_{-0.70}$&$19.1^{+2.1}_{-1.6}$&&&$1$\tablenotemark{a}&$1.07$\\
&&&\nodata&$52.7^{+4.7}_{-4.7}$&&&\nodata&\\
&&&\nodata&$2.09^{+0.07}_{-0.06}$&&&&\\
\enddata
\tablecomments{
(1) Galaxy name. 
(2) Hydrogen column density of the Galactic absorption in [$10^{20}$~cm$^{-2}$].
(3) Photon index of the direct component, and the normalization at 1~keV in [10$^{-2}$~photons~keV$^{-1}$~cm$^{-2}$~s$^{-1}$]. 
(4) Temperature of the \textsf{apec} model in [keV] and normalization of the \textsf{apec} model in [10$^{-19}/4\pi [D_{\mathrm{A}}(1+z)]^2 \int n_{\mathrm{e}} n_{\mathrm{H}}dV$], where $D_{\mathrm{A}}$ is the angular diameter distance to the source in [cm], $n_{\mathrm{e}}$ and $n_{\mathrm{H}}$ are the electron and hydrogen densities in [cm$^{-3}$].
(5) Hydrogen column density along the equatorial plane in [10$^{24}$~cm$^{-2}$], torus angular width in [degree], inclination angle of the observer in [degree], and hydrogen column density along the line of sight in [10$^{23}$~cm$^{-2}$].
(6) Scattering fraction in [\%]. 
(7) Relative normalization of the emission lines to the reflection component. 
(8) Time variability constant, relative normalization of \textit{NuSTAR}/FPMs to \textit{Suzaku}/XIS, and relative normalization of \textit{Suzaku}/HXD to \textit{Suzaku}/XIS.
}
\tablenotetext{a}{Fixed.}
\tablenotetext{b}{The parameter reaches a limit of its allowed range.}
\end{deluxetable*}
\end{longrotatetable}
\begin{deluxetable*}{llllll}
\tablecaption{X-ray Luminosity\label{lumi}}
\tablehead{\\
Object  &$\log L_{2-10}$\tablenotemark{a}   &$\log \lambda _{\mathrm{Edd}}$  &$\log M_{\mathrm{BH}}/M_\odot$  &Reference  
\\(1)&(2)&(3)&(4)&(5)
}
\startdata
NGC 2992 & 43.0 & $-1.52$ & 7.72 & a \\
MCG$-$5--23--16 & 43.4 & $-1.34$ & 7.98 & b \\
NGC 3783 & 42.9 & $-1.13$ & 7.27 & b \\
UGC 6728 & 42.3 & $-0.40$ & 5.85 & c \\
NGC 4051 & 41.9 & $-0.48$ & 5.60 & d \\
NGC 4395 & 40.0 & $-1.72$ & 4.88 & b \\
MCG$-$6--30--15 & 43 & $-1.24$ & 7.42 & b \\
IC 4329A & 43.9 & $-0.76$ & 7.84 & d \\
NGC 6814 & 41.8 & $-1.41$ & 6.46 & d \\
NGC 7213 & 42.2 & $-1.99$ & 7.37 & e \\
NGC 7314 & 42.5 & $-1.57$ & 7.24 & d \\
NGC 7469 & 43.3 & $-0.65$ & 7.11 & d \\
NGC 3081 & 42.6 & $-1.87$ & 7.70 & f \\
NGC 4388 & 43.0 & $-1.84$ & 8.00 & f \\
Centaurus A & 42.0 & $-2.70$ & 7.94 & b \\
NGC 6300 & 42.1 & $-1.99$ & 7.30 & f \\
\enddata
\tablecomments{
(1) Galaxy name. 
(2) Logarithmic intrinsic luminosity in the 2--10~keV band. 
(3) Logarithmic Eddington ratio ($\lambda_{\mathrm{Edd}} = L_{\mathrm{bol}}/L_{\mathrm{Edd}}$). Here we obtained the bolometric luminosity as $L_{\mathrm{bol}} = 20 L_{2-10}$ and defined the Eddington luminosity as $L_{\mathrm{Edd}} = 1.26 \times 10^{38} M_{\mathrm{BH}}/M_{\sun}$. 
(4) Logarithmic black hole mass. 
(5) Reference for the black hole mass.
}
\tablenotetext{a}{We calculate the X-ray luminosities from the redshifts except for
the very nearby objects NGC~4051, NGC~4395, and Centaurus~A, for which we adopt distances of
$17.6$ \citep{Yoshii2014}, $3.85$ \citep{Tully2009}, and $3.84$~Mpc \citep{Rejkuba2004}, respectively.}
\tablerefs{(a) \cite{Woo2002}. (b) \cite{Garcia2019}. (c) \cite{Bentz2016}. (d) \cite{Koss2017}. (e) \cite{Vasudevan2010}. (f) \cite{Kawamuro2016a}.}
\end{deluxetable*}

\subsection{Fitting Results}

We find that Model~1 and Model~2 give a fairly good description of the
broadband spectra of all objects ($\chi^2_{\mathrm{red}} < 1.3$).
Tables~\ref{Best_fit_UAGN} and~\ref{Best_fit_OAGN} summarize the
best-fit parameters of the unobscured and obscured AGNs, respectively.
On the left sides of Figures~\ref{fig1} and~\ref{fig2},
we overplot the best-fit models folded with the energy responses and 
their residuals from the data.  
The right sides of Figures~\ref{fig1} and~\ref{fig2} 
plot the best-fit models in units of $E I(E)$ (where
$I(E)$ is the energy flux) for unobscured and obscured AGNs,
respectively. Table~\ref{lumi} lists the best-estimated intrinsic
luminosities and Eddington ratios with adopted black hole masses. Here,
we define the Eddington luminosity as $L_\mathrm{Edd} = 1.26 \times
10^{38} (M_\mathrm{BH}/M_\odot) \mathrm{ \,erg \, s^{-1}}$, and convert
the 2--10~keV luminosities to bolometric ones with a correction factor
of 20.

Since all of our targets are nearby and bright AGNs, they have been
extensively studied by many authors. In Appendix~\ref{A1}, we
compare our results with previous works of broadband X-ray spectroscopy 
that utilized the same data as used here, when available.

\section{AGN Torus Properties}
\label{sec5}

We have shown that the torus reflection components in AGN X-ray spectra
can be well represented with the XCLUMPY model, confirming the previous
studies that applied it to broadband X-ray spectra of local AGNs
\citep{Ogawa2019,Tanimoto2019,Yamada2020,Tanimoto2020}. Combining the
16 objects newly analyzed in this paper and the 12 obscured AGNs studied
by \citet{Tanimoto2020}, we construct the largest sample whose X-ray and
infrared spectra are uniformly analyzed with the XCLUMPY and CLUMPY
models, respectively. Although our sample is not a statistically
well-defined sample, we may regard it as a representative one including
various types of AGNs. In particular, we have now included 12
unobscured AGNs in the sample, which gives us a new insight into the
torus structure.

\subsection{Torus Covering Factor as a Function of Eddington Ratio}
\label{sec5.1}

\begin{figure*}
\gridline{
\fig{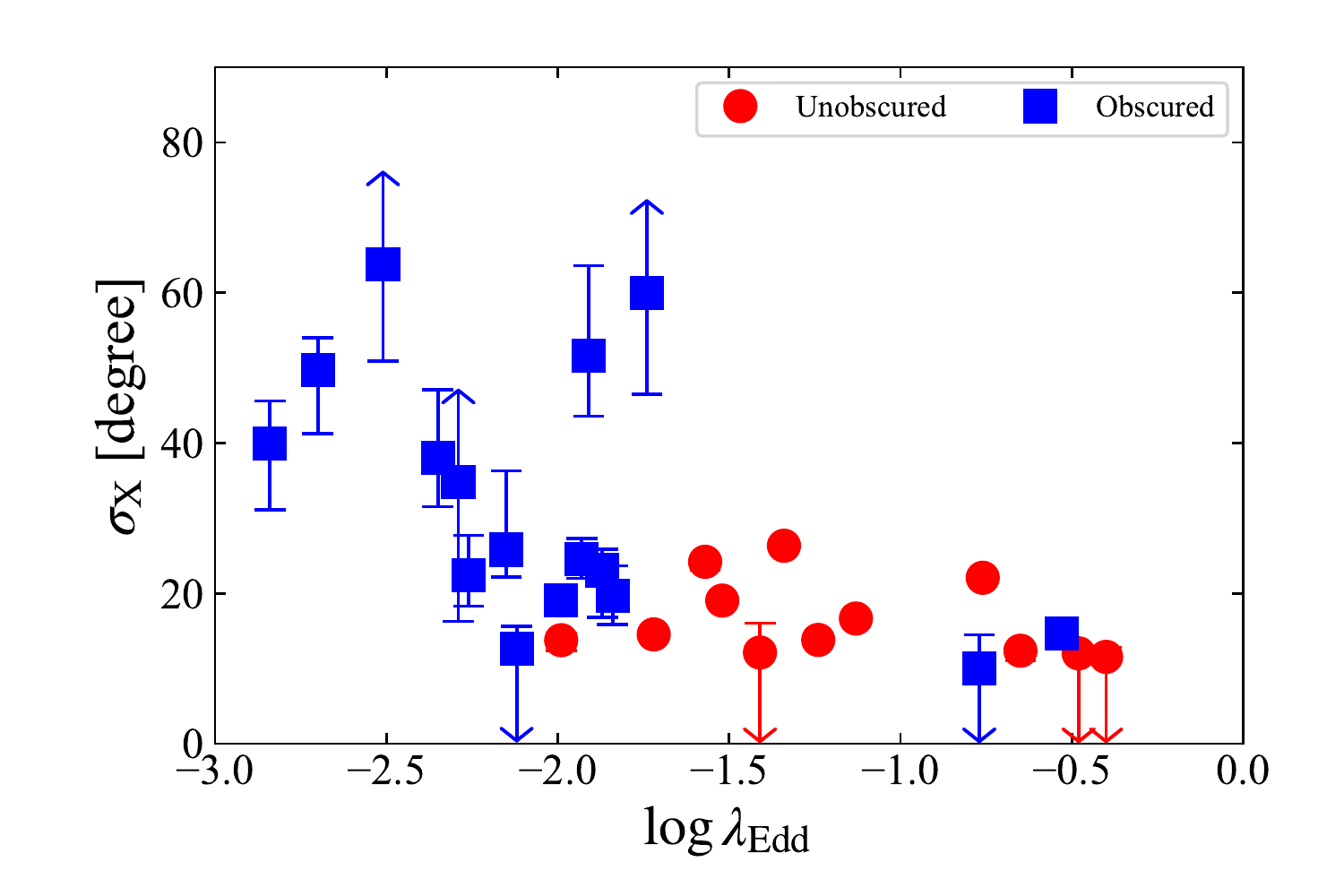}{0.52\textwidth}{(a)}
\fig{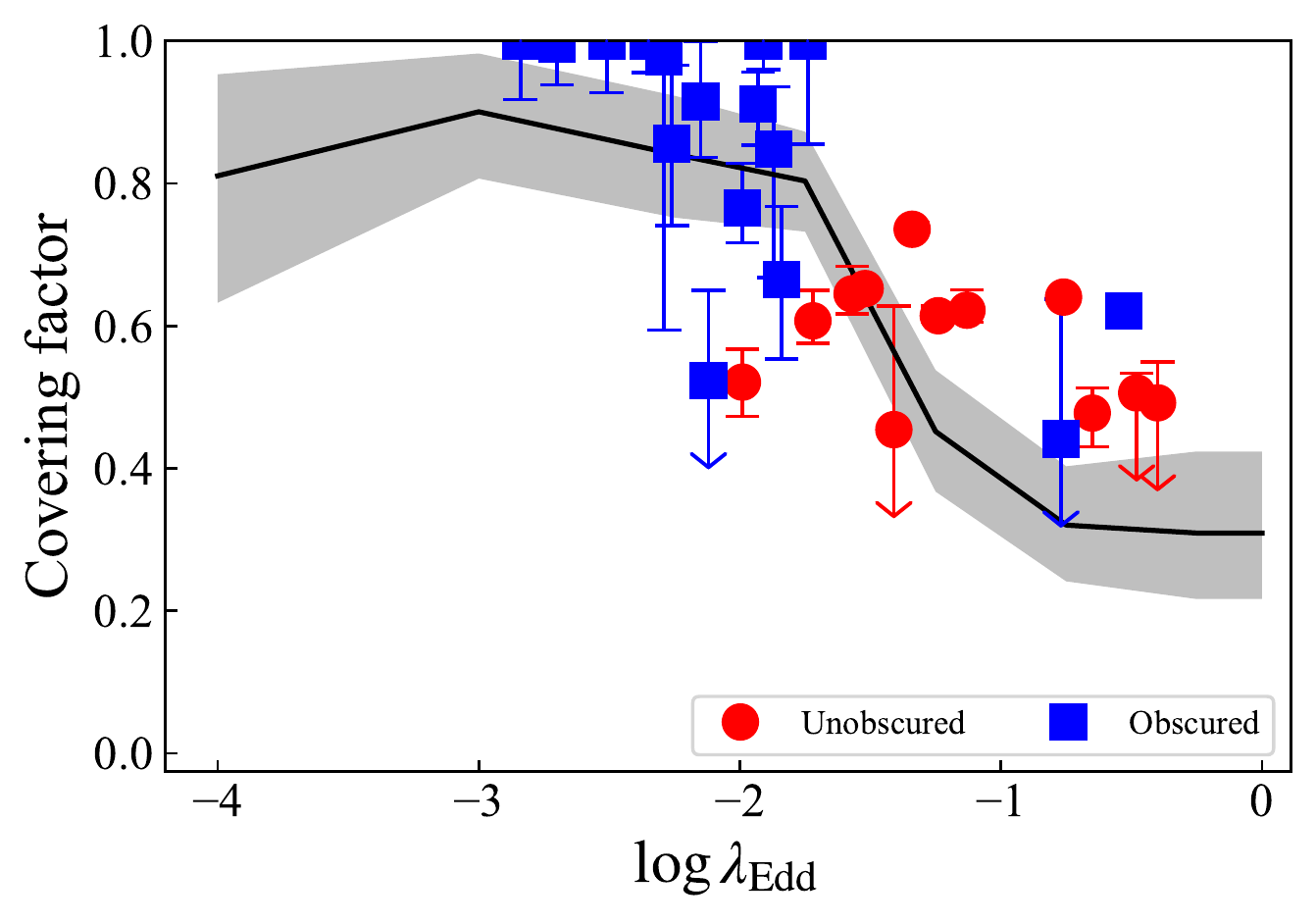}{0.48\textwidth}{(b)}
}
\caption{
(a): Relation between the torus angular width ($\sigma_{\rm X}$) and the
 Eddington ratio ($\lambda_{\mathrm{Edd}})$. 
(b): Relation between the torus covering factor ($C_\mathrm{T}$) and the
 Eddington ratio ($\lambda_{\mathrm{Edd}})$. The black curve and shaded
 region represent the best-fit and 1$\sigma$ error region obtained by
 \citet{Ricci2017Nature}.
The blue squares and red circles denote the obscured and unobscured AGNs,
respectively.
The arrows represent the results reach upper or lower boundary.
}
\label{fig-edd}
\end{figure*}

In this subsection, we summarize the torus properties of our sample
obtained from the X-ray spectral analysis described in
Section~\ref{sec4}. On the basis of the unobscured AGN fraction in a hard
X-ray selected sample, \citet{Ricci2017Nature} have shown that the Eddington ratio
(i.e., the luminosity normalized by the black hole mass) is a key
parameter that determines the torus geometry; since radiation pressure
works to expel dusty gas, the torus covering fraction rapidly decreases
when $\log \lambda_{\rm Edd} \gtrsim -1.5$. Thus, it is important to
check whether the torus structure we determine for the individual 
objects through X-ray spectroscopy is consistent with their prediction.

Figure~\ref{fig-edd}(a) plots 
the torus angular width determined from
the X-ray spectra ($\sigma_{\rm X}$) against Eddington ratio.
By including the unobscured AGNs, which 
show higher Eddington ratios than most of the obscured
AGNs in our sample, we confirm the trend reported by \citet{Tanimoto2020}
that AGNs with high Eddington ratio have smaller $\sigma_{\rm X}$.
To be more quantitative, we estimate the covering factor of the torus 
with $N_\mathrm{H} > 10^{22}$ cm$^{-2}$), $C_\mathrm{T}$, that is, the
fractional solid angle of torus material whose mean column density
exceeds $10^{22}$~cm$^{-2}$ in the XCLUMPY geometry.
In XCLUMPY, the mean hydrogen column density at the elevation angle 
$\theta_*$ ($\equiv$ $90^\circ -i_\mathrm{X}$) is given by 
\begin{eqnarray}
    N_{\rm H}\left(\theta_*\right) = N^{\rm Equ}_{\rm H} \exp\left(-\left(\frac{\theta_*}{\sigma_\mathrm{X}}\right)^2\right).
\end{eqnarray}
Defining $\theta_{\rm c}$ such that $N_{\rm H}\left(\theta_{\rm c}\right) = 10^{22}~\rm cm^{-2}$, 
$C_\mathrm{T}$ can be calculated from the torus parameters as 
\begin{eqnarray}
C_\mathrm{T} & = & \frac{1}{4\pi} \int_{0}^{2\pi} \!\!\!\int_{\frac{\pi}{2} - \theta_\mathrm{c}}^{\frac{\pi}{2} + \theta_\mathrm{c}} \sin \theta d\theta d\phi \nonumber\\
& = & \sin\left(\sigma_\mathrm{X} \sqrt{\ln\left(\frac{N_\mathrm{H}^\mathrm{Equ}}{10^{22}~\textrm{cm}^{-2}}\right)} \right).
\end{eqnarray}
Figure~\ref{fig-edd}(b) plots $C_\mathrm{T}$ against $\lambda_{\rm Edd}$.  
As noticed, our results follow the trend of the $C_\mathrm{T}$ vs
$\lambda_{\rm Edd}$ relation by \citet{Ricci2017ApJ} based on the
statistical study.
The good agreement
supports our assumption of $i_\mathrm{X}=45^{\circ}$ for unobscured AGNs in our
sample (see Appendix~\ref{A2} for the results with different $i_\mathrm{X}$ values).

\subsection{Comparison with the Infrared Results}
\label{sec5.2}

\begin{figure*}
\gridline{\fig{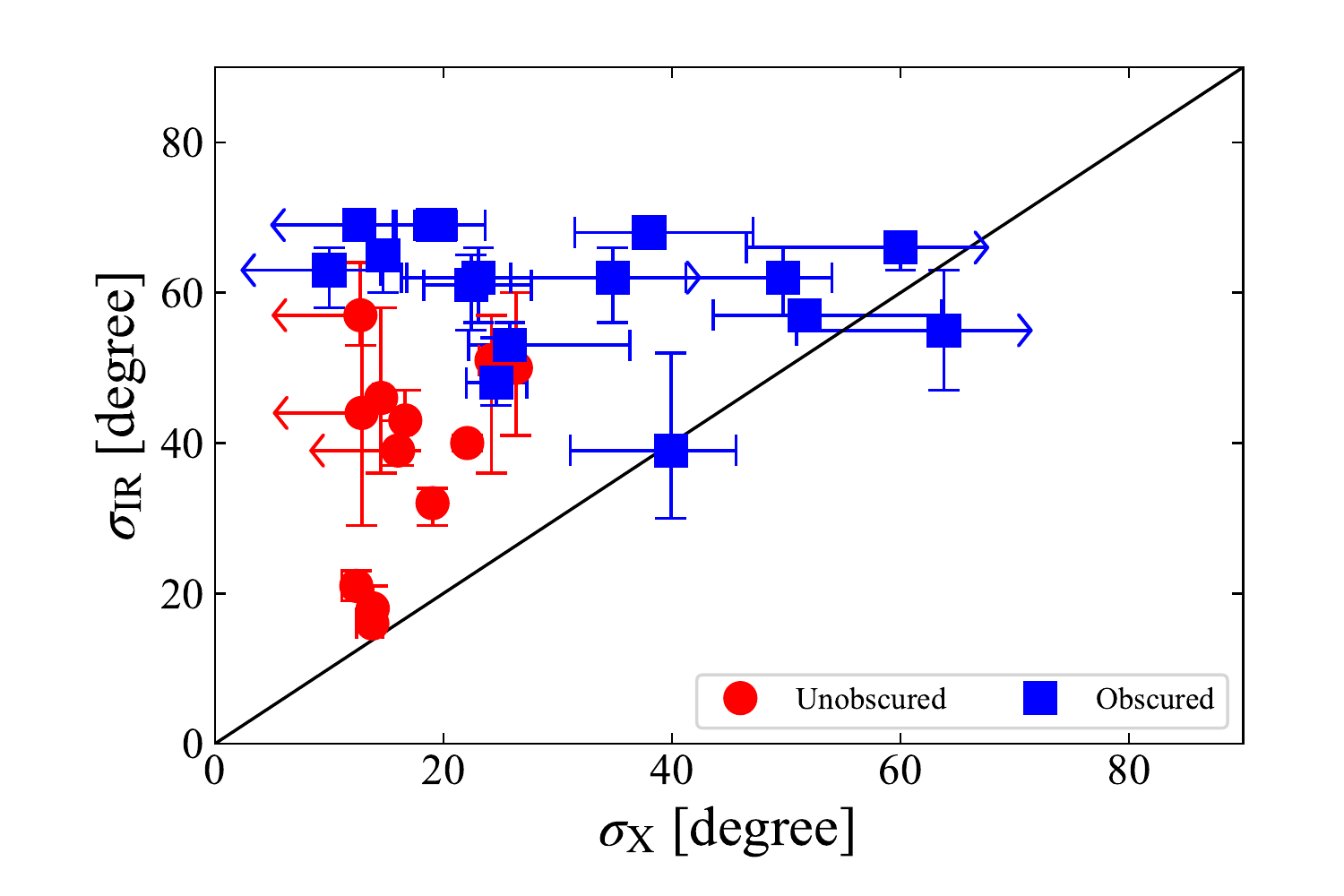}{0.5\textwidth}{(a)}
          \fig{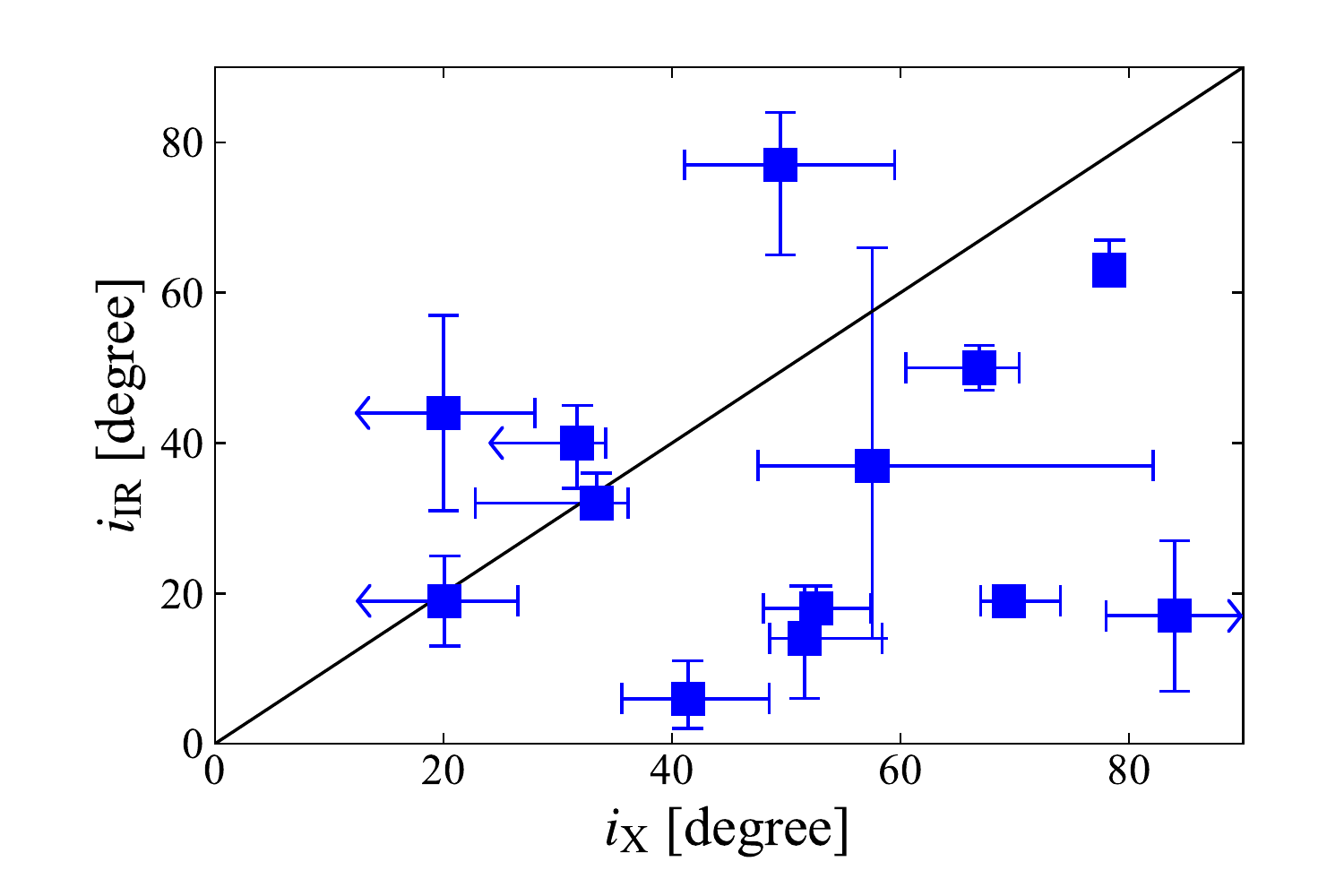}{0.5\textwidth}{(b)}}
\gridline{\fig{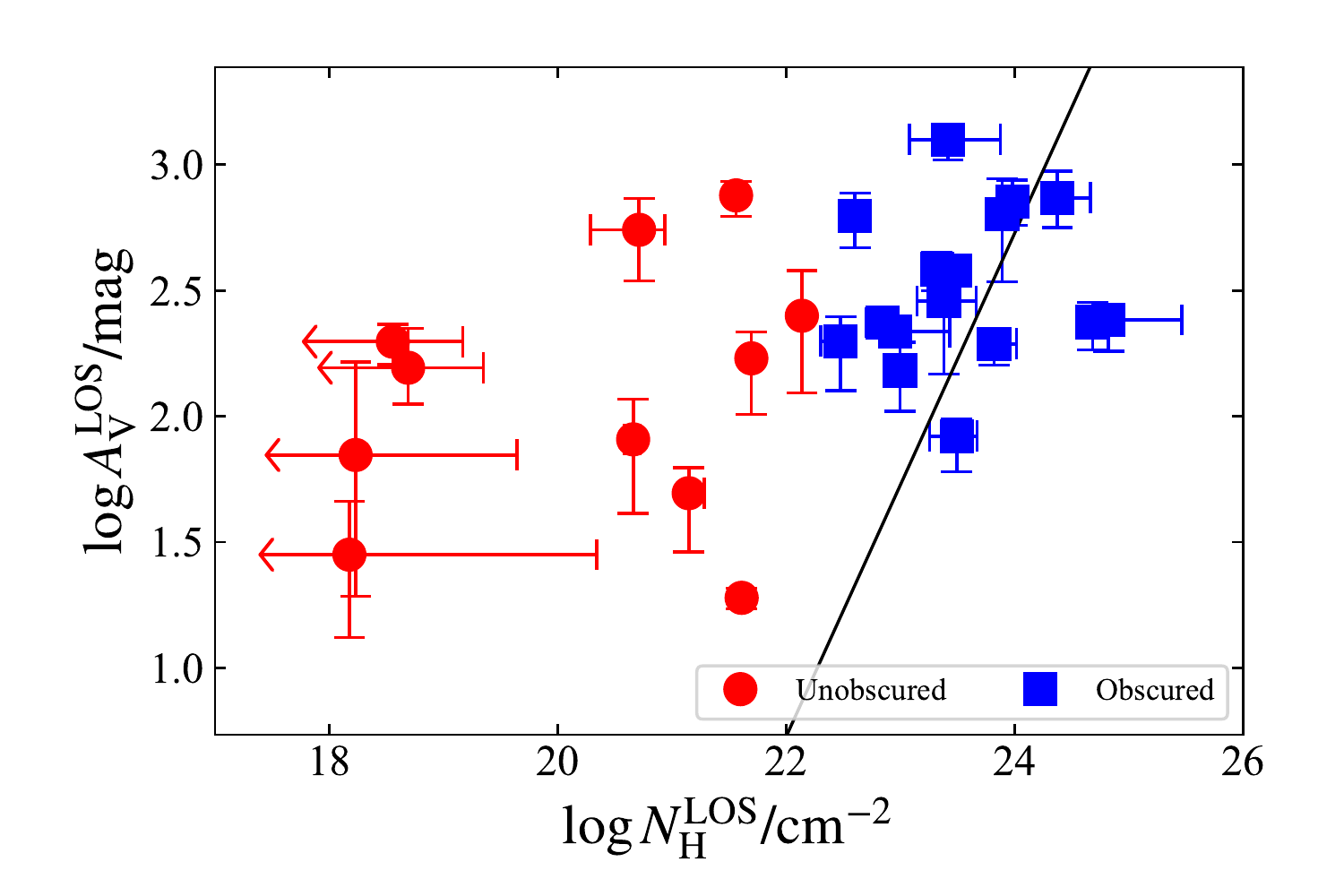}{0.5\textwidth}{(c)}
          \fig{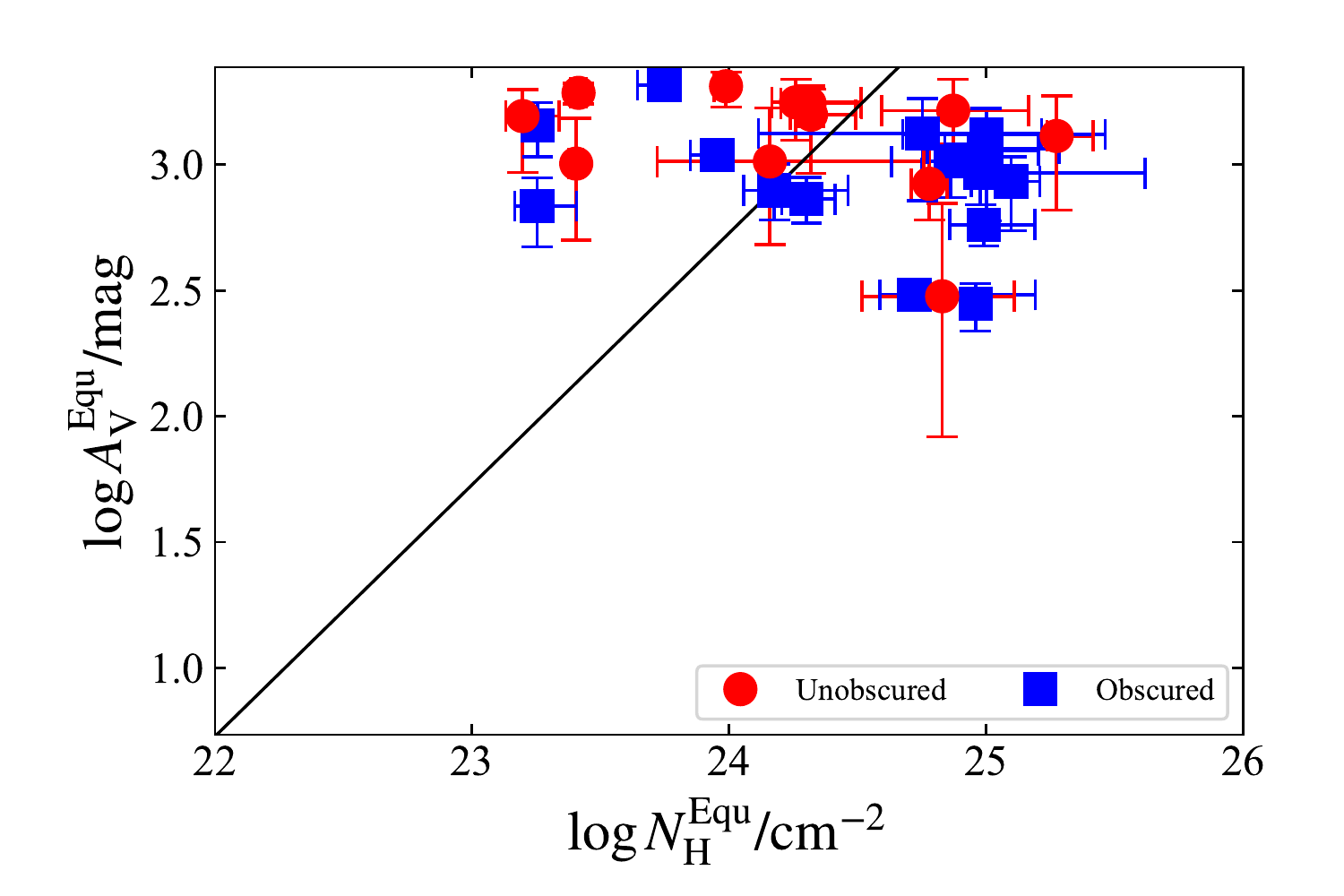}{0.5\textwidth}{(d)}}
\caption{
(a): Comparison between the torus angular width obtained from the infrared spectra ($\sigma_{\mathrm{IR}}$) and that from the X-ray ones ($\sigma_{\mathrm{X}}$).
The black line shows $\sigma_{\mathrm{X}} = \sigma_{\mathrm{IR}}$.
(b): Comparison between the inclination angle obtained from the infrared spectra ($i_{\mathrm{IR}}$) and that from the X-ray ones ($i_\mathrm{X}$).
The black line shows $i_\mathrm{X} = i_{\mathrm{IR}}$.
(c): Comparison between the V-band extinction along the line of sight ($A_{\mathrm{V}}^{\mathrm{LOS}}$) and the hydrogen column density along the line of sight ($N_{\mathrm{H}}^{\mathrm{LOS}}$). The black line corresponds to the Galactic value.
(d): Comparison between and the V-band extinction along the equatorial
 plane ($A_{\mathrm{V}}^{\mathrm{Equ}}$) and the hydrogen column density
 along the equatorial plane ($N_{\mathrm{H}}^{\mathrm{Equ}}$).
The ranges of the x- and y-axes 
correspond to the parameter 
boundaries of $N_{\mathrm{H}}^{\mathrm{Equ}}$ 
and $A_{\mathrm{V}}^{\mathrm{Equ}}$, respectively.
The black line corresponds to the Galactic value.
The blue squares and red circles denote the obscured and unobscured AGNs,
respectively.
The arrows represent the results reach upper or lower boundary.
}
\label{fig-ix}
\end{figure*}

Figure~\ref{fig-ix} compares the torus parameters determined from the
X-ray spectra and those from the infrared ones. We recall that the X-ray
results trace all material including gas and dust, whereas the infrared
ones only dust. Hence, if the spatial distribution of gas and dust is
not exactly the same and/or dust temperature is not spatially uniform,
it is possible that they give different solutions.

As noticed from Figure~\ref{fig-ix}(a), we confirm the finding by
\citet{Tanimoto2020} that the torus angular widths obtained from the
infrared spectra ($\sigma_\mathrm{IR}$) become systematically larger
than the those obtained from the X-ray ones ($\sigma_\mathrm{X}$), even
including unobscured AGNs (Figure~\ref{fig-ix}(a)). \citet{Tanimoto2020}
suggested that this apparent discrepancy may be explained by significant
contribution from polar dusty outflows to the observed infrared flux, as
found by infrared interferemetric observations of nearby AGNs
\citep{Tristram2014,LopezGonzaga2016,Leftley2018,Lyu2018}. \citet{Tanimoto2020}
argued that this effect became the most prominent at highest
inclinations because the flux from hot dust in the inner part of the
torus was reduced due to extinction by outer cooler dust. Our results
suggest that this effect is still present in our unobscured AGNs, which
might have inclinations of $\sim$45$^{\circ}$ on average. Although both
CLUMPY and XCLUMPY do not include such polar components, the effect is
more severe in the infrared spectra than in the X-ray ones. In X-rays,
because the total mass in such polar outflows is much smaller than that
contained in the torus itself, the reflection component from the polar
outflow is weak compared with the torus reflection component at energies
above a few keV \citep{Liu2019}.

In Figure~\ref{fig-ix}(b), we compare the inclination angles determined
from the X-ray ($i_\mathrm{X}$) and infrared spectra ($i_\mathrm{IR}$)
for obscured AGNs by adding 4 sources to the original
\citet{Tanimoto2020} sample. Here only objects whose $i_\mathrm{X}$ is
determined with an accuracy of $<$30$^{\circ}$ are included; unobscured
AGNs are not included because we cannot directly constrain their
$i_\mathrm{X}$ from the spectra. We confirm that $i_\mathrm{X}$ is
generally larger than $i_\mathrm{IR}$. If $\sigma_\mathrm{IR}$ is
overestimated as discussed above, then it could affect
$i_\mathrm{IR}$ to be underestimated due to their degeneracy
\citep{Nenkova2008b,RamosAlmeida2014}.

Figure~\ref{fig-ix}(c) confirms the results by \citet{Tanimoto2020} that
the $N_{\rm H}/A_{\rm V}$ ratios along the line of sight in obscured
AGNs are similar to the Galactic value ($ N_{\mathrm{H}}/A_{\mathrm{V}} = 1.87 \times 10^{21}~\rm{cm^{-2}~mag^{-1}}$; \citealt{Draine2003}) on average with a scatter of
$\sim$1 dex. By contrast, unobscured AGNs show systematically smaller
$N_{\rm H}/A_{\rm V}$ values than the Galactic one. 
In an unobscured AGN, the line-of-sight $A_{\rm
V}$ is not directly determined by extinction along the path 
toward the central region
but is constrained by the infrared emission in the torus
region (see next section). If the torus angular width
$\sigma_\mathrm{IR}$ is overestimated
by a larger factor than that in the elevation angle 
$(90^{\circ}-i_\mathrm{IR})$,  
it would make the line-of-sight $A_{\rm V}$ larger than the true value
(see Equation~\ref{nhlos}).

Figure~\ref{fig-ix}(d) also suggests that the $N_{\rm H}/A_{\rm V}$
ratios along the equatorial plane are similar to the Galactic value on
average for both obscured and unobscured AGNs\footnote{
Our new results of IC~4329A and NGC~7469 support the conclusion by
\citet{Ogawa2019}, who assumed the same inclination and torus angular width as
the infrared values, that their tori are dust rich compared with the
Galactic ISM.}.
It is consistent with the results of
$N_{\mathrm{H}}^{\mathrm{LOS}}$ obtained for obscured AGNs
(Figure~\ref{fig-ix}(c)), which are more directly constrained by the
data through X-ray absorption ($N_{\mathrm{H}}$) and infrared silicate
absorption feature ($A_{\mathrm{V}}$). Our results imply that the
gas-to-dust ratios of AGN tori are similar to the Galactic value on
average but may be variable object to object with $\sim$1 dex.

\section{Unified Picture of AGN Structure}
\label{sec6}

\begin{figure*}
\gridline{\fig{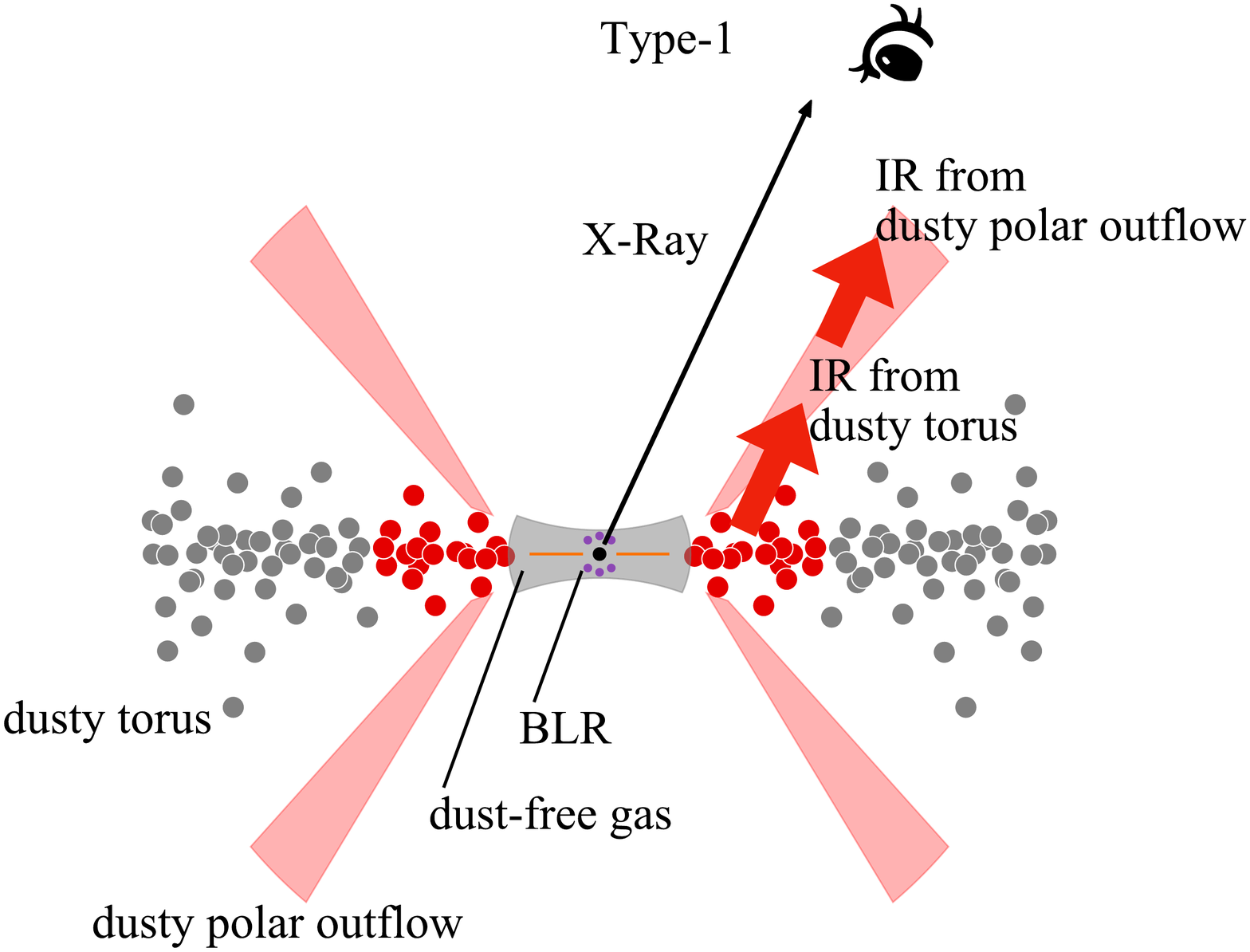}{0.5\textwidth}{(a)}
          \fig{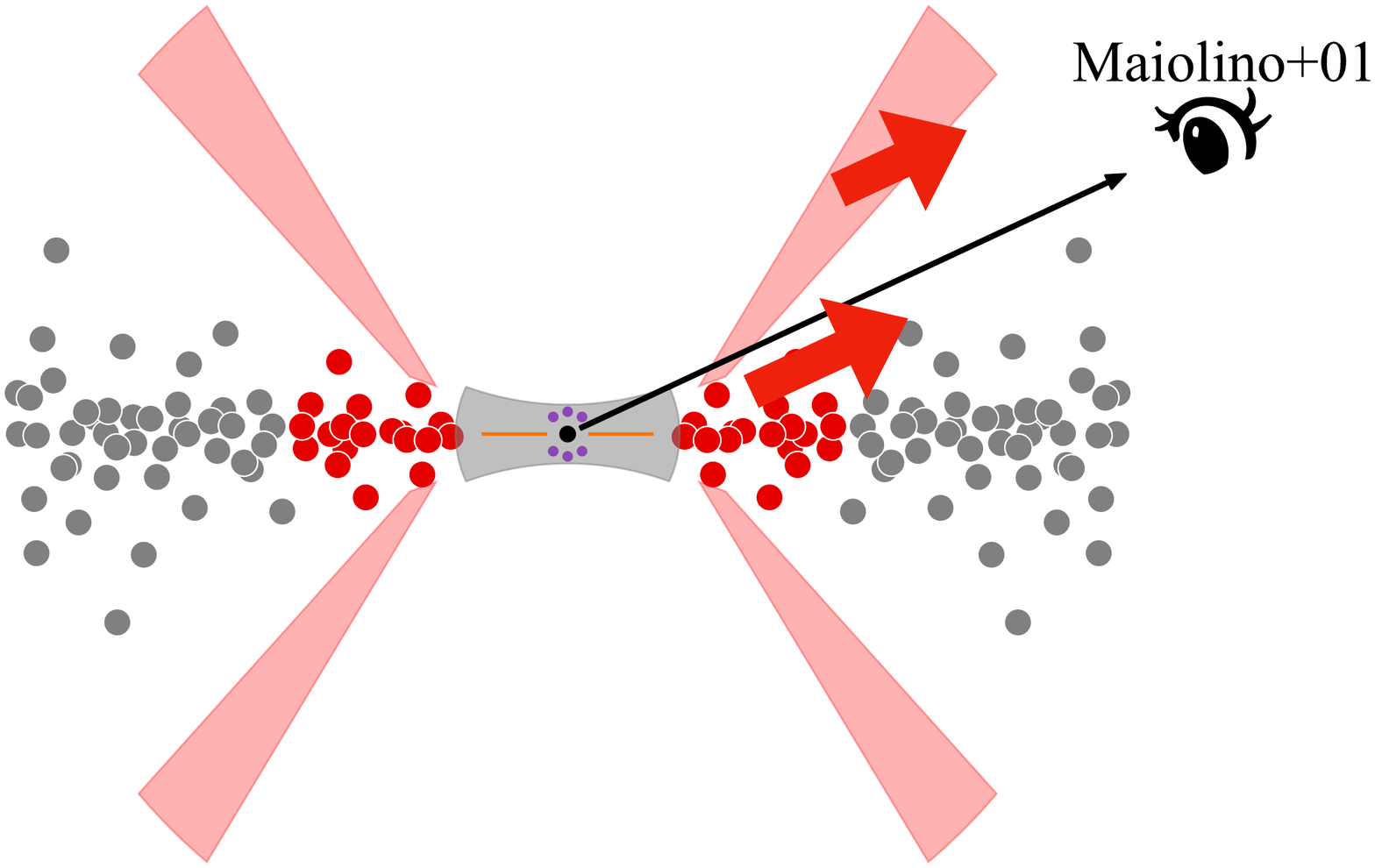}{0.5\textwidth}{(b)}}
\gridline{\fig{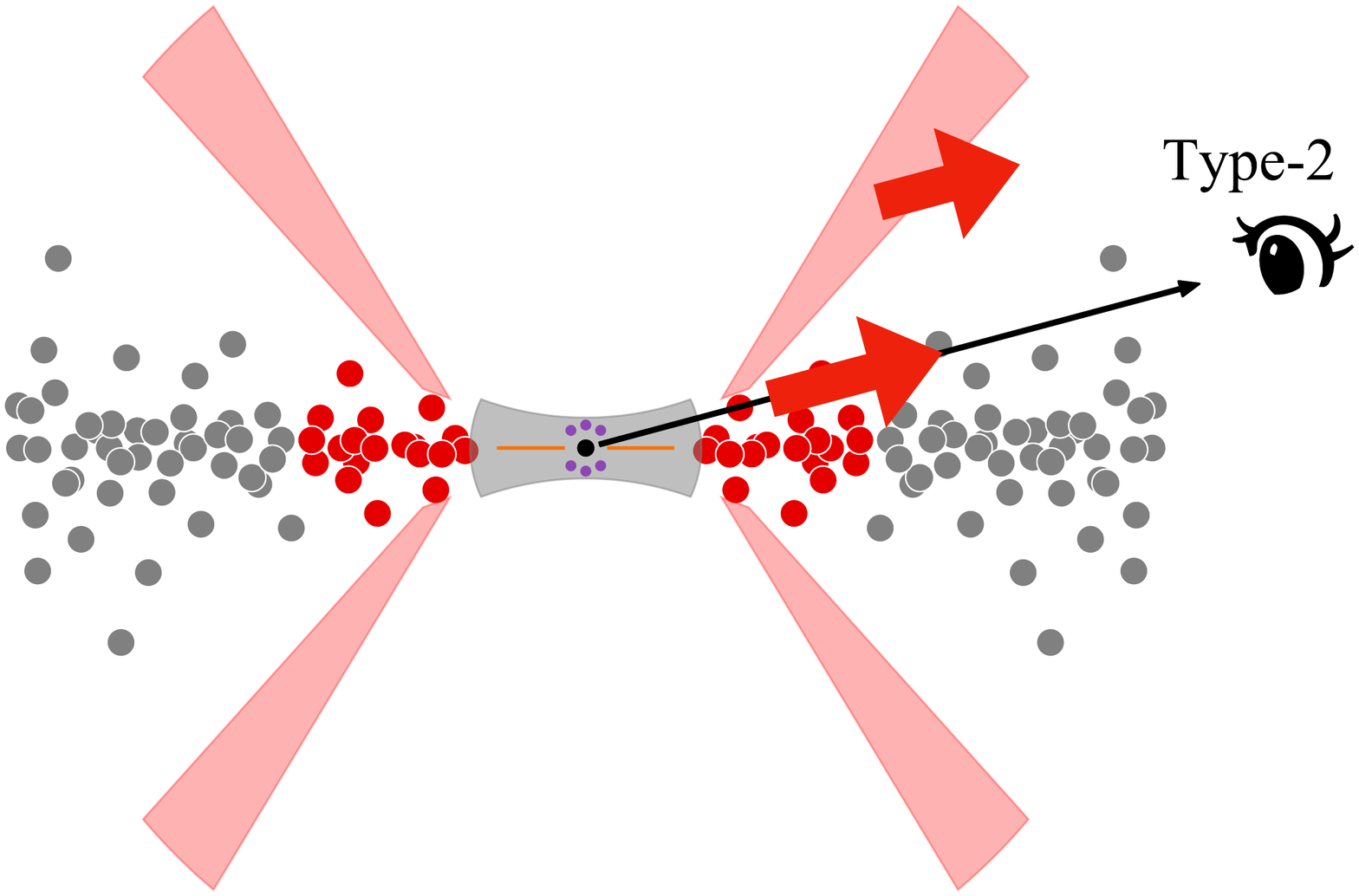}{0.5\textwidth}{(c)}
          \fig{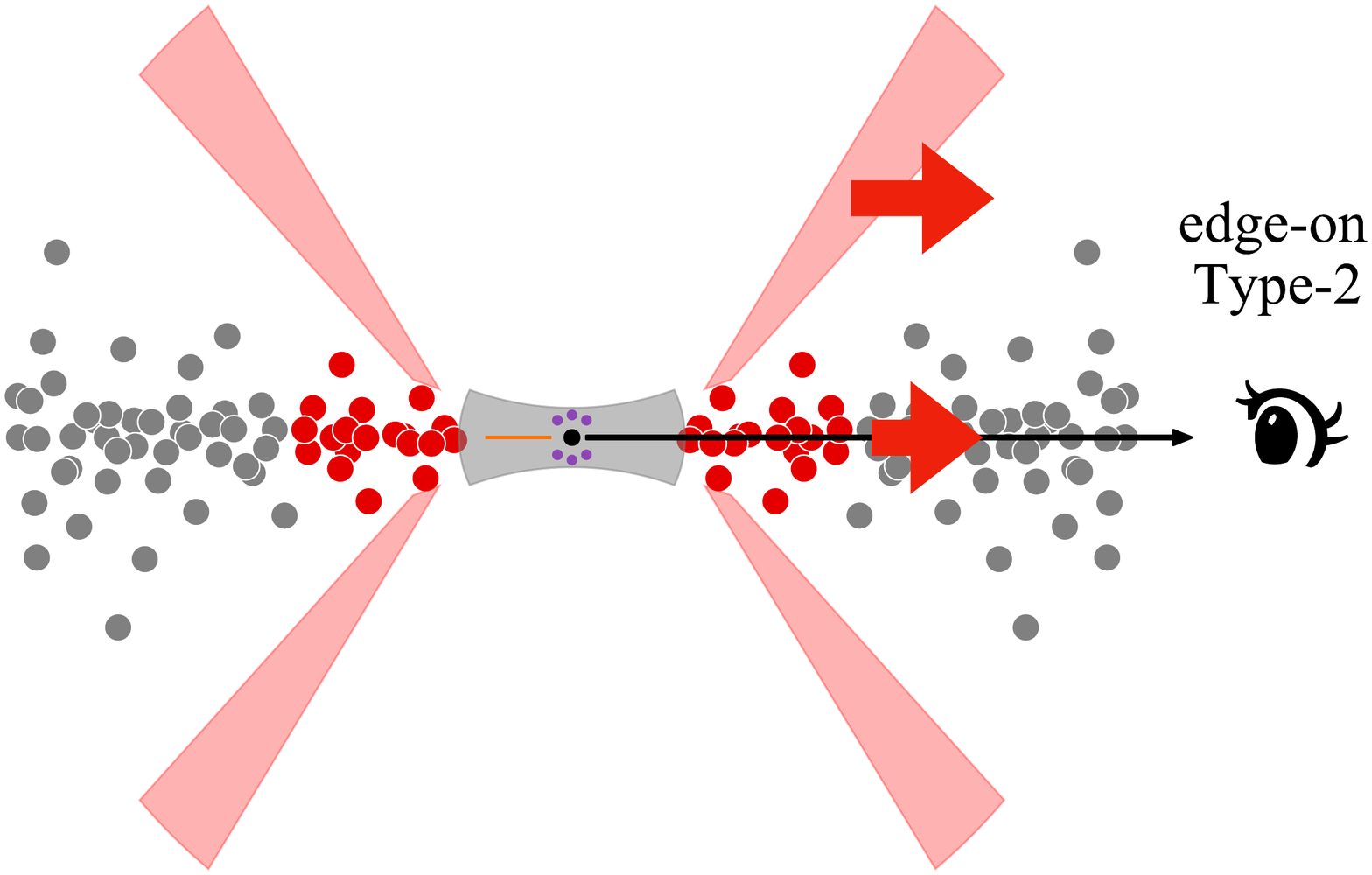}{0.5\textwidth}{(d)}}
\caption{
Unified picture of AGN sructure including a dusty torus (hot clumps in
red and cooler clumps in grey), dusty polar outflows (red shaded
region), and dust-free gas (grey shaded region inside the torus) where
the broad line region (purple circles) is located. The black and red arrows represent the light travel paths of the
X-ray power-law component and the infrared dust emission toward an observer, respectively. 
The inclination increases from (a) to (d). 
} \label{fig-uni}
\end{figure*}

Our findings can be well explained by
an updated unified picture of
AGN structure including (1) a dusty torus, (2) dusty polar outflows, and
(3) dust-free gas inside the dust sublimation radius, where the broad
line region (BLR) is located. As in the classical unified model of AGNs \citep{Antonucci1993}, 
we assume that only the inclination angle determines the X-ray
unobscured/obsured and optical type-1/type-2
classifications. Figures~\ref{fig-uni}(a)-(d) illustrate schematic views of AGN
structure with increasing order of inclination; (case a) X-ray
unobscured and optical type-1 AGNs, (case b) X-ray obscured and optical
type-1 AGNs, (case c) X-ray obscured and optical type-2 (not edge-on
case), and (case d) the same but close to the edge-on case. The X-ray
power-law component is emitted from hot corona close to the SMBH (whose
light path is illustrated by the the arrow), whereas the infrared emission
mainly comes from the inner dusty torus and polar outflows.  The spatial
distribution of the dust-free gas inside the dusty torus is unknown, and
we assume that it has a disk-like geometry (i.e., elongated along the
equatorial plane). Except for the dust-free gas region, the gas-to-dust
ratio is assumed to be close to the Galactic ISM value. Note that the
torus angular width and the direction of the polar outflows in the
figures are arbitrary, which could depend on the AGN parameters such as
the Eddington ratio.

In case (a), i.e., at low inclinations, the
paths along which X-ray and infrared emission travels are largely
different. 
In X-rays, $N_{\rm H}$ 
is constrained by absorption toward the central hot corona, whereas 
the $A_{\rm V}$ parameter in CLUMPY is determined by the infrared
emission from outer dusty regions, not by extinction toward the central
region.
In the CLUMPY model, 
dust optical depths at infrared
wavelengths is converted to $A_{\rm V}$ 
by assuming standard Galactic dust properties. 
The X-ray component shows
little absorption because the column density of the dust-free gas is
small, and optical emission lines from the BLR are not blocked by any
dusty material. Hence it is observed as an X-ray unobscured and optical
type-1 AGN. 
As stated in Section~\ref{sec5.2}, fitting the infrared spectrum 
with CLUMPY (i.e., by ignoring the dusty polar outflows) 
would largely overestimate the torus angular width ($\sigma_{\rm IR}$), 
leading to an overestimate of the line-of-sight $A_{\rm V}$.
Hence, the ratio of $N_{\rm
H}/A_{\rm V}$ {\it along the line of sight} becomes apparently 
much smaller than
the value in the torus region, which is assumed to be similar
to that of the Galactic ISM in our picture.

In case (b), where the inclination is higher than in case (a), the X-ray
spectrum can be absorbed with a moderate column density (e.g.,
$N_{\mathrm{H}} \sim 10^{22}$ cm$^{-2}$) by dust-free gas. In the optical band,
however, the BLR can be barely seen when the line-of-sight path crosses an
upper boundary region of the torus, assuming that extinction by the
dusty polar outflows is unimportant. Hence, it is observed as an X-ray
obscured but optical type-1 AGNs. This may explain (a part of) the AGN
population studied by \citet{Maiolino2001b,Maiolino2001a} that show X-ray absorption but broad
emission lines in the optical spectra, even without invoking dust
properties that are largely different from those in the Galactic ISM.

In case (c), the travel paths of X-rays and infrared emission become
close to each other, both are subject to significant extinction by the
dusty torus. Hence it is observed as an X-ray obscured and optical
type-2 AGN.  In this geometry, the absorption of the X-ray emission is
dominated by the dusty torus (with a column density of e.g.,
$N_{\mathrm{H}} \sim 10^{23}$~cm$^{-2}$), not by the dust-free gas. Hence, the
$N_{\rm H}^{\mathrm{LOS}}/A_{\rm V}^{\mathrm{LOS}}$ ratio represents the value in the torus
region, which is close to the Galactic value as assumed.

In case (d), the very edge-on case, it is also observed as an X-ray
obscured and optically type-2 AGN. In X-rays, it may become a Compton-thick 
AGN with a column density of $N_{\mathrm{H}} \gtrsim 10^{24}$ cm$^{-2}$. As
discussed in \citet{Tanimoto2020}, the infrared emission coming from the
innermost torus region is subject to extinction by cold dust in outer
parts of the torus or the circumnuclear disk, and its flux is more
reduced compared with the previous cases. Then, the relative
contribution of the infrared emission from the dusty polar outflows,
which is located above the torus region and hence is less subject to
extinction, becomes more prominent. The extinction ($A_{\mathrm{V}}$) determined from
the infrared spectrum represents an averaged value for the torus and
polar outflow regions, whereas the column density ($N_{\mathrm{H}}$) determined from
the X-ray spectrum represents that along close to the equatorial plane.
Hence, the $N_{\rm H}^{\mathrm{LOS}}/A_{\rm V}^{\mathrm{LOS}}$ ratio becomes larger than the
Galactic value.

We would like to remark that the proposed picture is still a toy model.
It is deduced by interpreting the results of currently available models
applied independently to the X-ray and infrared data.  For more
quantitative evolution, it is necessary to develop models that correctly
include the contribution from the dusty polar outflows, and
self-consistently apply them simultaneously to both infrared and X-ray
data. This will be left as an important future task.

\section{Conclusions}
\label{sec7}

\begin{enumerate}

\item We have successfully applied the X-ray clumpy torus model XCLUMPY
to the broadband spectra (0.3--70~keV) of 16 AGNs. Combining them with
previous works, we now have a sample of 28 AGNs whose torus properties
are independently estimated from the infrared and X-ray spectra with the
CLUMPY and XCLUMPY codes, respectively.

\item The relation between the Eddington ratio and the torus covering factor
determined from the X-ray torus parameters of each object follows
that derived by \citet{Ricci2017Nature} based on a statistical analysis of a hard
X-ray selected sample.

\item Comparing the torus parameters obtained from the X-ray and
infrared spectra, we confirm the results by \citet{Tanimoto2020} that
(1) the torus angular widths determined by the infrared data are systematically
larger those by the X-ray data, and (2) the $N_{\rm H}/A_{\rm V}$ ratios
along the line of sight in obscured AGNs are similar to the Galactic
value on average with a scatter of $\sim$1 dex. We find that unobscured
AGNs show apparently smaller line-of-sight $N_{\rm H}/A_{\rm V}$ ratios
than the Galactic one, 
which could be explained if the angular torus
widths are overestimated in the infrared spectral analysis.

\item 
Our findings can be well explained by
an updated unified
picture of AGN structure including a dusty torus, dusty polar outflows,
and dust-free gas (Figure~\ref{fig-uni}), where the X-ray and optical
classifications and observed torus properties in the X-ray
and infrared bands 
are determined by the inclination angle.

\end{enumerate}

\appendix

\section{Comparison of Spectral Fitting Results with Previous Studies}
\label{A1}

Since different
spectral models were used in earlier works, we focus on the photon
index of the intrinsic component and the line-of-sight column density
(for obscured AGNs) to check the overall consistency.
Section~\ref{A1}.1--10 and Section~\ref{A1}.11--14 describe the results
of unobscured AGNs and obscured AGNs, respectively.  The results of
IC~4329A and NGC~7469 are similar to those of \citet{Ogawa2019}, in
which comparison with previous studies were discussed in detail.  

\subsection{NGC 2992}

The simultaneous \textit{XMM-Newton} and \textit{NuSTAR} data observed in
2019 May are reported here for the first time. We obtain $\Gamma =
1.71^{+0.01}_{-0.01}$ and $N^{\mathrm{LOS}}_{\mathrm{H}} =
3.64^{+0.14}_{-0.14} \times10^{21}~\rm cm^{-2}$.

\subsection{MCG$-$5--23--16}

Jointly analyzing the \textit{NuSTAR} and \textit{Suzaku} spectra, we
obtain $\Gamma = 1.95^{+0.01}_{-0.01}$ and
$N^{\mathrm{LOS}}_{\mathrm{H}} = 1.37^{+0.01}_{-0.01} \times 10^{22}~\rm
cm^{-2}$. Our results are very similar to those using the same
\textit{Suzaku} data reported by \citet{Zoghbi2017}, $\Gamma =
1.90^{+0.01}_{-0.01}$ and $N^{\mathrm{LOS}}_{\mathrm{H}} =
1.41^{+0.01}_{-0.01} \times10^{22}~\rm cm^{-2}$, and to the
\textit{NuSTAR} results observed in the same epoch ($\Gamma =
1.87^{+0.01}_{-0.01}$ and $N^{\mathrm{LOS}}_{\mathrm{H}} =
1.39^{+0.01}_{-0.01} \times10^{22}~\rm cm^{-2}$). They utilized the
\textsf{xillver} \citep{Garcia2013} and \textsf{relxill} model
\citep{Dauser2014,Garcia2014} to reproduce the distant reflection and
relativistic reflection components, respectively.

\subsection{NGC 3783}

We obtain $\Gamma = 1.64^{+0.03}_{-0.02}$, which is smaller than the
result by \citet{Mao2019}
($\Gamma = 1.75^{+0.02}_{-0.02}$) using the same \textit{NuSTAR} and
\textit{XMM-Newton} data.
We infer that this is because they adopted a different spectral model.
They utilized the \textsf{REFL} model \citep{Magdziarz1995,Zycki1999} in
SPEX \citep{Kaastra1996} for a neutral reflection component, and took
9 warm absorbers into account.

\subsection{UGC 6728}

The photon index $\Gamma = 2.03^{+0.01}_{-0.02}$ well matches with the
result reported by \citet{Walton2013a} using the same \textit{Suzaku} data
($\Gamma = 2.00^{+0.04}_{-0.03}$) despite of the different spectral
modelling. They included a relativistic reflection component, using the
\textsf{REFLIONX} \citep{Ross2005} model convolved with \textsf{RELCONV}\citep{Dauser2013}.

\subsection{NGC 4051}

The simultaneous \textit{XMM-Newton} and \textit{NuSTAR} data observed in
2018 November are reported here for the first time. We obtain $\Gamma =
2.02^{+0.01}_{-0.01}$ and $N^{\mathrm{LOS}}_{\mathrm{H}} < 2.22 \times10^{19}~\rm cm^{-2}$. The source was in 
a high flux state \citep[e.g.,][]{Pounds2004}. 

\subsection{NGC 4395}

We obtain $\Gamma < 1.53$ and
$N^{\mathrm{LOS}}_{\mathrm{H}} = 5.15^{+3.53}_{-3.21} \times10^{20}~\rm
cm^{-2}$. The photon index is consistent with that reported by
\citet{Kawamuro2016LLAGN} using the same \textit{Suzaku} data ($\Gamma =
1.49^{+0.15}_{-0.10}$). The line-of-sight absorption by the torus is
smaller than their result ($N^{\mathrm{LOS}}_{\mathrm{H}} =
1.60^{+0.20}_{-0.19} \times10^{22}~\rm cm^{-2}$) because
\citet{Kawamuro2016LLAGN} did not consider an ionized absorber.

\subsection{MCG$-$6--30--15}

We obtain $\Gamma = 2.16^{+0.02}_{-0.01}$. This is in good agreement
with the result reported by \citet{Marinucci2014} using the same \textit{XMM-Newton} and \textit{NuSTAR}
data ($\Gamma = 2.16^{+0.01}_{-0.01}$). The slight difference comes
from the different spectral modelling; \citet{Marinucci2014} considered a distant
reflection component from ionized material using the \textsf{xillver}
model and 5~absorbers.

\subsection{NGC 6814}

We obtain $\Gamma = 1.60^{+0.05}_{-0.03}$ and
$N^{\mathrm{LOS}}_{\mathrm{H}} < 2.2 \times10^{20}~\rm
cm^{-2}$. 
These are different from the results based on the same
\textit{Suzaku} data reported by \citet{Walton2013b}, $\Gamma =
1.53^{+0.02}_{-0.02}$ and $N^{\mathrm{LOS}}_{\mathrm{H}} =
3.4^{+1.5}_{-1.5} \times10^{20}~\rm cm^{-2}$. 
The difference comes
from the different spectral modelling; they used the
\textsf{REFLIONX} code for a reflection component
and did not consider a partial absorber.

\subsection{NGC 7213}

We obtain $\Gamma = 1.63^{+0.10}_{-0.05}$. 
Our results are consistent with those reported by
\citet{Patrick2012} using the same \textit{Suzaku} data and \textit{swift}/BAT data, $\Gamma =
1.74^{+0.01}_{-0.01}$, within the errors. 
They used the \textsf{REFLIONX} model for a
reflection component.

\subsection{NGC 7314}

We obtain $\Gamma = 1.99^{+0.02}_{-0.01}$ and
$N^{\mathrm{LOS}}_{\mathrm{H}} = 4.95^{+0.20}_{-0.24} \times10^{21}~\rm
cm^{-2}$. These are slightly different with the 
\textit{NuSTAR} results \citep{Panagiotou2019}, $\Gamma = 2.09^{+0.01}_{-0.02}$
and $N^{\mathrm{LOS}}_{\mathrm{H}} = 1.1^{+0.2}_{-0.2} \times10^{22}~\rm
cm^{-2}$. We infer that because they 
analyzed only the \textit{NuSTAR} data 
and did not consider ionized absorbers, 
a larger absorption than ours was obtained.

\subsection{NGC 3081}

The model with two \textsf{apec} components well reproduces the broadband
X-ray spectra. We obtain $\Gamma > 1.56$ and
$N^{\mathrm{LOS}}_{\mathrm{H}} = 7.77^{+1.22}_{-0.76} \times10^{23}~\rm
cm^{-2}$. Our results are consistent with the results reported by
\citet{Kawamuro2016a} using the same \textit{Suzaku} data, $\Gamma =
1.73^{+0.05}_{-0.05}$ and $N^{\mathrm{LOS}}_{\mathrm{H}} =
8.25^{+0.40}_{-0.38} \times10^{23}~\rm cm^{-2}$.

\subsection{NGC 4388}

The model with two \textsf{apec} components well reproduces the
broadband \textit{NuSTAR}+\textit{Suzaku} spectrum. Our best fitting parameters are
$\Gamma < 1.51$ and $N^{\mathrm{LOS}}_{\mathrm{H}} = 3.02^{+0.05}_{-0.05}
\times10^{23}~\rm cm^{-2}$.  Analyzing the \textit{Suzaku} XIS$+$HXD spectra,
\citet{Kawamuro2016a} obtained $\Gamma = 1.65^{+0.01}_{-0.01}$ and
$N^{\mathrm{LOS}}_{\mathrm{H}} = 2.38^{+0.07}_{-0.07} \times10^{23}~\rm
cm^{-2}$. Using the \textit{NuSTAR} data, \citet{Masini2016} obtained $\Gamma =
1.65^{+0.08}_{-0.08}$ and $N^{\mathrm{LOS}}_{\mathrm{H}} =
4.4^{+0.6}_{-0.6} \times10^{23}~\rm cm^{-2}$.  \citet{Kawamuro2016a}
utilized the \textsf{pexrav} model \citep{Magdziarz1995}, and \citet{Masini2016} the MYTorus
model \citep{Murphy2009}.  Our photon index is slightly smaller than these results. We
infer that this is because the XCLUMPY model contains a strong
unabsorbed (hence soft) reflected continuum escaped through clumps in
the near-side torus, which works to make the intrinsic spectrum harder
(see the dicussion in \citealt{Tanimoto2019}).

\subsection{Centaurus A}

Results utilizing the \textit{Suzaku} data are reported here for the
first time. The model with one \textsf{apec} component well reproduces
the broadband \textit{NuSTAR}$+$\textit{Suzaku} spectrum. We obtain $\Gamma =
1.76^{+0.01}_{-0.01}$ and $N^{\mathrm{LOS}}_{\mathrm{H}} =
9.87^{+0.15}_{-0.08} \times10^{22}~\rm cm^{-2}$. Our photon index is
slightly smaller from the \textit{NuSTAR} result; 
\citet{Furst2016} obtained $\Gamma = 1.82^{+0.01}_{-0.01}$ and
$N^{\mathrm{LOS}}_{\mathrm{H}} = 1.11^{+0.15}_{-0.02} \times10^{23}~\rm 
cm^{-2}$, utilizing the MYTorus model. This may be explained by the same
reason as for NGC~4388.

\subsection{NGC 6300}

The model with one \textsf{apec} component well reproduces the broadband
\textit{NuSTAR}$+$\textit{Suzaku} spectrum.  We obtain $\Gamma = 1.79^{+0.05}_{-0.06}$ and
$N^{\mathrm{LOS}}_{\mathrm{H}} = 2.09^{+0.07}_{-0.06} \times10^{23}~\rm
cm^{-2}$, which are in good agreement with the \textit{Suzaku}
results reported by \citep{Kawamuro2016a}, $\Gamma =
1.86^{+0.02}_{-0.02}$ and $N^{\mathrm{LOS}}_{\mathrm{H}} =
2.22^{+0.04}_{-0.03} \times10^{23}~\rm cm^{-2}$.  Our photon index is
slightly smaller from that reported by \citep{Panagiotou2019} using the
\textit{NuSTAR} data, $\Gamma = 1.90^{+0.03}_{-0.03}$ and
$N^{\mathrm{LOS}}_{\mathrm{H}} = 2.07^{+0.05}_{-0.04} \times10^{23}~\rm
cm^{-2}$. 
Since they used the \textsf{pexrav} model for the reflection component,
the discrepancy may be also explained by the same reason as for NGC~4388.

\section{Inclination Angle Dependence of Torus Covering Factor in Unobscured AGNs}
\label{A2}

\begin{figure*}
\gridline{
\fig{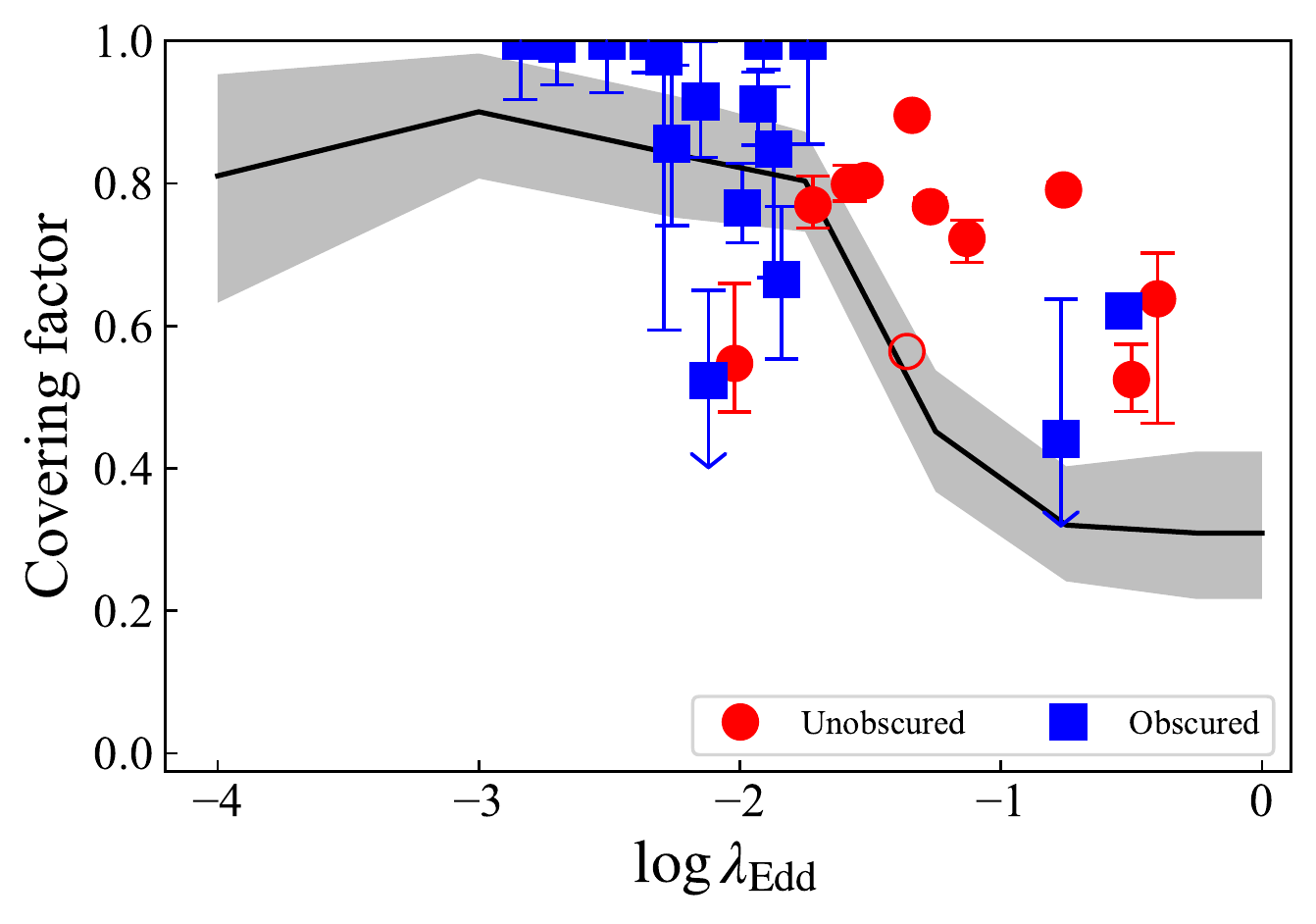}{0.5\textwidth}{(a)}
\fig{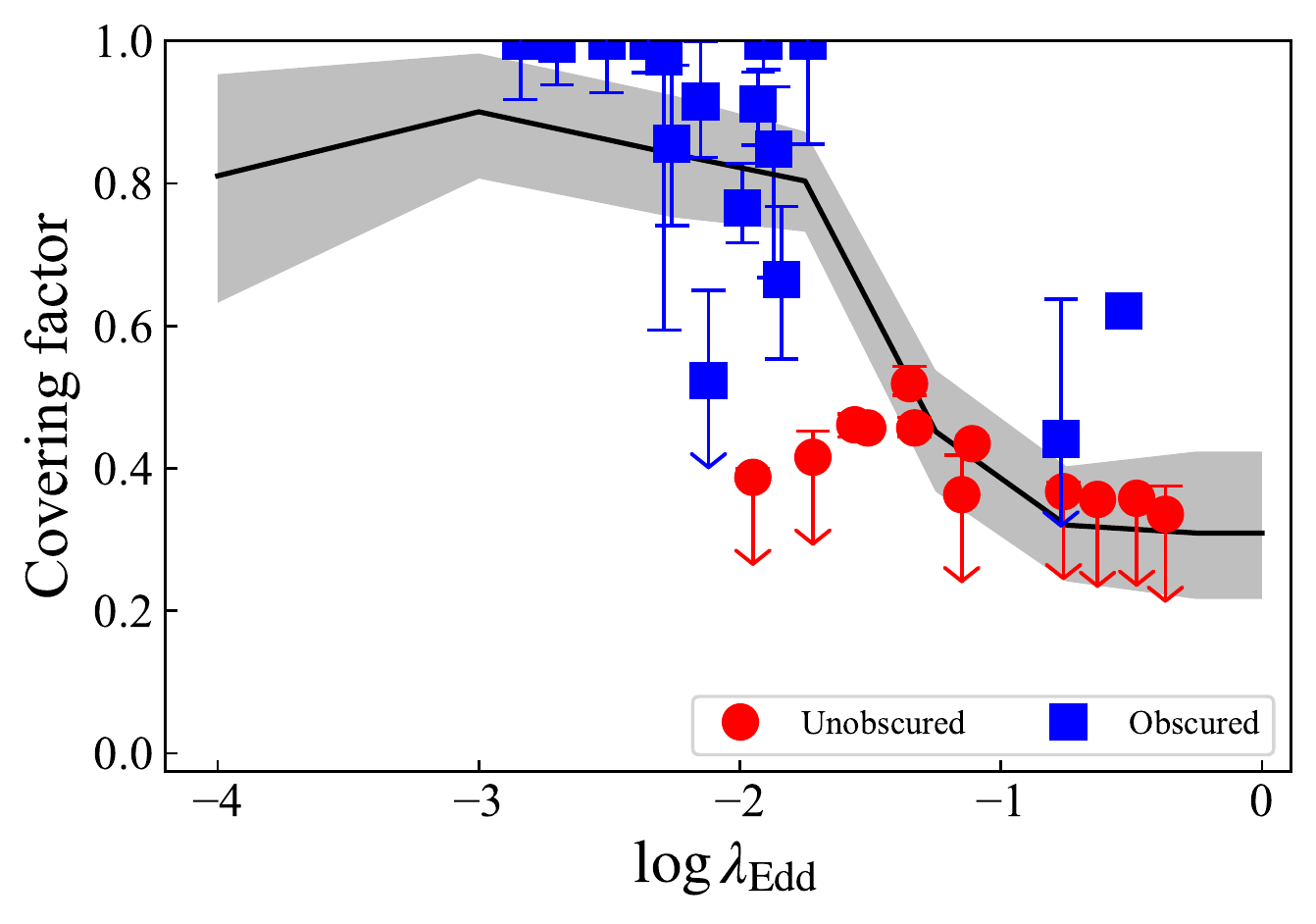}{0.5\textwidth}{(b)}
}
\caption{
Relation between the torus covering factor ($C_\mathrm{T}$) and the
Eddington ratio ($\lambda_{\mathrm{Edd}})$. The black curve and shaded
region represent the best-fit and 1$\sigma$ error region obtained by
\citet{Ricci2017Nature}.
The blue squares and red circles denote the obscured and unobscured AGNs,
respectively.
The arrows represent the results reaching the upper or lower boundary.
(a): The inclination angle ($i_\mathrm{X}$) of the unobsucred AGNs is set
 to 30$^{\circ}$. 
The empty circle denotes the best-fit $C_\mathrm{T}$ value of NGC~6814,
whose lower and upper limits reach the
boundaries (0 and 1) at $<$90\% confidence level.
(b): $i_\mathrm{X}$ is set to 60$^{\circ}$.
}
\label{figA-edd}
\end{figure*}

To investigate the dependence of the torus 
parameters on the assumed inclination angle
($i_\mathrm{X}$) for
unobscured AGNs, we fit the spectra by fixing $i_\mathrm{X}$ at 30$^{\circ}$ and
60$^{\circ}$. 
We find that the mean values of the torus angular width, logarithmic column
density along the equatorial plane, and torus covering factor are
($\langle \sigma_\mathrm{X} \rangle$, $\langle \log N^{\mathrm{Equ}}_{\mathrm{H}}/\mathrm{cm}^{-2} \rangle$, $ \langle C_\mathrm{T} \rangle$) 
$=$ ($22^{\circ}$, 24.2, 0.70) for $i_\mathrm{X}=30^{\circ}$,
($17^{\circ}$, 24.2, 0.58) for $i_\mathrm{X}=45^{\circ}$, and
($12^{\circ}$, 23.9, 0.41) for $i_\mathrm{X}=60^{\circ}$.
Figure~\ref{figA-edd}(a) and (b) plot 
$C_\mathrm{T}$ against $\lambda_{\rm Edd}$ for $i_\mathrm{X} =
30^{\circ}$ and $i_\mathrm{X} = 60^{\circ}$, respectively. 
As noticed, we find general trends that $C_\mathrm{T}$ are overestimated
for $i_\mathrm{X} = 30^{\circ}$ and underestimated for $i_\mathrm{X}
= 60^{\circ}$
assuming that the relation of \citet{Ricci2017Nature} holds for our sample.
Thus, we suggest that the assumption of $i_\mathrm{X} =45^{\circ}$ 
adopted in the main paper is the most reasonable among the three
cases.
It is remarkable that $i_\mathrm{X} = 30^{\circ}$, often adopted as a 
typical inclination for unobscured AGNs, 
cannot reproduce the \citet{Ricci2017Nature} relation.

\acknowledgements

We thank the referee for the useful comments and suggestions.
This work has been financially supported by the Grant-in-Aid for
Scientific Research 17K05384 and 20H01946 (Y.U.) and for JSPS Research
Fellowships (A.T. and S.Y.). This research has made use of the
\textit{NuSTAR} Data Analysis Software (NUSTARDAS) jointly developed by
the ASI Science Data Center (ASDC, Italy and the California Institute of
Technology (Caltech, USA).  This research has also made use of data
obtained with \textit{XMM-Newton}, an ESA science mission with
instruments and contributions directly funded by ESA Member States and
NASA, and use of public \textit{Suzaku} data obtained through the Data
ARchives and Transmission System (DARTS) provided by the Institute of
Space and Astronautical Science (ISAS) at the Japan Aerospace
Exploration Agency (JAXA), and use of the NASA/IPAC Extragalactic
Database (NED), which is operated by the Jet Propulsion Laboratory,
California Institute of Technology, under contract with the National
Aeronautics and Space Administration.  For data reduction, we used
software provided by the High Energy Astrophysics Science Archive
Research Center (HEASARC) at NASA/Goddard Space Flight Center.

\facilities{\textit{XMM-Newton}, \textit{Suzaku}, \textit{NuSTAR}.}

\software{HEAsoft \citep[v6.26.1;][]{HEAsoft2014}, SAS \citep[v17.0.0;][]{Gabriel2004}, NUSTARDAS, XSPEC \citep[v12.10.1f;][]{Arnaud1996}, XCLUMPY \citep{Tanimoto2019}.}

\clearpage
\bibliographystyle{aasjournal}
\bibliography{reference}

\end{document}